\documentclass[journal,twoside,web]{ieeecolor}
\usepackage{cite}

\usepackage{generic}
\usepackage{multicol}
\usepackage{multirow}
\usepackage{xcolor}
\usepackage{amsmath}
\usepackage{amsfonts}
\usepackage{amssymb}
\usepackage[export]{adjustbox}
\usepackage{subcaption}

\usepackage{amsthm}
\theoremstyle{definition}
\newtheorem{theorem}{Theorem}[section]

\newtheorem{assumption}[theorem]{\textbf{Assumption}}
\newtheorem{lemma}[theorem]{Lemma}

\theoremstyle{remark}
\newtheorem{claim}{Claim}[theorem]

\usepackage{graphicx}
\usepackage{accents}
\newcommand{\ubar}[1]{\underaccent{\bar}{#1}}
\usepackage{mathtools}
\usepackage[ruled,vlined,noend]{algorithm2e}
\usepackage{cite}
\usepackage{amsmath,amssymb,amsfonts}
\usepackage{graphicx}
\usepackage{textcomp}
\usepackage{algpseudocode}
\usepackage{makecell}
\newcolumntype{P}[1]{>{\centering\arraybackslash}p{#1}}
\algrenewcommand\textproc{}
\usepackage[english]{babel}
\usepackage[utf8]{inputenc}
\algnewcommand{\parState}[1]{\State%
    \parbox[t]{\dimexpr\linewidth-\algmargin}{\strut\hangindent= \algorithmicindent \hangafter=1 #1\strut}}

\def\BibTeX{{\rm B\kern-.05em{\sc i\kern-.025em b}\kern-.08em
    T\kern-.1667em\lower.7ex\hbox{E}\kern-.125emX}}
\begin{document}
\title{\LARGE \bf Distributed Safe Learning and Planning for Multi-robot Systems}

\author{Zhenyuan Yuan$^{1}$ \qquad Minghui Zhu$^{2}$
	\thanks{$^{1}$Zhenyuan Yuan is with the Virginia Tech Transportation Institute, Virginia Polytechnic Institute and State University, Blacksburg, VA 24061, USA (email:{\tt\small zyuan18@vt.edu}). }
		\thanks{$^{2}$Minghui Zhu is with School of Electrical Engineering and Computer Science, Pennsylvania State University, University Park, PA 16802, USA (email:
		{\tt\small \{zqy5086,muz16\}@psu.edu}).  }
		\thanks{This work was partially supported by NSF grants ECCS-1710859, CNS-1830390, ECCS-1846706 and the Penn State College of Engineering Multidisciplinary Research Seed Grant Program.}
}
\maketitle
\begin{abstract}
This paper considers the problem of online multi-robot motion planning with general nonlinear dynamics subject to unknown external disturbances. We propose dSLAP, a \textbf{d}istributed \textbf{S}afe \textbf{L}earning \textbf{A}nd \textbf{P}lanning framework that allows the robots to safely navigate through the environments by coupling online learning and motion planning. Gaussian process regression is used to online learn the  disturbances with uncertainty quantification. The planning algorithm ensures collision avoidance against the learning uncertainty and utilizes set-valued analysis  to achieve fast adaptation in response to the newly learned models. A set-valued model predictive control problem is formulated and solved to return a control policy that balances between actively exploring the unknown disturbances and reaching goal regions. Sufficient conditions are established to guarantee the safety of the robots. Monte Carlo simulations are conducted for evaluation.
\end{abstract}

\section{Introduction}
Intelligent multi-robot systems are becoming ubiquitous in our life, such as autonomous driving, delivery, precision agriculture and search-and-rescue. Robots operating in the real world are usually accompanied by unknown disturbances. In order to guarantee the safety of the robots as well as mission completion, it is crucial for the robots to be able to adapt to the disturbances online and update their motion plans accordingly.

Motion planning is a fundamental problem in robotics, and it aims to generate a series of low-level specifications for a robot to move from one point to another \cite{lavalle2006planning}. 
 In
 the real world, robots’ operations are usually subject to uncertainties, e.g., from the environments they
 operate in and from the errors in the modeling of robots’
 dynamics. To deal with the uncertainties and ensure safety, i.e., collision avoidance,  existing methods leverage techniques in robust control, e.g., \cite{lindemann2021robust,cohen2022robust,lakshmanan2020safe}, stochastic control, e.g., \cite{omainska2021gaussian,ono2015chance,castillo2020real}, and learning-based control, e.g., \cite{virani2018imitation,gupta2017cognitive,levine2016end,majumdar2018pac}. Robust control-based approaches model the uncertainties as bounded sets and synthesize control policies that tolerate all the uncertainties in the sets.
 Note that considering all possible events can result in over-conservative policies whereas extreme events may only take place rarely. Stochastic control-based approaches  model the uncertainties as known probability distributions.
 The generated motion plans enforce chance constraints, i.e., the probability of collision is less than a given threshold.  On the other hand, learning-based approaches relax the need of prior explicit uncertainty models by directly learning the best mapping from  sensory inputs to  control inputs from repetitive trials. Paper \cite{majumdar2018pac} leverages PAC-Bayes theory to provide guarantees on expected performances over a distribution of environments. 

The aforementioned approaches can all be classified as offline approaches where control policies are synthesized before the deployment of robots. When robots encounter significant changes of environments during online operation, online learning of the uncertainties is desired to ensure safe arrival to the goals. Recently, a class of methods on safe learning and control have been developed to safely steer a system to a goal region while  learning uncertainties online by adjusting learning-based controllers accordingly. For example, paper \cite{fisac2018general} uses Gaussian process regression (GPR) to learn about the disturbances and synthesizes the backup controller by solving a two-player zero-sum differential game between the  controller and the estimated disturbances; the overall control policy switches between the online learned controller and the online learned backup controller. Model predictive control (MPC) is leveraged in \cite{wabersich2021probabilistic} to  minimally adjust the learning-based controller to ensure all future states admit a safety controller  which can safely connect the system to a set of terminal states under (learned) stochastic uncertainties.
Reachability analysis using polynomial zonotopes is conducted in \cite{kochdumper2023provably} for a finite horizon under uncertainties with known bounds to identify safe control inputs, and safety is achieved by projecting learned control inputs to the closest safe control inputs. In \cite{brunke2022barrier}, the uncertain control barrier function (CBF) condition of the ground truth system is learned using Bayesian linear regression, and safety is achieved by projecting a given control input to the set of control inputs satisfying the learned CBF condition.
The aforementioned papers only consider single-robot systems and static state constraints (e.g., static obstacles). In multi-robot systems, from the perspective of each single robot and when centralized planning is not used, the state constraints are dynamic due to the motion of the other robots, analogous to moving obstacles.

 Motion planning problems are known to be computationally challenging even for single-robot systems. Paper \cite{reif1979complexity} shows that the generalized  mover's problem is PSPACE-hard in terms of degrees of freedom.  Multi-robot motion planning is an even more challenging problem as the computational complexity scales up by the number of robots.
Centralized planners \cite{schwartz1983piano}\cite{zhao2020pareto} consider all the robots as a single entity such that the methods for single-robot motion planning can be directly applied. However, as \cite{schwartz1983piano} points out, its worst-case computation complexity grows exponentially with the number of robots. Consequently, distributed methods are developed to address the scalability issue. Most of these methods are featured with each robot conducting a single-robot motion planning strategy but coupled with a coordination scheme to resolve conflicts. These works can be categorized as fully synthesized design or switching-based design.  A fully synthesized design \cite{dimarogonas2006feedback}\cite{dimarogonas2008connectedness} incorporates simple collision avoidance methods, such as artificial potential field, into the decoupled solution. 
Under switching-based design, a switching controller is developed such that each robot executes a nominal controller synthesized in a decoupled manner but switches to a local coordination controller when it is close to other robots \cite{wang2017safety,zhao2022scalable,bekris2012safe}. Prioritized planning, where a unique priority level is assigned to each robot such that the robots with lower priority levels make compromises with the robots with higher priority levels, is further adopted to reduce the need of coordination \cite{ma20173}\cite{van2011lqg}.  
There have been recent works which study dynamic and environmental uncertainties in multi-robot motion planning. For example, robust control-based approaches are studied in \cite{pan2020augmenting}\cite{zhou2018distributed}, and stochastic control-based approaches are studied in \cite{saravanos2021distributed}\cite{zhu2022decentralized}\cite{cheng2020safe}.
In \cite{long2018towards}\cite{fan2020distributed}, deep reinforcement learning is applied to train multiple robots in an offline manner to avoid collisions  when explicit uncertainty models are not available. 
 In this paper, we consider learning the uncertainties in an online fashion with data collected sequentially on the robots' trajectories.

\textbf{Contribution statement.} 
We consider the problem of online multi-robot motion planning with general nonlinear dynamics subject to unknown external disturbances.
We propose dSLAP, the \textbf{d}istributed \textbf{S}afe \textbf{L}earning \textbf{A}nd \textbf{P}lanning framework, where each robot collects streaming data to online learn about its own dynamics subject to the disturbances using GPR \cite{williams2006gaussian} and uses the learned models to synthesize a sequence of set-valued model predictive controllers (SVMPCs) \cite{risso2021set} to reach its goal taking all other robots into account as obstacles. A priorization strategy \cite{van2011lqg} is applied to reduce the complexity in multi-robot coordination, and the multi-robot planning problem is transformed into a
	planning problem for one robot under dynamic obstacles. In the communication each robot exchanges an overapproximation of its reachable sets. 
A major novelty of the proposed framework is the leverage of set-valued analysis \cite{cardaliaguet1999set} to compute the reachable sets of the robots, which are used to quickly identify  safe actions that avoid collisions for the synthesis of an  SVMPC for  motion planning. It is summarized as follows:
\begin{itemize}
\item   The  planner first utilizes discrete set-valued analysis to  construct a robot's  one-step forward reachable sets, and then obtains a set of safe control inputs by removing the control inputs leading to collisions using prioritized planning. 
    Then distributed SVMPCs are synthesized to select safe control inputs  balancing between moving towards the goals and actively learning the disturbances. 
    \item Our two-stage motion planning is in contrast to the classic formulation \cite{GZ-MZ:TAC18}\cite{cardaliaguet1999set} of optimal multi-robot motion planning, whose solutions solve collision avoidance and optimal arrival simultaneously and are known to be computationally challenging (PSPACE-hard \cite{reif1979complexity}). Instead, dSLAP first solves for collision avoidance and then for optimal arrival. The worst-case onboard computational complexity of each robot grows linearly with respect to the number of the robots.
    \item We derive the sufficient conditions to guarantee the safety of the robots in the absence of  backup policies. 
    \end{itemize}
Monte Carlo simulations are conducted to evaluate the proposed framework.

{\bf Distinction statement.}
This paper substantially distinguishes itself from the preliminary version \cite{Yuan2022dSLAP} in the following aspects: (i)
Detailed and structured proofs of theoretical results are included in Section \ref{sec: proof}. The proofs show how collision avoidance is achieved through the dSLAP framework,  how safety is related to the number of robots and the initial conditions of the robots.
(ii) A new set of theoretical results regarding the all-time safety of the system is detailed in Theorem \ref{thm2: system safe}. It provides guidance on the design of the algorithm, requirements on the computation speed of the robots and the implication on the density of the robots.  (iii) Section \ref{sec: discussion} includes additional discussions on the verification of the initial condition, the comparison of the two theorems and the computational complexity of the algorithm. (iv) Additional simulation results are provided in Fig. \ref{fig: safe grid vs safe region}, and comparisons with three benchmarks are conducted.

{\em Notations.} We use superscript $(\cdot)^{[i]}$ to distinguish the local values of robot $i$.
Define the distance metric $\rho(x,x')\triangleq\|x-x'\|_\infty$,  the point-to-set distance as $\rho(x, \mathcal{S})\triangleq\inf_{x'\in\mathcal{S}}\rho(x,x')$ for a set $\mathcal{S}$, the closed ball centered at $x\in\mathbb{R}^{n_x}$ with radius $r$ as $\mathcal{B}(x,r)\triangleq\{x'\in\mathbb{R}^{n_x}|\rho(x,x')\leqslant r\}$, and shorthand $\mathcal{B}$  the closed unit ball centered at $0$ with radius $1$.  Let $\mathbb{Z}$ be the space of integers and $\mathbb{N}$ the space of natural number.
Denote the cardinality of a set $\mathcal{S}$ as $|\mathcal{S}|$.

Below are the implementations of common procedures.
\textit{Element removal}:  Given a set $\mathcal{S}$ and an element $s$, procedure $\textsf{Remove}$  removes element $s$ from $\mathcal{S}$; i.e., $\textsf{Remove}(\mathcal{S},s)\triangleq \mathcal{S}\setminus\{s\}$. 
\textit{Element addition}: Given a set $\mathcal{S}$ and an element $s$, procedure \textsf{Add}  appends $s$ to $\mathcal{S}$, i.e., $\textsf{Add}(\mathcal{S},s)\triangleq \mathcal{S}\cup\{s\}$.
\textit{Nearest neighbor}: Given a state $s$ and a finite set $\mathcal{S}$, \textsf{Nearest}  chooses a state in $\mathcal{S}$ that is closest to $s$; i.e., $\textsf{Nearest}(s,\mathcal{S})$ picks $y\in\mathcal{S}$, where $\rho(s,y)=\rho(s,\mathcal{S})$.

\section{Problem Formulation}\label{sec:problem statement}
In this section, we introduce the model of a multi-robot system, describe the formulation of the motion planning problem, and state the objective of this paper. 

\emph{Mobile multi-robot system.} Consider a network of robots $\mathcal{V}\triangleq\{1,\cdots, n\}$. The dynamic system of each robot $i$ is given by the following differential equation:
\begin{equation}\label{eq: observation model}
    \dot{x}^{[i]}(t)=f^{[i]}(x^{[i]}(t),u^{[i]}(t))+g^{[i]}(x^{[i]}(t),u^{[i]}(t)),
\end{equation}
where $x^{[i]}(t)\in\mathcal{X}\subseteq  \mathbb{R}^{n_x}$ is the state of robot $i$ at time $t$,  $u^{[i]}(t)\in\mathcal{U}\subseteq\mathbb{R}^{n_u}$ is its control input, $f^{[i]}$ denotes the nominal system dynamics of robot $i$,  and $g^{[i]}$ represents the external unknown disturbance.  We impose the following assumptions:
\begin{assumption}\label{assmp: model}
\begin{enumerate}
    \item[\bf{(A1)}] {(\em Lipschitz continuity).} The system dynamics $f^{[i]}$ and the unknown disturbance $g^{[i]}$ are locally Lipschitz continuous  on $\mathcal{X}$ and $\mathcal{U}$.
    \item[\bf{(A2)}] {(\em Compactness).} Spaces $\mathcal{X}$ and  $\mathcal{U}$ are compact.
   $\hfill\blacksquare$
\end{enumerate}

\end{assumption}

Assumption {\bf (A1)} implies that $f^{[i]}+g^{[i]}$ is  Lipschitz continuous on $\mathcal{X}$ and  $\mathcal{U}$. Choose constant $\ell^{[i]}$, which is larger than the Lipschitz constant of $f^{[i]}+g^{[i]}$
and constant  $m^{[i]}$, 
which is larger than
the supremum of $f^{[i]}+g^{[i]}$ over $\mathcal{X}$ and $\mathcal{U}$.  These constants can typically be estimated using principles such as the Bernoulli equation \cite{batchelor2000introduction}, which bounds the variation and magnitude of wind speed based on limited variations in temperature and air pressure. 

{\em Motion planning}.  We denote  closed obstacle region by $\mathcal{X}_O\subseteq \mathcal{X}$, goal region by $\mathcal{X}^{[i]}_G \subseteq \mathcal{X}\setminus \mathcal{X}_O$, and free region at time $t$ by $\mathcal{X}^{[i]}_F(x^{[\neg i]}(t))\triangleq\mathcal{X}\setminus \big( \mathcal{X}_O\bigcup \cup_{j\neq i}\mathcal{B}(x^{[j]}(t),2\zeta)\big)$, where $\neg i\triangleq\mathcal{V}\setminus\{i\}$ and  $\zeta>0$ is the radius of an overestimation of the robot size. Each robot $i$ aims to synthesize a feedback policy $\pi^{[i]}:\mathcal{X}^n\to\mathcal{U}$ such that the solution to system \eqref{eq: observation model} under $\pi^{[i]}$
achieves safe arrival:
$x^{[i]}(t_*^{[i]})\in\mathcal{X}^{[i]}_G$, $x^{[i]}(\tau)\in\mathcal{X}^{[i]}_F(x^{[\neg i]}(\tau))$, $ 0\leqslant\tau\leqslant t_*^{[i]}<\infty$, 
where $t_*^{[i]}$ is the first time when robot $i$ reaches $\mathcal{X}_G^{[i]}$.
That is, each robot $i$ needs to reach the goal region within finite time and be free of collision. 


\emph{Problem statement.} This paper aims to solve the above multi-robot motion planning problem despite unknown external disturbance $g^{[i]}$.   Since $g^{[i]}$ is unknown, each robot needs to learn its own $g^{[i]}$ online to ensure reaching the goal safe and fast. The challenge of the problem stems from the need of fast  distributed planning with respect to a sequence of general nonlinear dynamic models resulting from online learning subject to dynamic constraints. Specifically, since the unknown  $g^{[i]}$ is learned online, each robot $i$ should quickly adapt its motion planner in response to a sequence of  newly learned 
models and the motion of the other robots.

\section{Distributed safe learning and planning}
\begin{figure}
    \centering
    \includegraphics[width=0.45\textwidth]{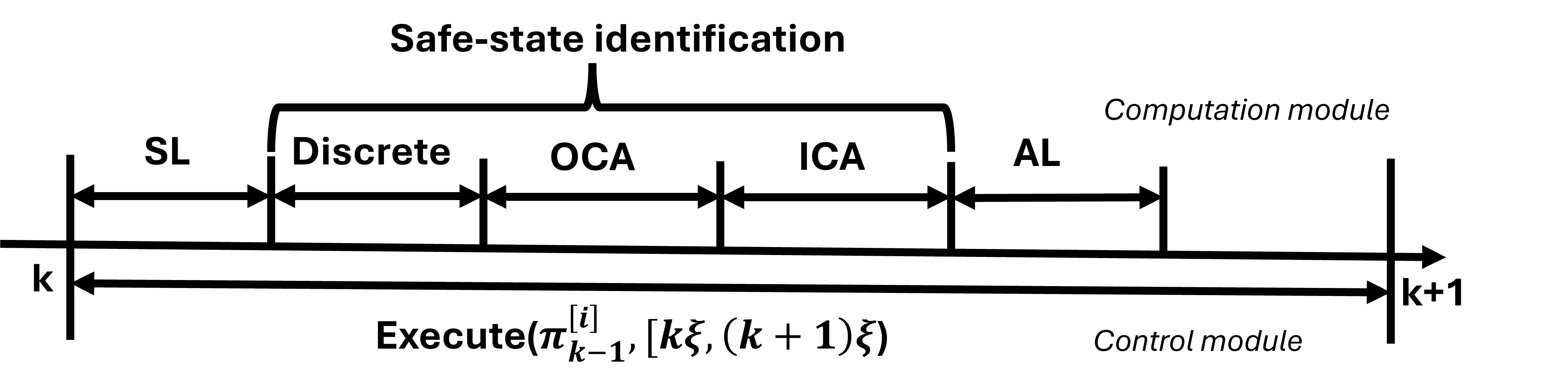}
    \caption{Implementation of dSLAP over one iteration}
    \label{fig:multithread}
\end{figure}

In this section, we propose the dSLAP framework.
Fig. \ref{fig:multithread}  illustrates one iteration of the dSLAP framework for 
 robot $i$. 
In each iteration $k$, the robot executes two modules  {\em in parallel}: the computation module and the control module. The computation module synthesizes a control policy $\pi^{[i]}_k$, which is executed by the continuous system \eqref{eq: observation model} in the next iteration $k+1$. Specifically, robot $i$ first collects a new dataset $\mathcal{D}^{[i]}_k$ and performs system learning (\textsf{SL}) using all the collected data to update the predictive mean $\mu^{[i]}_k$ and standard deviation $\sigma^{[i]}_k$ of the external disturbance $g^{[i]}$. 
Safe-state identification is then performed over discrete set-valued dynamics $\textsf{FR}^{[i]}_k$ obtained from dynamics discretization (\textsf{Discrete}) over the whole state space and control space. It first conducts  obstacle collision avoidance (\textsf{OCA}) to output a preliminary set of safe states $\mathcal{X}^{[i]}_{\text{safe},k}$ and then inter-robot collision avoidance (\textsf{ICA}) to output the final set $\mathcal{X}^{[i]}_{\text{safe},k}$. 
Finally,  active learning (\textsf{AL}) synthesizes the control policy $\pi^{[i]}_{k}$. The discretization parameter increases in each iteration to achieve finer granularity for  $\textsf{FR}^{[i]}_k$. The control module executes the control policy $\pi^{[i]}_{k-1}$, computed in iteration $k-1$,  for all $t\in[k\xi,(k+1)\xi)$, where $\xi$ is the duration of each iteration. 
Algorithm \ref{alg:overall} provides a formal description of the dSLAP framework.

\begin{algorithm}
\caption{The dSLAP framework }\label{alg:overall}
\SetAlgoLined
\textbf{Input}: State space: $\mathcal{X}$; Control input space: $\mathcal{U}$;  Obstacle: $\mathcal{X}_O$; Indices of robots: $\mathcal{V}$; Goal: ${\mathcal{X}^{[i]}_G}, i\in\mathcal{V}$; 
Initial discretization parameter: $p_{init}$;
Termination iteration: $\tilde{k}$; 
Period of iteration: $\xi$

 \textbf{Init:} $p_1\leftarrow p_{init}$;  $\pi^{[i]}_0, \forall i\in\mathcal{V}$\; 
\For{$k=0,1, 2,\cdots,\tilde{k}$}{
\For{$i\in\mathcal{V}$  (Computation module)}{

$\mathcal{D}^{[i]}_k\leftarrow \textsf{CollectData}$\;

$\mu_k^{[i]}, \sigma_k^{[i]}
\leftarrow\textsf{SL}(\mathcal{D}^{[i]}_k)$\;

$\textsf{FR}_k^{[i]}\leftarrow\textsf{Discrete}(p_k,\mathcal{X},\mathcal{U})$\;

$\mathcal{X}^{[i]}_{safe,k} \leftarrow \textsf{OCA}(\textsf{FR}_k^{[i]},\mathcal{X}_O)$\;

\label{ln: Overall: oca}

$\mathcal{X}^{[i]}_{safe,k}\leftarrow \textsf{ICA}(\mathcal{X}^{[i]}_{safe,k},\mathcal{V})$\;

${\pi}^{[i]}_{k} \leftarrow \textsf{AL}(\mathcal{X}^{[i]}_{safe,k},\mathcal{X}^{[i]}_G)$\;

$p_{k+1}\leftarrow p_k+1$\;
}

\For{$i\in\mathcal{V}$ (Control module)}{ 
$\textsf{Execute}(\pi^{[i]}_{k-1}, [k\xi,(k+1)\xi)) $\;
}
}

\end{algorithm}
\setlength{\textfloatsep}{0pt}

\subsection{System learning}\label{sec: system learning}
In this section, we introduce the $\textsf{SL}$ procedure for learning the external disturbance $g^{[i]}$. 
In each iteration $k$, each robot $i$ first collects a  new dataset  $\mathcal{D}_k^{[i]}$ through the \textsf{CollectData} procedure,
which returns $$
\mathcal{D}^{[i]}_k\triangleq\{g^{[i]}(x^{[i]}(\tau), u^{[i]}(\tau))+e^{[i]}(\tau), x^{[i]}(\tau),u^{[i]}(\tau)\}_{\tau=(k-1)\xi}^{(k-1)\xi+\delta\bar{\tau}},$$ where $e^{[i]}(\tau)\sim\mathcal{N}(0,(\sigma_e^{[i]})^2I_{n_x})$ is robot $i$'s local observation error, $\delta$ is the sampling period,  and $\bar{\tau}$ is the number of samples to be obtained.
Then robot $i$ independently estimates $g^{[i]}$ through GPR \cite{williams2006gaussian} using all the collected data $\cup_{k'=1}^k\mathcal{D}^{[i]}_{k'}$. 
By specifying prior mean function $\mu_0:\mathcal{X}\times\mathcal{U}\to\mathbb{R}^{n_x}$, and prior covariance function $\kappa:[\mathcal{X}\times\mathcal{U}]\times[\mathcal{X}\times\mathcal{U}]\to\mathbb{R}_{>0}$, GPR models external disturbance $g^{[i]}$ as a sample from a Gaussian process prior $\mathcal{GP}(\mu_0,\kappa)$ and  predicts $g^{[i]}(x^{[i]}, u^{[i]})\sim\mathcal{N}(\mu^{[i]}_k(x^{[i]}, u^{[i]}),(\sigma_k^{[i]}(x^{[i]}, u^{[i]}))^2)$.
 We use recursive GPR \cite{huber2014recursive} to maintain constant complexity.

\subsection{Safe-state identification}\label{sec: motion planning}
Safe-state identification is a novel multi-grid algorithm utilizing set-valued  analysis \cite{cardaliaguet1999set} to compute reachable sets and  quickly identify a set of safe actions that avoid collisions. 
We successively discretize the state space and the control space of system \eqref{eq: observation model} and build a new discrete set-valued dynamics incorporating the estimate of external disturbance $g^{[i]}$ in \textsf{SL} above. The set-valued dynamics is used to construct   one-step forward sets and approximate the  reachable sets of system \eqref{eq: observation model} for a fixed duration for each state-control pair. For each state, we then use its one-step forward sets to identify unsafe control inputs,  which inevitably lead to collision with the obstacles and the other robots. The unsafe control inputs are then removed from the set of admissible control inputs on the state. A state is labeled as a safe state if the set of control inputs on the state is non-empty. All the above procedures are conducted on the discrete set-valued dynamics.

{\em  Dynamics discretization.} 
In each iteration $k$, we discretize the whole state space $\mathcal{X}$ into  $\mathcal{X}_{p_k}$ and the whole control input space $\mathcal{U}$ into $\mathcal{U}_{p_k}$ as below:
\begin{align}\label{discrete}
	h_{p_k}\triangleq2^{-p_k}, \mathcal{X}_{p_k}\triangleq h_{p_k}\mathbb{Z}^{n_x}\cap \mathcal{X}, \mathcal{U}_{p_k}\triangleq h_{p_k}\mathbb{Z}^{n_u}\cap \mathcal{U}.
\end{align}
Then we use the above discretized spaces to obtain a discrete set-valued dynamics of \eqref{eq: observation model} similar to \cite{GZ-MZ:TAC18}\cite{cardaliaguet1999set}. Note that different from \cite{GZ-MZ:TAC18}\cite{cardaliaguet1999set}, part of the dynamics in \eqref{eq: observation model}, $g^{[i]}$, is unknown but learned as $g^{[i]}\sim\mathcal{N}(\mu^{[i]}_k,\sigma^{[i]}_k)$ in each iteration $k$ through \textsf{SL}. Therefore, we use confidence interval $\mu^{[i]}_k(x^{[i]},u^{[i]})+\gamma\sigma_k^{[i]}(x^{[i]},u^{[i]})\mathcal{B}$, where $\gamma$ is the reliability factor, to estimate $g^{[i]}(x^{[i]},u^{[i]})$ for each $x^{[i]}\in \mathcal{X}_{p_k}$ and $u^{[i]}\in\mathcal{U}_{p_k}$. Then  the discrete set-valued dynamics of \eqref{eq: observation model} for  $x^{[i]}\in \mathcal{X}_{p_k}$ and $u^{[i]}\in\mathcal{U}_{p_k}$ is given by
\begin{align*}
	\textsf{FR}_k^{[i]}\big(x^{[i]},u^{[i]}\big)\triangleq&\Big[x^{[i]}+\epsilon^{[i]} (f^{[i]}(x^{[i]},u^{[i]})+\mu^{[i]}_k(x^{[i]},u^{[i]}))+\\
	&(\epsilon^{[i]}\gamma\bar{\sigma}_k^{[i]}+\alpha^{[i]}_{p_k}+h_{p_k})\mathcal{B}\Big]\cap\mathcal{X}_{p_k},
\end{align*}
where $\epsilon^{[i]}$ is the temporal resolution for the discretization,  $\bar{\sigma}_k^{[i]}\triangleq  \sup_{x^{[i]}\in\mathcal{X},u^{[i]}\in\mathcal{U}}\sigma^{[i]}_k(x^{[i]},u^{[i]})$ is the maximum standard deviation in the estimate of $g^{[i]}$, and $\alpha^{[i]}_{p_k}\triangleq 2h_{p_k}+2\epsilon^{[i]} h_{p_k}\ell^{[i]}+(\epsilon^{[i]})^2\ell^{[i]}m^{[i]}$ is the dilation term accounting for discretization error \cite{cardaliaguet1999set}. We refer $\textsf{FR}_k^{[i]}{\big(x^{[i]},u^{[i]}\big)}$ as one-step forward set as it also approximates the  $\epsilon^{[i]}$-forward reachable set of system \eqref{eq: observation model} with respect to  uncertainty learned in iteration $k$ starting from state $x^{[i]}\in\mathcal{X}_{p_k}$ under control  $u^{[i]}\in\mathcal{U}_{p_k}$.
Furthermore, we impose that the duration of each iteration  can be partitioned into a multiple number of small intervals with duration $\epsilon^{[i]}$: $\xi=\bar{n}^{[i]}\epsilon^{[i]}$, where $\bar{n}^{[i]}$ is a positive integer.

Notice that, by \eqref{discrete}, finer discretization, corresponding to a larger discretization parameter $p_k$, provides a tighter approximation of the dynamic model, while coarser discretization results in faster solutions. Hence, we increment the discretization parameter, i.e., $p_{k+1}=p_k+1$, at the end of computation module in each iteration $k$.  This refines the discretization in the next iteration and reduces the spatial resolution by half to uncover less conservative actions incrementally.

\begin{algorithm}[t]

\caption{Procedure \textsf{OCA}}\label{alg:safety_control}

$\mathcal{X}^{[i]}_{\text{unsafe},k,0}\leftarrow\emptyset$\;
\For{$x^{[i]}\in\mathcal{X}_{p_k}$ \label{ln: Graphcon:  for all x}}{
 
\eIf {$\rho(x^{[i]},\mathcal{X}_O)\leqslant m^{[i]}\epsilon^{[i]}+h_{p_k}$\label{ln: GraphCon: U empty}}{
	$\textsf{Add}(\mathcal{X}^{[i]}_{\text{unsafe},k,0},x^{[i]})$\label{ln: GraphCon: add unsafe x}\;}
{
$\mathcal{U}^{[i]}_{p_k}(x^{[i]})\leftarrow \mathcal{U}_{p_k}$\;
    \For{$u^{[i]}\in \mathcal{U}_{p_k}$}{
    
%
            \For{
            	$y^{[i]}\in\textsf{FR}^{[i]}_k(x^{[i]}, u^{[i]})$}{
                 $\textsf{Add}(\textsf{BR}^{[i]}_k(y^{[i]},u^{[i]}),x^{[i]})$\;\label{ln: Graphcon: parents add}
            }
        }
    }
    }

\label{ln:end for all x}

$\bar{\mathcal{X}}^{[i]}_{\text{unsafe},k,0}\leftarrow \textsf{UnsafeUpdate}(\mathcal{X}^{[i]}_{\text{unsafe},k,0})$ \label{ln: OCA: poupdate}\;
$\mathcal{X}^{[i]}_{\text{safe},k}\leftarrow\mathcal{X}_{p_k}\setminus\bar{\mathcal{X}}^{[i]}_{\text{unsafe},k,0}$\;

\textbf{Return }
 $\mathcal{X}^{[i]}_{\text{safe},k}$\;


\end{algorithm}

{\em Obstacle collision avoidance (Algorithm \ref{alg:safety_control}).} 
Procedure \textsf{OCA}  leverages $\textsf{FR}_k^{[i]}\big(x^{[i]},u^{[i]}\big)$ to identify a subset of safe states from all admissible states in $\mathcal{X}_{p_k}$ by removing the control inputs (in $\mathcal{U}_{p_k}$) which lead to collision with the obstacles. Informally, a state is safe if there is a controller which can prevent the robot from colliding with the obstacles when the robot starts from the state. Otherwise, the state is unsafe. \textsf{OCA} is composed of two steps as follows. 

First, each robot identifies a preliminary set of unsafe states $\mathcal{X}^{[i]}_{\text{unsafe},k,0}$. A state  $x^{[i]}\in\mathcal{X}^{[i]}_{\text{unsafe},k,0}\subset \mathcal{X}_{p_k}$ if the distance $\rho(x^{[i]},\mathcal{X}_O)$ between state $x^{[i]}$ and the obstacle $\mathcal{X}_O$ is less than $ m^{[i]}\epsilon^{[i]}+h_{p_k}$. The distance $m^{[i]}\epsilon^{[i]}+h_{p_k}$ represents an over-approximation of the distance the robot can travel within one time step  $\epsilon^{[i]}$ on $\mathcal{X}_{p_k}$. This distance
prevents the robot from cutting corners around the obstacles
due to the discretization.  
If state $x^{[i]}$ satisfies $\rho(x^{[i]},\mathcal{X}_O)>m^{[i]}\epsilon^{[i]}+h_{p_k}$,  then for  all the states $y^{[i]}\in\textsf{FR}^{[i]}_k(x^{[i]}, u^{[i]})$, $x^{[i]}$ is recorded in the one-step backward set  $\textsf{BR}^{[i]}_k(y^{[i]},u^{[i]})$ for all $u^{[i]}\in \mathcal{U}_{p_k}$, where $\textsf{BR}^{[i]}_k(y^{[i]},u^{[i]})$ is the set of  states which  can reach $y^{[i]}$ via control $u^{[i]}$ through their one-step forward sets:
\begin{align*}
	\textsf{BR}^{[i]}_k(y^{[i]},u^{[i]})\triangleq \{&x^{[i]}\in\mathcal{X}_{p_x}\mid 
	y^{[i]}\in\textsf{FR}^{[i]}_k(x^{[i]}, u^{[i]}),\\
	& \rho(x^{[i]},\mathcal{X}_O)>m^{[i]}\epsilon^{[i]}+h_{p_k}\}.
\end{align*}

\begin{algorithm}[t]
\caption{\textsf{UnsafeUpdate}$(\mathcal{X}^{[i]}_{\text{unsafe},k,j})$}\label{alg:parent_update}
$\bar{\mathcal{X}}^{[i]}_{\text{unsafe},k,j}\leftarrow\mathcal{X}^{[i]}_{\text{unsafe},k,j}$\;
$Flag\leftarrow 1$\;
\While{$Flag==1$}{
$Flag\leftarrow 0$\;
\For{$y^{[i]}\in\bar{\mathcal{X}}^{[i]}_{\text{unsafe},k,j}$ \label{ln: PoUp: x in unsafe}}{
    \For{$u^{[i]}\in\mathcal{U}_{p_k}$ \label{ln: PoUp all u}}{
        \For { $x^{[i]}\in$ $\textsf{BR}^{[i]}_k(y^{[i]},u^{[i]})$\label{ln: PoUp: parents of x}}{
        $\textsf{Remove}(\mathcal{U}^{[i]}_{p_k}(x^{[i]}),u^{[i]})$\label{ln: PoUp: remove u}\;
            \If{$\mathcal{U}^{[i]}_{p_k}(x^{[i]})==\emptyset$ and $x^{[i]}\not\in \bar{\mathcal{X}}^{[i]}_{\text{unsafe},k,j}$ \label{ln: PoUp: if U emty}}{
            $\textsf{Add}(\bar{\mathcal{X}}^{[i]}_{\text{unsafe},k,j},x^{[i]})$\;
           $Flag\leftarrow1$\;
            }
        }
    }
}
}
\textbf{Return } $\bar{\mathcal{X}}^{[i]}_{\text{unsafe},k,j}$\;

\end{algorithm}
\setlength{\textfloatsep}{0pt}

In the second step,
robot $i$ runs the procedure $\textsf{UnsafeUpdate}$ (Algorithm \ref{alg:parent_update}) to iteratively remove all the control inputs leading to  unsafe states $\mathcal{X}^{[i]}_{\text{unsafe},k,0}$. 
Specifically, if $x^{[i]}\in	\textsf{BR}^{[i]}_k(y^{[i]},u^{[i]})$ and $y^{[i]}\in\mathcal{X}^{[i]}_{\text{unsafe},k,0}$, then control $u^{[i]}$ is removed from $\mathcal{U}^{[i]}_{p_k}(x^{[i]})$, where $\mathcal{U}^{[i]}_{p_k}(x^{[i]})$ is the set of available control inputs at state $x^{[i]}$ and is initialized as $\mathcal{U}_{p_k}$.
  If  $\mathcal{U}^{[i]}_{p_k}(x^{[i]})$ becomes empty, state $x^{[i]}$ is identified as unsafe and included in set $\bar{\mathcal{X}}^{[i]}_{\text{unsafe},k,0}$, which is initialized as $\mathcal{X}^{[i]}_{\text{unsafe},k,0}$.
Robot $i$'s set of unsafe states $\bar{\mathcal{X}}^{[i]}_{\text{unsafe},k,0}$ is then completed when the procedure terminates. For each state $x^{[i]}$ in the initial set of safe states $\mathcal{X}^{[i]}_{\text{safe},k}=\mathcal{X}_{p_k}\setminus\bar{\mathcal{X}}^{[i]}_{\text{unsafe},k,0}$, we have $\mathcal{U}^{[i]}_{p_k}(x^{[i]})\neq\emptyset$ and control $u^{[i]}\in\mathcal{U}^{[i]}_{p_k}(x^{[i]}) $ ensures collision avoidance with the obstacles.

\begin{figure*}[t]
	\centering
	\begin{subfigure}[t]{0.235\textwidth}
		\includegraphics[width=1\textwidth]{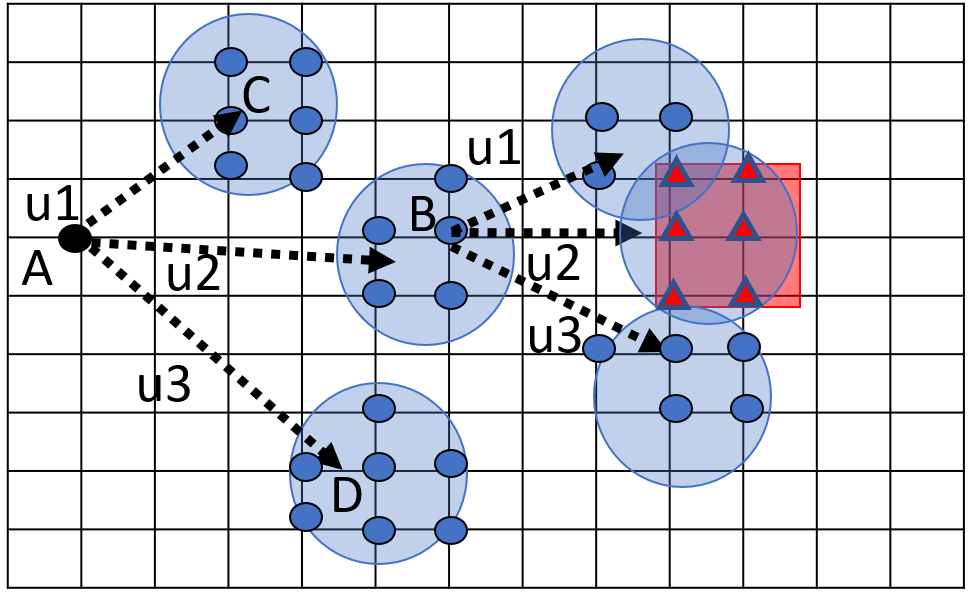}
		\centering
		\caption{}
		\label{fig: graph 1}
	\end{subfigure}
	\hfil
	\begin{subfigure}[t]{0.235\textwidth}
		\includegraphics[width=1\textwidth]{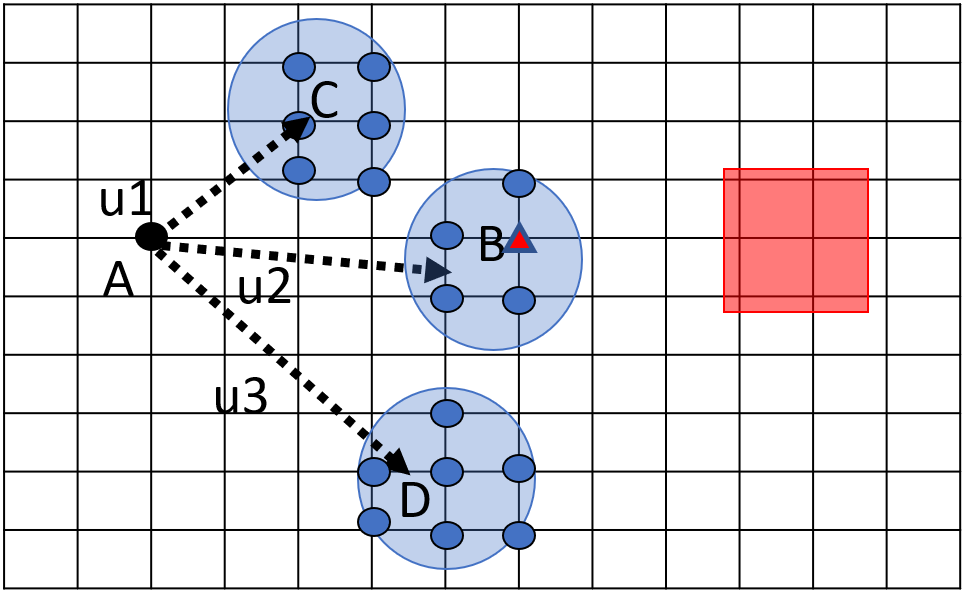}
		\centering
		\caption{}
		\label{fig: graph 2}
	\end{subfigure}
	\hfil
	\begin{subfigure}[t]{0.235\textwidth}
		\includegraphics[width=1\textwidth]{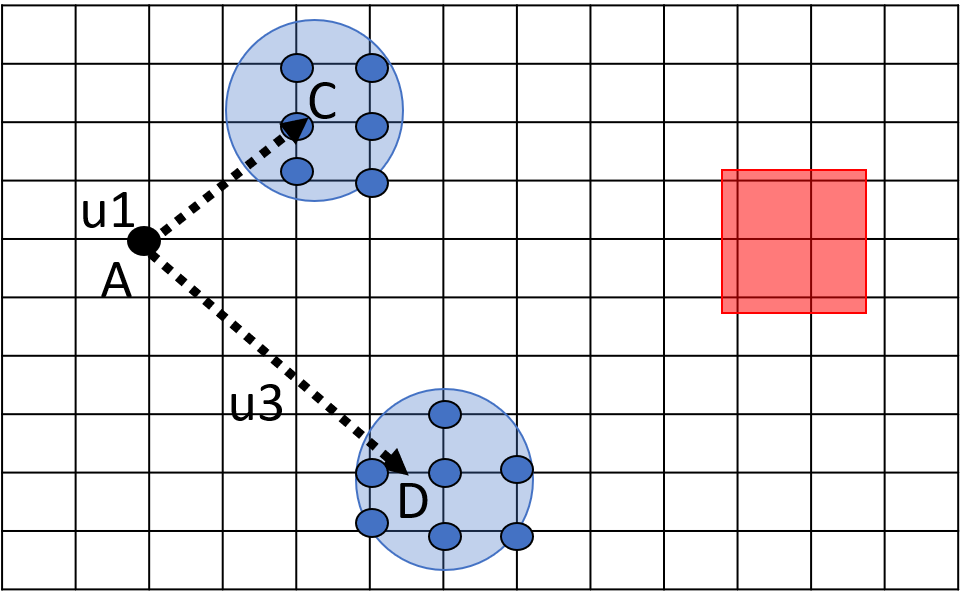}
		\caption{}
		\label{fig: graph 3}
	\end{subfigure}
	\caption{A graphical illustration of obstacle collision avoidance}
	\label{fig:graph: oca}
\end{figure*}

 A graphical illustration of \textsf{OCA} is shown in Fig. \ref{fig:graph: oca}. The square denotes the obstacle, and the intersections on the grid denote the states on discretized state space $\mathcal{X}_{p_k}$ for some $k$. The triangle states are unsafe. The arrows show the state transitions given the control inputs, and the blue circles represent the corresponding one-step forward set $\textsf{FR}_k^{[i]}$.  Starting from Fig. \ref{fig: graph 1}, the one-step forward sets of state $B$ under all the control inputs, $u1$, $u2$ and $u3$, have intersections with the obstacle, and hence these are unsafe control inputs and  removed from state $B$. Since there is no  (safe) control input left for state $B$, i.e., $\mathcal{U}_{p_k}^{[i]}(\textrm{state }B)=\emptyset$, it is labeled as unsafe as in Fig. \ref{fig: graph 2}. Since state $B$ is unsafe, control input $u_2$ is removed from state $A$. This gives Fig. \ref{fig: graph 3}, where state $A$ is safe with control inputs $u_1$ and $u_3$, and state $A$ belongs to the one-step backward sets $\textsf{BR}^{[i]}_k(\textrm{state }D, u_3)$ and $\textsf{BR}^{[i]}_k(\textrm{state }C, u_1)$.

{\em Inter-robot collision avoidance (Algorithm \ref{alg:inter_collision})}.
Procedure \textsf{ICA} adopts a prioritized planning scheme \cite{erdmann1987multiple} and aims to further remove the control inputs which lead to inter-robot collision.
 Each robot is assigned with a unique priority level. 
The robots with higher priority are treated as moving obstacles, and the robots with lower priority remove all the control inputs that lead to these obstacles.
Specifically, each robot $i$ first broadcasts a set which accounts for an overapproximation of its reachable set in two iterations starting from the beginning of each iteration $k$, the robot size and the discretization errors: $\mathcal{X}^{[i]}_k\triangleq x_q^{[i]}(k\xi)+(2\xi m^{[i]}+2\zeta+m^{[i]}\epsilon^{[i]}+2h_{p_k})\mathcal{B}$. Upon receiving the messages from each robot $j$ with higher priority, i.e., $j<i$,  robot $i$ identifies a preliminary set of unsafe states which accounts for $\mathcal{X}^{[j]}_k$ and the variation in the estimate of $g^{[i]}$: $\mathcal{X}^{[i]}_{\text{unsafe},k,j}\triangleq\big[\mathcal{X}^{[j]}_{k}+\epsilon^{[i]}\gamma\bar{\sigma}_k^{[i]}\mathcal{B}\big]\cap\mathcal{X}_{p_k}$. Then robot $i$ invokes the procedure $\textsf{UnsafeUpdate}$ to remove all the control inputs leading to the states in $\mathcal{X}^{[i]}_{\text{unsafe},k,j}$ and obtain the new set of unsafe states $\bar{\mathcal{X}}^{[i]}_{\text{unsafe},k,j}$. Robot $i$ then updates the set of safe states  $\mathcal{X}^{[i]}_{\text{safe},k}$ by removing $\bar{\mathcal{X}}^{[i]}_{\text{unsafe},k,j}$. 
For each state $x^{[i]}\in\mathcal{X}^{[i]}_{\text{safe},k}$ output from \textsf{ICA},  $\mathcal{U}_k^{[i]}(x^{[i]})\neq \emptyset$ and any control $u^{[i]}\in\mathcal{U}_k^{[i]}(x^{[i]})$ can ensure collision avoidance with the obstacles and the robots with higher priority until the end of iteration $k+1$.

\begin{algorithm}[t]
\caption{Procedure \textsf{ICA}}\label{alg:inter_collision}
$\mathcal{X}^{[i]}_{k}\leftarrow 
x_q^{[i]}(k\xi)+(2\xi m^{[i]}+2\zeta+m^{[i]}\epsilon^{[i]}+2h_{p_k})\mathcal{B}$\;
$\textsf{Broadcast}(\mathcal{X}^{[i]}_{k})$\label{ln:ICA:broadcast}\;
\For{$j\in\mathcal{V}$, $j\neq i$\label{ln: ICA: message available}}{
\If {{$j<i$}\label{ln: ICA: priority}}{
 $\mathcal{X}^{[i]}_{unsafe,k,j}\leftarrow\big[\mathcal{X}^{[j]}_{k}+\epsilon^{[i]}\gamma\bar{\sigma}_k^{[i]}\mathcal{B}\big]\cap\mathcal{X}_{p_k}$\; \label{ln:ICA: unsafe safe}

$\bar{\mathcal{X}}^{[i]}_{unsafe,k,j}\leftarrow\textsf{UnsafeUpdate}(\mathcal{X}^{[i]}_{unsafe,k,j},k)$\;

$\mathcal{X}^{[i]}_{safe,k}\leftarrow\mathcal{X}^{[i]}_{safe,k}\setminus \bar{\mathcal{X}}^{[i]}_{unsafe,k,j}$\label{ln:ICA: new safe}\;
}
}
\textbf{Return}  $\mathcal{X}^{[i]}_{safe,k}$\;
\end{algorithm}

 \setlength{\textfloatsep}{0pt}
\begin{algorithm}[t]
\caption{\textsf{AL}}\label{alg:active learning}
{\bf Procedure }$\pi_k^{[i]}(x^{[i]}(t))$\;
\quad $w[k]\leftarrow e^{-\psi k}$\label{ln:AL: weight}\;
 \quad $\hat{x}^{[i]}(t)\leftarrow\textsf{Nearest}(x^{[i]}(t),\mathcal{X}^{[i]}_{safe,k})$\;
\quad$(u^{[i]}_*(t), \cdots, u^{[i]}_*(t+\varphi\epsilon^{[i]}))\leftarrow$ solve MPC in  \eqref{opt: multi-objective2}\;
\textbf{Return} $u^{[i]}_*(t)$\;

\end{algorithm}
\subsection{Active learning and real-time control (Algorithm \ref{alg:active learning})}\label{sec:active learning}
In this section, we utilize the safe control inputs obtained above and synthesize an SVMPC \cite{risso2021set} to actively learn the external disturbance $g^{[i]}$ and approach the goal. 


First, 
the current state $x^{[i]}(t)$ of robot $i$ is projected onto  $\mathcal{X}^{[i]}_{\text{safe},k}$; the projection is denoted by $\hat{x}^{[i]}(t)\triangleq\textsf{Nearest}(x^{[i]}(t),\mathcal{X}^{[i]}_{\text{safe},k})$. Second, 
we capture the objective of goal reaching using distance $\rho(\hat{x}^{[i]}(t+\varphi\epsilon^{[i]}),\mathcal{X}_G^{[i]})$, where $\varphi\in\mathbb{N}$ is the discrete horizon of the SVMPC formulated below.  Then the objective of exploration is described by a utility function $r^{[i]}_k(\hat{x}^{[i]}(t),u^{[i]}(t))$; candidate utility functions, e.g., $r^{[i]}_k(\hat{x}^{[i]}(t),u^{[i]}(t))=\sigma^{[i]}_k(\hat{x}^{[i]}(t),u^{[i]}(t))$,  are available in \cite{settles2009active}.
 Next, the safety constraint is honored by choosing control inputs from the safe control set $\mathcal{U}^{[i]}_{k}(\hat{x}^{[i]}(t))$. Lastly, the dynamics of the SVMPC is given by the discrete set-valued dynamics $\textsf{FR}^{[i]}_k$, which is a set-valued approximation of dynamics \eqref{eq: observation model}. Formally, the  controller $\pi^{[i]}_k:\mathcal{X}\to\mathcal{U}$ returns control inputs by solving the  finite-horizon optimal control problem:
 \begin{align}
    \min\quad &(1-w[k])\rho(\hat{x}^{[i]}(t+\varphi{\epsilon^{[i]}}),\mathcal{X}_G^{[i]})\nonumber\\
    &~+w[k] {\sum_{\tau=t}^{t+\varphi{\epsilon^{[i]}}}}\displaystyle r^{[i]}_{k}(\hat{x}^{[i]}(\tau),u^{[i]}(\tau)), \label{opt: multi-objective2}
 \end{align}
where the decision variables are $u^{[i]}(t)\in\mathcal{U}^{[i]}_{k}(\hat{x}^{[i]}(t)),\cdots, u^{[i]}(t+\varphi\epsilon^{[i]})\in\mathcal{U}^{[i]}_{k}(\hat{x}^{[i]}(t+\varphi\epsilon^{[i]}))$,  subject to $\hat{x}^{[i]}(\tau+\epsilon^{[i]})\in\textsf{FR}_k^{[i]}(\hat{x}^{[i]}(\tau),u^{[i]}(\tau))$ for all $
    \tau\in\{t,t+\epsilon^{[i]},\cdots, t+(\varphi-1)\epsilon^{[i]}\}$. To ensure the robot eventually reach the goal, we select the weight $w[k]\triangleq e^{-\psi k}$ for some $\psi>0$ such that $w[k]$ diminishes. 

The above finite-horizon optimal control problem is solved once for every time duration $\epsilon^{[i]}$, and the returned control input is executed with zero-order hold for duration $\epsilon^{[i]}$. Specifically, consider a set of time instants $\{t^{[i]}_{k+1,n}\}_{n=0}^{\bar{n}^{[i]}-1}\subset [(k+1)\xi,(k+2)\xi]$, where 
 $t^{[i]}_{{k+1},0}=(k+1)\xi$ and $t^{[i]}_{{k+1},n}=t^{[i]}_{{k+1},n-1}+\epsilon^{[i]}$.  For each $n=0,1,\cdots, \bar{n}^{[i]}-1$,  Procedure $\pi^{[i]}_k(x^{[i]}(t^{[i]}_{{k+1},n}))$ is invoked solving the corresponding problem \eqref{opt: multi-objective2}. The solution has the form $(u^{[i]}_*(t^{[i]}_{{k+1},n}), \cdots, u^{[i]}_*(t^{[i]}_{{k+1},n}+\varphi\epsilon^{[i]}))$,  and control input $\pi^{[i]}_k(x^{[i]}(t^{[i]}_{{k+1},n}))=u^{[i]}_*(t^{[i]}_{{k+1},n})$ is returned. For all $t\in[t^{[i]}_{k+1,n},t^{[i]}_{k+1,n+1})$, robot $i$ executes $u^{[i]}(t)=u^{[i]}_*(t^{[i]}_{k+1,n})$ for system \eqref{eq: observation model}. The controller execution is denoted as procedure \textsf{Execute}  in Algorithm \ref{alg:overall}.

\setlength{\textfloatsep}{0pt}

\subsection{Performance guarantees}\label{sec: guarantees}
In this section, we provide the performance guarantees for dSLAP. To obtain theoretic guarantees, we assume that $g^{[i]}$ is a realization of a known Gaussian process.
For notational simplicity, we assume  $g^{[i]}\in\mathbb{R}$. Generalizing $g^{[i]}$ to multi-dimensional can be  done by applying union bound.

\begin{assumption}\label{assmp: ||g||_kappa}
{\em(Realization of process). }It satisfies that $g^{[i]}\in\mathbb{R}$ and $g^{[i]}\sim\mathcal{GP}(\mu_0,\kappa)$. $\hfill\blacksquare$
\end{assumption}
That is, function $g^{[i]}$ is a realization of Gaussian process with prior mean $\mu_0$ and kernel $\kappa$. This assumption is common in the analysis of GPR (Theorem 1, \cite{srinivas2012information}).  
Theorem \ref{thm: system safe} below provides the probability of the robots being safe until the end of an iteration if they are around the set of safe states at the beginning of the iteration.

\begin{theorem}\label{thm: system safe}
		{\em(One-iteration safety).}
Suppose Assumptions \ref{assmp: model} and \ref{assmp: ||g||_kappa} hold.  If 
$
    \mathcal{B}(x^{[i]}(k\xi), h_{p_{k-1}}) \cap\mathcal{X}^{[i]}_{safe, k-1}\neq \emptyset
$
, $k\geqslant1$,
for all $i\in\mathcal{V}$, then dSLAP renders $x_q^{[i]}(t)\in\mathcal{X}^{[i]}_F(x_q^{[\neg i]}(t))$ for all time $t\in[k\xi,k\xi+\xi)$ with probability at least $1-|\mathcal{V}||\mathcal{X}_p||\mathcal{U}_p|e^{-\gamma^2/2}$.
 $\hfill\blacksquare$
\end{theorem}
The proof of Theorem \ref{thm: system safe} can be found in Section \ref{proof: system safe}.

Denote $\bar{\sigma}\triangleq\|\kappa\|_{[\mathcal{X}\times\mathcal{U}]\times[\mathcal{X}\times\mathcal{U}]}$.
Then Theorem \ref{thm2: system safe} below provides the probability as well as the requirement on discretization and the computation speed of the robots such that they can be safe throughout the entire mission.
\begin{theorem}\label{thm2: system safe}
	{\em(All-time safety).}
Suppose $4\gamma\bar{\sigma}\epsilon^{[i]}\leqslant h_{p_{\tilde{k}}}$ 
and Assumptions \ref{assmp: model}and  \ref{assmp: ||g||_kappa} hold.  Suppose $\xi\leqslant\frac{h_{p_{\tilde{k}}}}{\max_{j\in\mathcal{V}}m^{[j]}}$. For all $i\in\mathcal{V}$, if $
    \mathcal{B}(x^{[i]}(k\xi), h_{p_{k-1}}) \cap\mathcal{X}^{[i]}_{\text{safe}, k-1}\neq \emptyset
$, for some $k\geqslant 1$,   then the dSLAP algorithm renders that $x_q^{[i]}(t)\in\mathcal{X}^{[i]}_F(x_q^{[\neg i]}(t))$,  for all $t\geqslant k\xi$ with probability at least $1-\tilde{k}|\mathcal{V}||\mathcal{X}_{p_{\tilde{k}}}||\mathcal{U}_{p_{\tilde{k}}}|e^{-\gamma^2/2}$.
 $\hfill\blacksquare$
\end{theorem}

The proof of Theorem \ref{thm2: system safe} can be found in Section \ref{proof: safe in k+1}.

The requirement for $\xi$ indicates
that the robots' onboard computation should be fast with respect to the speed of robots, while the computation can be relieved with coarse discretization. This provides the required update frequency for the decoupled controller.

The sufficient condition $4\gamma\bar{\sigma}\epsilon^{[i]}\leqslant h_{p_{\tilde{k}}}$  imposes a requirement in designing set-valued dynamics to discretize robot dynamics \eqref{eq: observation model} using $\textsf{FR}^{[i]}_k$. 
Given Assumption \ref{assmp: ||g||_kappa}, $\gamma\bar{\sigma}$ represents an upper bound over the variability of the disturbances the robots {\color{blue}can} tolerate. On the right hand side, $h_{p_{\tilde{k}}}$ represents the  minimal spatial resolution. Then the sufficient condition implies that the product between the variability of the disturbances and the temporal resolution   should be small with respect to and the spatial resolution. 

\subsection{Discussion}\label{sec: discussion}

{\em (Probabilistic safety).} The probability of the safety guarantees stems from the  analysis of  GPR  estimates being able to capture the
ground truth dynamics over the whole state-control space (and over all the iterations). The analysis can be conservative but is independent of the other components in the proposed algorithm. In order to reduce the probability of unsafe execution, or increase the probability of safe
execution, the robots can increase $\gamma$ according to the theorems. This may
cause conservative actions as $\gamma$ is a  factor for constructing the one-step forward set $\textsf{FR}^{[i]}_k$ and could 
lead to no solution at all if $\gamma$ is too large. However, this can be addressed by having the robots collecting more data online to train the GPR such that $\sigma_k^{[i]}(x^{[i]},u^{[i]})$
becomes small. 

{\em (Verification of initial safety}
$\mathcal{B}(x^{[i]}(0), h_{p_{-1}})\cap\mathcal{X}^{[i]}_{\text{safe}, -1}\neq\emptyset$).   To ensure the robots are safe for all the time, Theorem \ref{thm2: system safe} implies that it suffices to  satisfy the sufficient condition $ \mathcal{B}(x^{[i]}(0), h_{p_{-1}})\cap\mathcal{X}^{[i]}_{\text{safe}, -1}\neq\emptyset$ {\em a priori}.  To achieve this, one can compute $\mathcal{X}^{[i]}_{\text{safe}, -1}$ using data collected a priori or a prior conservative estimates of the  disturbances. 
This prior knowledge can be obtained in most situations by, e.g., using historical data. Examples include wind speed, water current, and road texture in a local area. 
In addition, smaller $h_{p_{-1}}$ can enlarge the set $\mathcal{X}^{[i]}_{\text{safe}, -1}$ such that more initial states $x^{[i]}(0)$ can satisfy the condition.

{\em (Computational complexity).}
The algorithms in \cite{GZ-MZ:TAC18}\cite{cardaliaguet1999set} aim to solve optimal arrival and collision avoidance simultaneously in a centralized manner. The computational complexity scale as $\mathcal{O}((|\mathcal{X}_{p_k}||\mathcal{U}_{p_k}|)^n)$, which grows exponentially in the number of robots. In order to reduce the computational complexity, dSLAP has two stages. The first stage includes Procedures \textsf{OCA} and \textsf{ICA}, which are distributed and  remove unsafe control inputs  on the discretized state-control space of each robot. \textsf{OCA} is independent of the other robots. \textsf{ICA} combines the reachable sets of higher priority robots and correspondly removes the unsafe control inputs. Its worst-case onboard computational complexity of each robot   scales linearly with $n$.  The local safe control inputs enable decoupled planning through \textsf{AL} in the second  stage, whose computational complexity is also independent of $n$. 

{\em (Strength and weakness)}.
The proposed framework dSLAP is able to compute safe control inputs for a multi-robot system with general nonlinear dynamics in a distributed manner amid online uncertainty learning. Nevertheless, dSLAP can be conservative
for the following two reasons. First, it  overapproximates  the continuous dynamics using discretized set-valued dynamics. To enable fast computation, the discretization is usually coarse and hence the approximation error can be large, which leads to conservative actions. However, this conservativeness can be reduced via finer discretization provided sufficient computation power.  Second,  the coordination among the robots is simple.  The application of prioritized planning is suboptimal and can lose completeness \cite{ma2019searching}. Furthermore, higher priority robots are viewed as moving obstacles by lower priority robots. The overapproximation of the reachable sets in two iterations of the higher priority robots can be conservatives since the overapproximation is determined by the maximum speed of the robots multiplied by the duration of two iterations. This conservativeness can be reduced by developing a more sophisticated scheme of coordination among the robots, optimizing the assignment of priority levels, and/or shortening the duration of one iteration. We leave this for future work. Furthermore, dSLAP can suffer from the curse-of-dimensionality for each individual system.
\section{Proof}\label{sec: proof}
In this section, we prove Theorems \ref{thm: system safe} and \ref{thm2: system safe}. Below is a roadmap for the proofs.
\begin{enumerate}
    \item We present the concentration inequality resulting from GPR in Section \ref{appendix: GP}. This provides the probability of the event that $g^{[i]}$ belongs to the tube $\mu^{[i]}_k\pm\gamma\sigma^{[i]}_k$. The rest of the analysis is performed under this event.
    \item Section \ref{proof: notations} introduces a set of preliminary notations for set-valued mappings and the related properties. The properties examine the approximation of system dynamics \eqref{eq: observation model} through the set-valued mappings.  
    \item Utilizing the set-valued approximations of dynamics \eqref{eq: observation model}, Section \ref{proof: system safe} derives that $\mathcal{B}(x^{[i]}(k\xi), h_{p_{k-1}}) \cap\mathcal{X}^{[i]}_{\text{safe}, k-1}\neq \emptyset$ for some $k\geqslant 1$ is a sufficient condition  to ensure that robot $i$ being collision-free during iteration $k$ and hence proves Theorem \ref{thm: system safe}. 
    \item Given one-iteration safety in  iteration $k$, Section \ref{appdix: proof of obstacle change} examines the distance between $x^{[i]}((k+1)\xi)$ and $\bar{\mathcal{X}}^{[i]}_{\text{unsafe},k-1,j}$ as well as the inclusion of $\bar{\mathcal{X}}^{[i]}_{\text{unsafe},k,j}$ in terms of $\bar{\mathcal{X}}^{[i]}_{\text{unsafe},k-1,j}$.
    \item   Utilizing the two relations in (iv), the distance  $\rho(x^{[i]}((k+1)\xi),\bar{\mathcal{X}}^{[i]}_{\text{unsafe},k,j})$ can be characterized. This is further used in Section \ref{proof: safe in k+1} to establish the sufficient condition   for $\mathcal{B}(x^{[i]}((k+1)\xi), h_{p_{k}}) \cap\mathcal{X}^{[i]}_{\text{safe}, k}\neq\emptyset$ to hold, given  $\mathcal{B}(x^{[i]}(k\xi), h_{p_{k-1}}) \cap\mathcal{X}^{[i]}_{\text{safe}, k-1}\neq\emptyset$.
    This  ensures one-iteration safety hold in iteration $k+1$ and completes the proof of
Theorem \ref{thm2: system safe}.
\end{enumerate}

\subsection{Concentration inequality of Gaussian process}\label{appendix: GP}

    

The concentration inequality resulted from GPR is presented  in Lemma \ref{lmm: concentration GPR} below.

\begin{lemma}\label{lmm: concentration GPR} 
Under Assumption \ref{assmp: ||g||_kappa}, for any discretization parameter $p$ and each robot $i$,  the following holds  with probability at least $1-|\mathcal{X}_{p}||\mathcal{U}_{p}|e^{-\gamma^2/2}$: $\forall x^{[i]}\in\mathcal{X}_{p}, u^{[i]}\in\mathcal{U}_{p}$, 
\begin{align}\label{eq: g in mu}
&|\mu^{[i]}_{k}(x^{[i]},u^{[i]})-g^{[i]}(x^{[i]},u^{[i]})|\leqslant \gamma\sigma^{[i]}_k(x^{[i]},u^{[i]}).
\end{align}

{\bf Proof:} The proof mainly follows the proof of Lemma 5.1 in \cite{srinivas2012information}. 
At iteration $k$, we  have the input dataset $Z^{[i]}_{1:k}$ and output dataset $y^{[i]}_{1:k}$, where
\begin{align*}
    y^{[i]}_{1:k}&\triangleq [[y^{[i]}_{1}]^T,\cdots,[y^{[i]}_{k}]^T]^T,\\
    y^{[i]}_{l}&\triangleq[g(x^{[i]}(\tau),u^{[i]}(\tau))+e^{[i]}(\tau)]_{\tau=(l-1)\xi}^{(l-1)\xi+\delta\bar{\tau}},\\
    Z^{[i]}_{l}&\triangleq  \{x^{[i]}(\tau),u^{[i]}(\tau)\}_{\tau=(l-1)\xi}^{(l-1)\xi+\delta\bar{\tau}}, Z^{[i]}_{1:k}\triangleq \{Z^{[i]}_{l}\}_{l=1}^k, ~
\end{align*}
Let test input $z^{[i]}_*\in\mathcal{Z}_p\triangleq \mathcal{X}_p\times\mathcal{U}_p$. 
 Assumption \ref{assmp: ||g||_kappa} gives
\begin{align*}
    &\begin{bmatrix}    y^{[i]}_{1:k}\\    g^{[i]}(z^{[i]}_*)    \end{bmatrix}
    =\\
    &
    \mathcal{N}(  \begin{bmatrix}    \mu_0(Z^{[i]}_{1:k})\\    \mu_0(z^{[i]}_*)    \end{bmatrix},\begin{bmatrix}    \kappa(Z^{[i]}_{1:k},Z^{[i]}_{1:k})+(\sigma^{[i]}_e)^2I&\kappa(Z^{[i]}_{1:k},z^{[i]}_*)\\    \kappa(z^{[i]}_*,Z^{[i]}_{1:k})&\kappa(z^{[i]}_*,z^{[i]}_*) \end{bmatrix}),
\end{align*}
where 
$
    \kappa(Z^{[i]}_{1:k},z^{[i]}_*)\triangleq[\kappa(z^{[i]},z^{[i]}_*)]_{z^{[i]}\in Z^{[i]}_{1:k}}$, and $    \kappa(Z^{[i]}_{1:k},Z^{[i]}_{1:k})\triangleq[\kappa(z^{[i]},\tilde{z}^{[i]})]_{z^{[i]},\tilde{z}^{[i]}\in Z^{[i]}_{1:k}}.
$

Applying identities of joint Gaussian distribution (page 200, \cite{williams2006gaussian}), we obtain the posterior distribution $g^{[i]}(z^{[i]}_*)\sim\mathcal{N}(\mu^{[i]}_k(z^{[i]}_*),(\sigma^{[i]}_k(z^{[i]}_*))^2)$, where
\begin{align*}
   &\mu^{[i]}_k(z^{[i]}_*)\triangleq \mu^{[i]}_0(z^{[i]}_*)+\kappa(z^{[i]}_*,Z^{[i]}_{1:k})(\kappa(Z^{[i]}_{1:k},Z^{[i]}_{1:k})\\
   &\qquad\qquad+(\sigma_e^{[i]})^2)^{-1}\cdot(y^{[i]}_{1:k}-\mu_0(Z^{[i]}_{1:k})), \\
   &(\sigma_k^{[i]}(z^{[i]}_*))^2\triangleq\kappa(z^{[i]}_*,z^{[i]}_*)\\
   &\quad+\kappa(z^{[i]}_*,Z^{[i]}_{1:k},)(\kappa(Z^{[i]}_{1:k},Z^{[i]}_{1:k})+(\sigma_e^{[i]})^2)^{-1}\kappa(Z^{[i]}_{1:k},z^{[i]}_*).
\end{align*}


Consider $r\sim\mathcal{N}(0,1)$. It holds that for $c>0$,
\begin{align*}
    Pr\{r>c\}&=e^{-c^2/2}(2\pi)^{-1/2}\int_{c}^\infty e^{-\frac{(r-c)^2}{2}-c(r-c)}dr\\
    &\leqslant e^{-c^2/2}Pr\{r>0\}=\frac{1}{2}e^{-c^2/2}
\end{align*}
where the inequality uses the fact $e^{-c(r-c)}\leqslant 1$ for $r\geqslant c$. Analogously, $Pr\{r<-c\}\leqslant \frac{1}{2}e^{-c^2/2}$. Therefore, let $r=\frac{g^{[i]}(z^{[i]}_*)-\mu^{[i]}_k(z^{[i]}_*)}{\sigma^{[i]}_k(z^{[i]}_*)}$ and $c=\gamma$, we have $Pr\{ |g^{[i]}(z^{[i]}_*)-\mu^{[i]}_k(z^{[i]}_*)|> \gamma\sigma^{[i]}_k(z^{[i]}_*)\}\leqslant e^{-\gamma^2/2}$. 
Denote event $E_{z^{[i]}_*}\triangleq \{|g^{[i]}(z^{[i]}_*)-\mu^{[i]}_k(z^{[i]}_*)|> \gamma\sigma^{[i]}_k(z^{[i]}_*)\}$.
Applying the union bound (Theorem 2-3, \cite{papoulis2002probability}), we have 
$$
Pr\{ \cup_{z^{[i]}_*\in\mathcal{Z}_p}E_{z^{[i]}_*}\}\leqslant |\mathcal{Z}_p|e^{-\gamma^2/2}.
$$
Note that $Pr\{ \cap_{z^{[i]}_*\in\mathcal{Z}_p}E_{z^{[i]}_*}\}=1-Pr\{ \cup_{z^{[i]}_*\in\mathcal{Z}_p}E_{z^{[i]}_*}\}$. Hence,
\begin{align*}
    |g^{[i]}(z^{[i]}_*)-\mu^{[i]}_k(z^{[i]}_*)|\leqslant \gamma\sigma^{[i]}_k(z^{[i]}_*),
\end{align*}
simultaneously for all $z^{[i]}_*\in\mathcal{Z}_p$ with probability at least $1-|\mathcal{Z}_p|e^{-\gamma^2/2}=1-|\mathcal{X}_{p}||\mathcal{U}_{p}|e^{-\gamma^2/2}$.
$\hfill\blacksquare$

\end{lemma}

\subsection{Set-valued approximation}\label{proof: notations}
In this section, we first introduce a collection of set-valued notations from \cite{cardaliaguet1999set} to discretize system \eqref{eq: observation model} in the time and state spaces. Lemma \ref{lmm: phi in G} shows that the set-valued discretization is a good approximation of the continuous system \eqref{eq: observation model}. Subsequently, we discuss additional properties in the discretized space.

Define
\begin{align*}
&F^{[i]}(x^{[i]},u^{[i]})\triangleq f^{[i]}(x^{[i]},u^{[i]})+g^{[i]}(x^{[i]},u^{[i]}),\\ &F^{[i]}_\epsilon(x^{[i]},u^{[i]})\triangleq F^{[i]}(x^{[i]},u^{[i]})+m^{[i]}\ell^{[i]}\epsilon\mathcal{B},\\
&G^{[i]}_\epsilon(x^{[i]},u^{[i]})\triangleq x^{[i]}+\epsilon F^{[i]}_\epsilon(x^{[i]},u^{[i]}).
\end{align*}
Page 222 of \cite{cardaliaguet1999set} uses the following discrete set-valued map 
\begin{align}\label{eq: Gamma}
    \Gamma^{[i]}_{\epsilon,h}(x^{[i]},u^{[i]})\triangleq [G^{[i]}_{\epsilon}(x^{[i]},u^{[i]})+2(1+\ell^{[i]}\epsilon)h\mathcal{B}]\cap\mathcal{X}_p,
\end{align}
which is discrete in time and state, to approximate system \eqref{eq: observation model}, which is continuous in time and state.
Let 
\begin{align*}
x^{[i]}(t_0, t_0+\epsilon,u^{[i]})\triangleq& x^{[i]}(t_0)+\int_{0}^\epsilon f^{[i]}(x^{[i]}(t_0+\tau),u^{[i]})\\
&+g^{[i]}(x^{[i]}(t_0+\tau),u^{[i]})d\tau
\end{align*}
 be the state at time $t_0+\epsilon$ when system \eqref{eq: observation model} starts from $x^{[i]}(t_0)$ at time $t_0$ and applies constant input $u^{[i]}\in\mathcal{U}$ within the time interval $[t_0,t_0+\epsilon]$. Lemma \ref{lmm: phi in G} below shows that $G^{[i]}_{\epsilon}(x^{[i]}(t_0),u^{[i]})$ contains the trajectory of system \eqref{eq: observation model} under constant control  for any duration $\epsilon$.
\begin{lemma}\label{lmm: phi in G}
Under Assumption \ref{assmp: model}, for any $x^{[i]}(t_0)\in\mathcal{X}$, $t_0\geqslant 0$, $u^{[i]}\in\mathcal{U}$ and $\epsilon>0$, it holds that $x^{[i]}(t_0, t_0+\epsilon,u^{[i]})\in G^{[i]}_{\epsilon}(x^{[i]}(t_0),u^{[i]})$.

{\bf Proof: }
The proof is part of the proof on page 194 in \cite{cardaliaguet1999set}. 
Let $\tau\in[0,\epsilon]$.
 By {\bf (A1)} in Assumption \ref{assmp: model} and the definitions of $m^{[i]}$ and $x^{[i]}(t_0, t_0+\epsilon,u^{[i]})$, we have 
$
     \|x^{[i]}(t_0, t_0+\tau,u^{[i]})-x^{[i]}(t_0)\|
     \leqslant\tau m^{[i]}\leqslant\epsilon m^{[i]}.
$
Lipschitz continuity further gives that 
\begin{align*}
F^{[i]}(x^{[i]}(t_0, t_0+\tau,u^{[i]}),u^{[i]})&\in F^{[i]}(x^{[i]}(t_0),u^{[i]})+m^{[i]}\ell^{[i]}\epsilon\mathcal{B}\\
&=F^{[i]}_{\epsilon}(x^{[i]}(t_0),u^{[i]}).
\end{align*}
Then 
\begin{align}
    &\dot{x}^{[i]}(t_0, t_0+\tau,u^{[i]})=(f^{[i]}+g^{[i]})(x^{[i]}(t_0, t_0+\tau,u^{[i]}),u^{[i]})\nonumber\\
    &= F^{[i]}(x^{[i]}(t_0, t_0+\tau,u^{[i]}),u^{[i]})\in  F^{[i]}_{\epsilon}(x^{[i]}(t_0),u^{[i]}).\label{eq: dot phi in F}
\end{align}

By definition, $F^{[i]}_{\epsilon}(x^{[i]},u^{[i]})$ is compact and convex since it is just a closed ball, which indicates that  $G^{[i]}_{\epsilon}(x^{[i]},u^{[i]})$ is also convex and compact. For any state $ \lambda\in\mathcal{X}$, \eqref{eq: dot phi in F} renders
$$
    \sup_{y^{[i]}\in F^{[i]}_{\epsilon}(x^{[i]}(t_0),u^{[i]})}\langle y^{[i]},\lambda\rangle\geqslant\langle \dot{x}^{[i]}(t_0, t_0+\tau,u^{[i]}),\lambda\rangle, \tau\in[0,\epsilon].
$$ 
Applying integration gives $$ 
\sup_{y^{[i]}\in x^{[i]}(t_0)+\epsilon F^{[i]}_{\epsilon}(x^{[i]}(t_0),u^{[i]})}\langle y^{[i]},\lambda\rangle\geqslant \langle x^{[i]}(t_0, t_0+\epsilon,u^{[i]}), \lambda\rangle.
$$ Therefore, applying the Separating Hyperplane Theorem on page 46 in \cite{boyd2004convex}, we prove the lemma.
$\hfill\blacksquare$

\end{lemma}


Similarly, let
\begin{align}
&\tilde{F}^{[i]}_k(x^{[i]},u^{[i]})\triangleq f^{[i]}(x^{[i]},u^{[i]})+\mu^{[i]}_k(x^{[i]},u^{[i]})\nonumber\\
&\qquad\qquad\qquad\quad+\gamma\sigma^{[i]}_k(x^{[i]},u^{[i]})\mathcal{B},\nonumber\\
&\tilde{F}^{[i]}_{\epsilon,k}(x^{[i]},u^{[i]})\triangleq \tilde{F}^{[i]}_k(x^{[i]},u^{[i]})+m^{[i]}\ell^{[i]}\epsilon\mathcal{B},\nonumber\\
&\tilde{G}^{[i]}_{\epsilon,k}(x^{[i]},u^{[i]})\triangleq x^{[i]}+\epsilon \tilde{F}^{[i]}_{\epsilon,k}(x^{[i]},u^{[i]}),\nonumber\\
    &\tilde{\textsf{FR}}_k^{[i]}(x^{[i]}, u^{[i]}, p,\epsilon)\triangleq[\tilde{G}^{[i]}_{\epsilon,k}(x^{[i]},u^{[i]})+2(1+\ell^{[i]}\epsilon)h_p\mathcal{B}]\cap\mathcal{X}_p.\label{eq: Reach=}
\end{align}
Denote $\delta_{\gamma,p}\triangleq |\mathcal{X}_{p}||\mathcal{U}_{p}|e^{-\gamma^2/2}$. By Lemma \ref{lmm: concentration GPR}, we have
\begin{align}\label{eq: g in interval}
    g^{[i]}(x^{[i]},u^{[i]})\in \mu_k^{[i]}(x^{[i]},u^{[i]})+\gamma\sigma_k^{[i]}(x^{[i]},u^{[i]})\mathcal{B}
\end{align}
for all $x^{[i]}\in\mathcal{X}_p$ and $u^{[i]}\in\mathcal{U}_p$ with probability at least $1-\delta_{\gamma,p}$. This gives each of the followings  holds with probability at least $1-\delta_{\gamma,p}$ for all  $x^{[i]}\in\mathcal{X}_{p}, u^{[i]}\in\mathcal{U}_{p}$:
\begin{align*}
& F^{[i]}(x^{[i]},u^{[i]})\subseteq 
\tilde{F}^{[i]}_k(x^{[i]},u^{[i]}),~ F^{[i]}_{\epsilon}(x^{[i]},u^{[i]})\subseteq \tilde{F}^{[i]}_{\epsilon,k}(x^{[i]},u^{[i]})\\
&G^{[i]}_{\epsilon}(x^{[i]},u^{[i]})\subseteq\tilde{G}^{[i]}_{\epsilon,k}(x^{[i]},u^{[i]}),\\
&  \Gamma^{[i]}_{\epsilon,h_p}(x^{[i]},u^{[i]})\subseteq\tilde{\textsf{FR}}_k^{[i]}(x^{[i]}, u^{[i]}, p,\epsilon).
\end{align*}

Lemma \ref{lemma: reach*} characterizes the relation between $\tilde{\textsf{FR}}$ and $\textsf{FR}$.
\begin{lemma}\label{lemma: reach*}
For any $x^{[i]}\in\mathcal{X}$ and $u^{[i]}\in\mathcal{U}$, it holds that $[\tilde{\textsf{FR}}_k^{[i]}\big(x^{[i]},u^{[i]},p_k,\epsilon^{[i]}\big)+h_{p_k}\mathcal{B}]\cap\mathcal{X}_{p_k}\subset \textsf{FR}_k^{[i]}\big(x^{[i]},u^{[i]}\big)
$.

{\bf Proof: } 
Consider state $y^{[i]}\in\tilde{\textsf{FR}}_k^{[i]}\big(x^{[i]},u^{[i]},p_k,\epsilon^{[i]}\big)$. 
Then   
\begin{align*}
    y^{[i]}\in& [x^{[i]}+\epsilon^{[i]}(f^{[i]}(x^{[i]},u^{[i]})+\mu^{[i]}_k(x^{[i]},u^{[i]}))\\
    &+(\gamma\sigma^{[i]}_k(x^{[i]},u^{[i]})\epsilon^{[i]}+\alpha^{[i]}_{p_k})\mathcal{B}].
\end{align*}
Hence 
$
y^{[i]}+h_{p_k}\mathcal{B}\subset[x^{[i]}+\epsilon^{[i]}(f^{[i]}(x^{[i]},u^{[i]})+\mu^{[i]}_k(x^{[i]},u^{[i]}))+(\gamma\bar{\sigma}_k^{[i]}\epsilon^{[i]}+\alpha^{[i]}_{p_k}+h_{p_k})\mathcal{B}].
$
This  gives
\begin{align*}
    &[y^{[i]}+h_{p_k}\mathcal{B}]\cap\mathcal{X}_{p_k}\subset\\
    &[x^{[i]}+\epsilon^{[i]}(x^{[i]}(f^{[i]}(x^{[i]},u^{[i]})+\mu^{[i]}_k(x^{[i]},u^{[i]}))\\
    &+(\gamma\bar{\sigma}_k^{[i]}\epsilon^{[i]}+\alpha^{[i]}_{p_k}+h_{p_k})\mathcal{B}]\cap\mathcal{X}_{p_k},
\end{align*}
where the right hand side is equivalent to $\textsf{FR}_k^{[i]}\big(x^{[i]},u^{[i]}\big)$.$\hfill\blacksquare$

\end{lemma}

Recall that Assumption \ref{assmp: model} implies that the system dynamics $f^{[i]}+g^{[i]}$, or $F^{[i]}$, is Lipschitz continuous. In particular,  $F^{[i]}_\epsilon$ is identical to the example in the Lipschitz case on page 191 in \cite{cardaliaguet1999set}, and hence it satisfies Assumptions  H0,  H1, and  H2 in \cite{cardaliaguet1999set}. Therefore, applying Lemma 4.13 \cite{cardaliaguet1999set} gives
\begin{align}\label{eq: discrete property}
   \underset{\rho(\tilde{x}^{[i]}, x^{[i]})\leqslant h_p}{\cup} [G^{[i]}_{\epsilon}(\tilde{x}^{[i]},u^{[i]})+h_p\mathcal{B}]\cap \mathcal{X}_p\subset \Gamma^{[i]}_{\epsilon,h_p}(x^{[i]},u^{[i]}). 
\end{align}

Below are the other properties in the discretized space. Lemmas \ref{lmm: nearest in reach} and \ref{lmm: rho(G,x_unsafe)}
show that  $\tilde{\textsf{FR}}_k^{[i]}(x^{[i]}, u^{[i]}, p,\epsilon)$ almost contains the union of $G^{[i]}_{\epsilon}(x^{[i]},u^{[i]})$.
\begin{lemma}\label{lmm: nearest in reach}
Suppose that Assumptions \ref{assmp: model} and \eqref{eq: g in interval} hold $\forall x^{[i]}\in\mathcal{X}_{p_{k-1}}, u^{[i]}\in\mathcal{U}_{p_{k-1}}$, $i\in\mathcal{V}$. For any $x^{[i]}\in\mathcal{X}$, $u^{[i]}\in\mathcal{U}$ and $\epsilon>0$, for all $y^{[i]}\in\cup_{\rho(\tilde{x}^{[i]},x^{[i]})\leqslant h_p} G^{[i]}_{\epsilon}(\tilde{x}^{[i]},u^{[i]})$, it holds that $\textsf{Nearest}(y^{[i]},\mathcal{X}_p)\in \tilde{\textsf{FR}}_k^{[i]}(x^{[i]}, u^{[i]}, p,\epsilon)$.

{\bf Proof: } 
Since $y^{[i]}\in \cup_{\rho(\tilde{x}^{[i]},x^{[i]})\leqslant h_p} G^{[i]}_{\epsilon}(\tilde{x}^{[i]},u^{[i]})$, we have
 $$
 y^{[i]}+h_p\mathcal{B}\subset \cup_{\rho(\tilde{x}^{[i]},x^{[i]})\leqslant h_p} G^{[i]}_{\epsilon}(\tilde{x}^{[i]},u^{[i]})+h_p\mathcal{B}.
 $$ Combining this inclusion property  with \eqref{eq: discrete property}, we have $$
 [y^{[i]}+h_p\mathcal{B}]\cap\mathcal{X}_p\subset \Gamma^{[i]}_{\epsilon,h_p}(x^{[i]},u^{[i]}).
 $$
Since \eqref{eq: g in interval} holds,  we have $$
[y^{[i]}+h_p\mathcal{B}]\cap\mathcal{X}_p\subset\tilde{\textsf{FR}}_k^{[i]}(x^{[i]}, u^{[i]}, p,\epsilon).
$$ 
Note that the \textsf{Discrete} operation ensures $[y^{[i]}+h_p\mathcal{B}]\cap\mathcal{X}_p\neq\emptyset$. 
Hence  $\textsf{Nearest}(y,\mathcal{X}_p)\in [y^{[i]}+h_p\mathcal{B}]\cap\mathcal{X}_p\subset \tilde{\textsf{FR}}_k^{[i]}(x^{[i]}, u^{[i]}, p,\epsilon)$. 
$\hfill\blacksquare$
\end{lemma}

\begin{lemma}\label{lmm: rho(G,x_unsafe)}
Consider closed set $\mathcal{A}\subset\mathcal{X}$ and $\Delta>\frac{1}{2}h_{p_k}$. Suppose event \eqref{eq: g in interval} is true $\forall x^{[i]}\in\mathcal{X}_{p_{k-1}}, u^{[i]}\in\mathcal{U}_{p_{k-1}}$, $i\in\mathcal{V}$. For any $\tilde{x}^{[i]}\in\mathcal{X}$, $u^{[i]}\in\mathcal{U}$ and $\epsilon>0$, if
\begin{align}\label{eq: rho( Reach, x_unsafe) >x}
    \rho\big(x^{[i]},\mathcal{A}\big)\geqslant \Delta, \forall x^{[i]}\in\tilde{\textsf{FR}}_k^{[i]}(\tilde{x}^{[i]}, u^{[i]},p_k,\epsilon),
\end{align}
then  $\forall y^{[i]}\in \cup_{\rho(\ring{x}^{[i]},\tilde{x}^{[i]})\leqslant h_p}G^{[i]}_{\epsilon}(\ring{x}^{[i]},u^{[i]})$,   $\rho(y^{[i]},\mathcal{A})\geqslant\Delta$.

{\bf Proof: }
We prove the claim by contradiction. Suppose \eqref{eq: rho( Reach, x_unsafe) >x} is true but there exists $y^{[i]}\in G^{[i]}_{\epsilon}(\ring{x}^{[i]},u^{[i]})$, $\ring{x}^{[i]}\in\mathcal{B}(\tilde{x}^{[i]},h_{p_k})$ such that $\rho(y^{[i]},\mathcal{A})<\Delta$.


Since $\mathcal{A}$ is closed, there exists $\hat{x}^{[i]}\in\mathcal{A}$ such that $\rho(y^{[i]},\hat{x}^{[i]})=\rho(y^{[i]},\mathcal{A})<\Delta$.
Let $y^{[i]}_{j}$ be the $j$-th element of $y^{[i]}$. Define operation $\textsf{Floor}(z)\triangleq\max\{z'\in\mathbb{Z}|z'\leqslant z\}$ that finds the largest integer no greater than $z\in\mathbb{R}$ and recall $\textsf{Ceil}(z)=\min\{z'\in\mathbb{Z}|z'\geqslant z\}$.  Denote real values $\ubar{y}^{[i]}_{j}\triangleq h_{p_k}\textsf{Floor}(\frac{y^{[i]}_{j}}{h_{p_k}})$ and  $\bar{y}^{[i]}_{j}\triangleq h_{p_k}\textsf{Ceil}(\frac{y^{[i]}_{j}}{h_{p_k}})$.  
Then for each dimension $j$, we have two cases:

1). $\hat{x}^{[i]}_{j}\in[\ubar{y}^{[i]}_{j},\bar{y}^{[i]}_{j}]$. Notice that $0\leqslant\bar{y}^{[i]}_{j}-\ubar{y}^{[i]}_{j}\leqslant h_{p_k}$. Choose $\check{x}^{[i]}_{j}=\frac{1}{2}(\bar{y}^{[i]}_{j}+\ubar{y}^{[i]}_{j})$. It is easy to see  that $|\check{x}^{[i]}_{j}-\hat{x}^{[i]}_{j}|\leqslant\frac{1}{2}h_{p_k}<\Delta$.
 
2). $\hat{x}^{[i]}_{j}\not\in[\ubar{y}^{[i]}_{j},\bar{y}^{[i]}_{j}]$.  We can select $\check{x}^{[i]}_{j}=\ubar{y}^{[i]}_{j}$ if $\hat{x}^{[i]}_{j}<\ubar{y}^{[i]}_{j}$; otherwise, we have $\hat{x}^{[i]}_{j}>\bar{y}^{[i]}_{j}$, and  we can select $\check{x}^{[i]}_{j}=\bar{y}^{[i]}_{j}$. Note that $y^{[i]}_{j}\in[\ubar{y}^{[i]}_{j},\bar{y}^{[i]}_{j}]$. Therefore under this selection, we have $|\check{x}^{[i]}_{j}-\hat{x}^{[i]}_{j}|\leqslant |y^{[i]}_{j}-\hat{x}^{[i]}_{j}|< \Delta$.

Since $\hat{x}^{[i]}\in\mathcal{A}$, the above two cases imply that $\rho(\check{x}^{[i]},\mathcal{A})\leqslant\rho(\check{x}^{[i]},\hat{x}^{[i]})< \Delta$ and $\check{x}^{[i]}\in\mathcal{X}_{p_k}$. 
Note that 
$$\check{x}^{[i]}\in\mathcal{B}(y^{[i]},h_{p_k})= y^{[i]}+h_{p_k}\mathcal{B}\subset[G^{[i]}_{\epsilon}(\ring{x}^{[i]},u^{[i]})+h_{p_k}\mathcal{B}].$$
Combining this with \eqref{eq: discrete property} renders $\check{x}^{[i]}\in \Gamma^{[i]}_{\epsilon,h_{p_k}}(\tilde{x}^{[i]},u^{[i]})$. Since \eqref{eq: g in interval} is true, we have $\check{x}^{[i]}\in\tilde{\textsf{FR}}_k^{[i]}(\tilde{x}^{[i]}, u^{[i]},p_k,\epsilon)$,  which contradicts \eqref{eq: rho( Reach, x_unsafe) >x}. $\hfill\blacksquare$
\end{lemma}

 Let $x^{[i]}\in\mathcal{X}_{p_k}$ and $u^{[i]}\in\mathcal{U}_{p_k}$.
By definition, we can write 
\begin{align}\label{eq: rewrite reach*}
	\textsf{FR}^{[i]}_{k}(x^{[i]},u^{[i]})= \mathcal{B}^{[i]}_{x^{[i]},u^{[i]},k}\cap\mathcal{X}_{p_k},
\end{align}
where $\mathcal{B}^{[i]}_{x^{[i]},u^{[i]},k}\triangleq x^{[i]}+\epsilon^{[i]}(f(x^{[i]},u^{[i]})+\mu^{[i]}_k(x^{[i]},u^{[i]}))+ (\gamma\bar{\sigma}_k^{[i]}\epsilon^{[i]}+\alpha^{[i]}_{p_k}+h_{p_k})\mathcal{B}$. 
 Let $y^{[i]}\in[x^{[i]}+h_{p_k}\mathcal{B}]\cap\mathcal{X}_{p_{k-1}}$. Lemma \ref{lemma: By k+1 in Bx} below characterizes the relation between $\mathcal{B}_{x^{[i]},u^{[i]},k}^{[i]}$ and $\mathcal{B}_{y^{[i]},u^{[i]},k-1}^{[i]}$.
\begin{lemma}\label{lemma: By k+1 in Bx}
	Suppose $4\gamma\bar{\sigma}\epsilon^{[i]}\leqslant h_{p_{\tilde{k}}}$ and \eqref{eq: g in interval} holds.
It holds that	$\mathcal{B}_{x^{[i]},u^{[i]},k}^{[i]}+h_{p_{k}}\mathcal{B}\subset\mathcal{B}_{y^{[i]},u^{[i]},k-1}^{[i]}$.

	\textbf{Proof:}
	Let $\check{x}^{[i]}\in\mathcal{B}_{x^{[i]},u^{[i]},k}^{[i]}+h_{p_{k}}\mathcal{B}$. 
	Denote $\tilde{f}^{[i]}_k(x^{[i]},u^{[i]})\triangleq f^{[i]}(x^{[i]},u^{[i]})+\mu^{[i]}_{k}(x^{[i]},u^{[i]})$.
	Then applying triangular inequality, we have
	\begin{align}\label{ineq: rho 0}
		&\rho\big(\check{x}^{[i]}, y^{[i]}+\epsilon^{[i]} \tilde{f}^{[i]}_{k-1}(y^{[i]},u^{[i]})\big)\nonumber\\
		&\leqslant \rho\big(\check{x}^{[i]}, x^{[i]}+\epsilon^{[i]} \tilde{f}^{[i]}_{k}(x^{[i]},u^{[i]})\big)\nonumber\\
		&+\rho\big(x^{[i]}+\epsilon^{[i]} \tilde{f}^{[i]}_{k}(x^{[i]},u^{[i]}), y^{[i]}+\epsilon^{[i]} \tilde{f}^{[i]}_{k}(y^{[i]},u^{[i]})\big)\nonumber\\
		&+ \rho\big(y^{[i]}+\epsilon^{[i]} \tilde{f}^{[i]}_{k}(y^{[i]},u^{[i]}), y^{[i]}+\epsilon^{[i]}\tilde{f}^{[i]}_{k-1}(y^{[i]},u^{[i]})\big).
	\end{align}
	Next we find the upper bound of each term on the right hand side of \eqref{ineq: rho 0}.
	Since  $\check{x}^{[i]}\in\mathcal{B}_{x^{[i]},u^{[i]},k}^{[i]}+h_{p_{k}}\mathcal{B}$, we have
	\begin{align}\label{ineq: rho 1}
		\rho\big(\check{x}^{[i]}, x^{[i]}+\epsilon^{[i]} \tilde{f}^{[i]}_{k}(x^{[i]},u^{[i]})\big)\leqslant \gamma\bar{\sigma}_{k}^{[i]}\epsilon^{[i]}+\alpha^{[i]}_{p_{k}}+2h_{p_k}.
	\end{align}

	 Since $\rho(x^{[i]},y^{[i]})\leqslant h_{p_{k}}$, Lipschitz continuity  yields $\|f^{[i]}(x^{[i]},u^{[i]})-f^{[i]}(y,u^{[i]})\|\leqslant\ell^{[i]}h_{p_{k}}$. Then applying triangular inequality gives
	\begin{align}
		&\rho\big(x^{[i]}+\epsilon^{[i]} \tilde{f}^{[i]}_{k}(x^{[i]},u^{[i]}), y^{[i]}+\epsilon^{[i]} \tilde{f}^{[i]}_{k}(y^{[i]},u^{[i]})\big)\nonumber\\
		&\leqslant\|x^{[i]}-y^{[i]}\|+\|\epsilon^{[i]}\tilde{f}^{[i]}_{k}(x^{[i]},u^{[i]})-\epsilon^{[i]} \tilde{f}^{[i]}_{k}(y^{[i]},u^{[i]})\|\nonumber\\
		&\leqslant h_{p_{k}}+\epsilon^{[i]}\|f^{[i]}(x^{[i]},u^{[i]})-f^{[i]}(y^{[i]},u^{[i]})\|\nonumber\\
		&\quad+\epsilon^{[i]}\|\mu^{[i]}_{k}(x^{[i]},u^{[i]})- \mu^{[i]}_{k}(y^{[i]},u^{[i]})\|\nonumber\\
		&\leqslant h_{p_{k}}+\epsilon^{[i]}\ell^{[i]}h_{p_k}+\epsilon^{[i]}\|\mu^{[i]}_{k}(x^{[i]},u^{[i]})- \mu^{[i]}_{k}(y^{[i]},u^{[i]})\|.\label{ineq: rho x+fk+1, y+fk+1}
	\end{align}
	Since \eqref{eq: g in interval} holds for all $x^{[i]}\in\mathcal{X}$ and $u^{[i]}\in\mathcal{U}$, we have
	\begin{align}\label{eq: ||g-mu||}
		&\|g^{[i]}(x^{[i]},u^{[i]})-\mu^{[i]}_{k}(x^{[i]},u^{[i]})\|\leqslant \gamma\sigma^{[i]}_k(x^{[i]},u^{[i]})\leqslant  \gamma\bar{\sigma}_{k}^{[i]} \nonumber\\
		&\textrm{and } \|g^{[i]}(y^{[i]},u^{[i]})-\mu^{[i]}_{k-1}(y^{[i]},u^{[i]})\|\leqslant   \gamma\bar{\sigma}_{k-1}^{[i]}. 
	\end{align}
	Since $\rho(x^{[i]},y^{[i]})\leqslant h_{p_{k}}$, the Lipschitz continuity of $g^{[i]}$ renders $\|g^{[i]}(x^{[i]},u^{[i]})-g^{[i]}(y^{[i]},u^{[i]})\|\leqslant \ell^{[i]} h_{p_{k}}$. Then applying triangular inequality gives
	\begin{align*}
		&\|\mu^{[i]}_{k}(x^{[i]},u^{[i]})- \mu^{[i]}_{k}(y^{[i]},u^{[i]})\|\\
		&= \|\mu^{[i]}_{k}(x^{[i]},u^{[i]})- g^{[i]}(x^{[i]},u^{[i]})+g^{[i]}(x^{[i]},u^{[i]})\\
		&\quad- g^{[i]}(y^{[i]},u^{[i]})+g^{[i]}(y^{[i]},u^{[i]})-\mu^{[i]}_{k}(y^{[i]},u^{[i]})\|\\
		&\leqslant  \|\mu^{[i]}_{k}(x^{[i]},u^{[i]})- g^{[i]}(x^{[i]},u^{[i]})\|\\
		&\quad+\|g^{[i]}(x^{[i]},u^{[i]})- g^{[i]}(y^{[i]},u^{[i]})\|\\
		&\quad+\|g^{[i]}(y^{[i]},u^{[i]})- \mu^{[i]}_{k}(y^{[i]},u^{[i]})\|\leqslant \ell^{[i]}h_{p_{k}}+2\gamma\bar{\sigma}_{k}^{[i]}.
	\end{align*}
	Combining the above inequality with \eqref{ineq: rho x+fk+1, y+fk+1} gives
	\begin{align}\label{ineq: rho 2}
		&\rho\big(x^{[i]}+\epsilon^{[i]} \tilde{f}^{[i]}_{k}(x^{[i]},u^{[i]}), y^{[i]}+\epsilon^{[i]} \tilde{f}^{[i]}_{k}(y^{[i]},u^{[i]})\big)\nonumber\\
		&\leqslant h_{p_{k}}+2\epsilon^{[i]}\ell^{[i]}h_{p_k}+2\gamma\bar{\sigma}_{k}^{[i]}\epsilon^{[i]}.
	\end{align}
	
	Applying triangular inequality,  we can further write 
	\begin{align}\label{ineq: rho 3}
		&\rho\big(y^{[i]}+\epsilon^{[i]} \tilde{f}^{[i]}_{k}(y^{[i]},u^{[i]}), y^{[i]}+\epsilon^{[i]} \tilde{f}^{[i]}_{k-1}(y^{[i]},u^{[i]})\big)\nonumber\\
		&=\|\epsilon^{[i]} \tilde{f}^{[i]}_{k}(y^{[i]},u^{[i]})-\epsilon^{[i]} \tilde{f}^{[i]}_{k-1}(y^{[i]},u^{[i]})\|
		\nonumber \\
		&=\|\epsilon^{[i]}(f+g^{[i]}-g^{[i]}+\mu^{[i]}_k)(y^{[i]},u^{[i]})\nonumber\\
		&\quad-\epsilon^{[i]}(f+g^{[i]}-g^{[i]}+\mu^{[i]}_{k-1})(y^{[i]},u^{[i]})\|\nonumber\\
		&\leqslant \epsilon^{[i]}\|(\mu^{[i]}_k-g^{[i]})(y^{[i]},u^{[i]})\|\nonumber\\
		&\quad+\epsilon^{[i]}\|(\mu^{[i]}_{k-1}-g^{[i]})(y^{[i]},u^{[i]})\|\nonumber\\
		&\leqslant \epsilon^{[i]}\gamma\bar{\sigma}_{k}^{[i]}+\epsilon^{[i]}\gamma\bar{\sigma}_{k-1}^{[i]},
	\end{align}
	where the last inequality follows from \eqref{eq: ||g-mu||}.
	
	Returning to \eqref{ineq: rho 0}, combining \eqref{ineq: rho 1}, \eqref{ineq: rho 2} and \eqref{ineq: rho 3} gives
	\begin{align}\label{ineq: rho (x'', y +)}
		&\rho\big(\check{x}^{[i]}, y^{[i]}+\epsilon^{[i]} \tilde{f}^{[i]}_{k-1}(y^{[i]},u^{[i]})\big)\nonumber\\
		&\leqslant4\gamma\bar{\sigma}_{k}^{[i]}\epsilon^{[i]}+\alpha^{[i]}_{p_{k}}+3h_{p_{k}}+2\epsilon^{[i]}\ell^{[i]}h_{p_k}+\gamma\bar{\sigma}_{k-1}^{[i]}\epsilon^{[i]}
		.
	\end{align}
	Note that  $4\gamma\bar{\sigma}\epsilon^{[i]}\leqslant h_{p_{\tilde{k}}}\leqslant h_{p_k}$. By equation (2.24) in \cite{williams2006gaussian}, $\bar{\sigma}\geqslant \bar{\sigma}^{[i]}_{k'}$ for all $k'\geqslant 0$.
	Note that $\alpha^{[i]}_{p_k}= 2h_{p_k}+2\epsilon^{[i]} h_{p_k}\ell^{[i]}+(\epsilon^{[i]})^2\ell^{[i]}m^{[i]}$
	and $h_{p_{k-1}}=2h_{p_{k}}$.
	Combining these gives
	\begin{align*}
		&\rho\big(\check{x}^{[i]}, y^{[i]}+\epsilon^{[i]} \tilde{f}^{[i]}_{k-1}(y^{[i]},u^{[i]})\big)\\
		&\leqslant \alpha^{[i]}_{p_{k}}+4h_{p_{k}}+2\epsilon^{[i]}\ell^{[i]}h_{p_k}+\gamma\bar{\sigma}_{k-1}^{[i]}\epsilon^{[i]}\\
		&= \alpha^{[i]}_{p_{k-1}}+h_{p_{k-1}}+\gamma\bar{\sigma}_{k-1}^{[i]}\epsilon^{[i]},
	\end{align*}
	which implies $\check{x}^{[i]}\in\mathcal{B}_{y^{[i]},u^{[i]},k-1}^{[i]}$. $\hfill\blacksquare$
\end{lemma}

\subsection{Proof of Theorem \ref{thm: system safe}}\label{proof: system safe}

Denote $\bar{\mathcal{X}}^{[i]}_O[k_0,k_1)\triangleq\mathcal{X}_O\bigcup\cup_{j< i,t\in[k_0\xi,k_1\xi)}\mathcal{B}(x_q^{[j]}(t),2\zeta)$ the obstacle regions (i.e., static obstacles and the robots with higher priority) for robot $i$ from time $t_0$ to $t_1$. Denote shorthand $x^{[i]}[k]\triangleq x^{[i]}(k\xi)$ for the state in discrete time.

{\em Roadmap of the proof:}  First, in Lemma \ref{lmm: reach in safe}, we establish  that $\textsf{FR}_k^{[i]}(x^{[i]},u^{[i]})$ is invariant in  $\mathcal{X}^{[i]}_{\text{safe},k}$ under inputs in $\mathcal{U}^{[i]}_{p_k}(x^{[i]})$. Then we examine the distance between $\mathcal{X}^{[i]}_{\text{safe},k}$ and  $\bar{\mathcal{X}}^{[i]}_O[k,k+2)$  in Lemma \ref{lemma: from obstacle}. Next Lemma  \ref{lmm: x(t) in X free} utilizes the previous two lemmas to show that system \eqref{eq: observation model} stays safe throughout for a duration $[t^{[i]}_{k+1,n}, t^{[i]}_{k+1,n+1}]$ under constant control, and Lemma  \ref{lemma: safe in iteration always} extends the safety result to the piecewise constant control law rendered by dSLAP for one iteration. Finally, we incorporate the concentration inequality in Lemma \ref{lmm: concentration GPR} and prove Theorem \ref{thm: system safe}.

\begin{lemma}\label{lmm: reach in safe}
For all $x^{[i]}\in\mathcal{X}^{[i]}_{\text{safe},k}$,  it holds that $\mathcal{U}^{[i]}_{p_k}(x^{[i]})\neq\emptyset$ and $\textsf{FR}_k^{[i]}(x^{[i]},u^{[i]})\subset\mathcal{X}^{[i]}_{\text{safe},k}$ for all  $u^{[i]}\in\mathcal{U}^{[i]}_{p_k}(x^{[i]})$.

{\bf Proof: }
We prove the lemma by contradiction. 

Suppose $\mathcal{U}^{[i]}_{p_k}(x^{[i]})=\emptyset$. 
Control input removal only takes place in
 the \textsf{OCA} procedure and the \textsf{UnsafeUpdate} procedure. In both procedures, when it reduces to $\mathcal{U}^{[i]}_{p_k}(x^{[i]})=\emptyset$, the procedures rule that $x^{[i]}\in\mathcal{X}^{[i]}_{\text{unsafe},k,0}$ or $x^{[i]}\in\bar{\mathcal{X}}^{[i]}_{\text{unsafe},k,j}$, $j<i$, through the \textsf{Add} procedure. Therefore,  $x^{[i]}\not\in\mathcal{X}^{[i]}_{\text{safe},k}$.

Suppose there exists $u^{[i]}\in\mathcal{U}^{[i]}_{p_k}(x^{[i]})$, such that $\textsf{FR}_k^{[i]}(x^{[i]},u^{[i]}) \cap[\cup_{j<i}\bar{\mathcal{X}}^{[i]}_{\text{unsafe},k,j}]\neq\emptyset$. Note that the \textsf{OCA} procedure constructs the $\textsf{BR}^{[i]}_k$ set such that $x^{[i]}\in\textsf{BR}^{[i]}_k(\tilde{x}^{[i]},u^{[i]})$ for all $\tilde{x}^{[i]}\in \textsf{FR}_k^{[i]}(x^{[i]}, u^{[i]})$. Due to the \textsf{UnsafeUpdate} procedure, if $\tilde{x}^{[i]}\in[\cup_{j<i}\bar{\mathcal{X}}^{[i]}_{\text{unsafe},k,j}]$ and $x^{[i]}\in\textsf{BR}^{[i]}(\tilde{x}^{[i]},u^{[i]})$, we must have $u^{[i]}\not\in\mathcal{U}^{[i]}_{p_k}(x^{[i]})$.
Hence, this case is impossible. 
$\hfill\blacksquare$
\end{lemma}

 The following lemma characterizes the distance between $\mathcal{X}^{[i]}_{\text{safe},k}$ and $\bar{\mathcal{X}}^{[i]}_O[k,k+2)$.
\begin{lemma}\label{lemma: from obstacle}
It holds that $\rho(x^{[i]},\bar{\mathcal{X}}^{[i]}_O[k,k+2))>m^{[i]}\epsilon^{[i]}+h_{p_k}$ for all $x^{[i]}\in\mathcal{X}^{[i]}_{\text{safe},k}$.

{\bf Proof: } 
Let $x^{[i]}\in\mathcal{X}^{[i]}_{\text{safe},k}$.
In \textsf{OCA}, when $\rho(x^{[i]},\mathcal{X}_O)\leqslant m^{[i]}\epsilon^{[i]}+h_{p_k}$,   $\textsf{Add}(\mathcal{X}^{[i]}_{\text{unsafe},k,0},x^{[i]})$ is executed such that $x^{[i]}\in\mathcal{X}^{[i]}_{\text{unsafe},k,0}$. Since $\mathcal{X}^{[i]}_{safe,k}\subset \mathcal{X}_{p_k}\setminus \mathcal{X}^{[i]}_{\text{unsafe},k,0}\neq\emptyset$,
\begin{align}\label{eq: rho X safe X O}
   \rho(x^{[i]},\mathcal{X}_O)>m^{[i]}\epsilon^{[i]}+h_{p_k}. 
\end{align}

In \textsf{ICA}, when $j<i$, $\mathcal{X}^{[i]}_{\text{unsafe},k,j}
\supset \mathcal{X}^{[j]}_{k}\cap\mathcal{X}_{p_k}$. Therefore, by definition of $ \mathcal{X}^{[j]}_{k}$,  if $$
\rho(x^{[i]}, x^{[j]}[k]+2\xi m^{[j]}\mathcal{B})\leqslant 2\zeta+m^{[i]}\epsilon^{[i]}+h_{p_k},$$
 we must have $x^{[i]}\in\mathcal{X}^{[i]}_{\text{unsafe},k,j}$. 
Hence 
\begin{align}\label{ineq: rho(x_q, x_q+2B)}
\rho(x^{[i]},x^{[j]}[k]+2\xi m^{[j]}\mathcal{B})>2\zeta+m^{[i]}\epsilon^{[i]}+h_{p_k}.    
\end{align}

Note that {\bf (A1)} in Assumption \ref{assmp: model} implies $$
\dot{x}^{[i]}(\tau)=f^{[i]}(x^{[i]}(\tau),u^{[i]}(\tau))+g^{[i]}(x^{[i]}(\tau),u^{[i]}(\tau))\in m^{[i]}\mathcal{B},$$
 for all $u^{[i]}(\tau)\in\mathcal{U}$. 
This implies 
$$
x^{[j]}(t)\in x^{[j]}[k]+2\xi m^{[j]}\mathcal{B},~\forall t\in[k\xi,(k+2)\xi).$$
  Combining this with \eqref{ineq: rho(x_q, x_q+2B)} gives,
$\forall t\in[k\xi,(k+2)\xi)$,
\begin{align}\label{ineq: rho X safe j}
  \rho(x^{[i]},\mathcal{B}(x^{[j]}(t),2\zeta))>m^{[i]}\epsilon^{[i]}+h_{p_k}.  
\end{align}

By definition,  obstacle region $\bar{\mathcal{X}}^{[i]}_O[k,k+2)=\mathcal{X}_O\bigcup\cup_{j<i,t\in[k\xi,(k+2)\xi)}\mathcal{B}(x^{[j]}(t),2\zeta)$. Hence the lemma directly follows from \eqref{eq: rho X safe X O} and  \eqref{ineq: rho X safe j}.$\hfill\blacksquare$

\end{lemma}


Recall that, for all iterations $k\geqslant 1$, \textsf{AL} and controller execution render $\{u^{[i]}(t^{[i]}_{k,n})\}_{n=0}^{\bar{n}^{[i]}-1}$, 
$u^{[i]}(t^{[i]}_{k,n})\in\mathcal{U}^{[i]}_{p_{k-1}}$, as control inputs, each executed for a duration $\epsilon^{[i]}$  by robot $i$ such that $u^{[i]}(t)=u^{[i]}(t^{[i]}_{k,n})$ for all $t\in[t^{[i]}_{k,n},t^{[i]}_{k,n+1})$. Then we have $x^{[i]}(t^{[i]}_{k,n})= x^{[i]}(t^{[i]}_{k,n-1}, t^{[i]}_{k,n-1}+\epsilon^{[i]},u^{[i]}(t^{[i]}_{k,n-1}))$.  The following lemma gives the sufficient conditions such that the robots are near the safe states within one iteration.
\begin{lemma}
\label{lemma: near safe states}

Suppose Assumption \ref{assmp: model} holds and \eqref{eq: g in interval} is true $\forall x^{[i]}\in\mathcal{X}_{p_{k-1}}$, $i\in\mathcal{V}$.
If  
$\mathcal{B}(x^{[i]}[k], h_{p_{k-1}}) \cap\mathcal{X}_{\text{safe},k-1}^{[i]}\neq\emptyset$
for all $i\in\mathcal{V}$ for some $k>1$, then $ \mathcal{B}(x^{[i]}(t^{[i]}_{k,n}) ,h_{p_{k-1}})\cap\mathcal{X}^{[i]}_{\text{safe},k-1}\neq\emptyset$ for all $n=0,1,\cdots, \bar{n}^{[i]}$, $i\in\mathcal{V}$.

{\bf Proof: }
We prove the lemma using induction.
For the base case $n=0$, the condition in the lemma statement and the definition $t^{[i]}_{k,0}=k\xi$  indicate that $\mathcal{B}(x^{[i]}(t^{[i]}_{k,0}), h_{p_{k-1}}) \cap\mathcal{X}_{\text{safe},k-1}^{[i]}\neq\emptyset$.

Now suppose that $ \mathcal{B}(x^{[i]}(t^{[i]}_{k,n}) ,h_{p_{k-1}})\cap\mathcal{X}^{[i]}_{\text{safe},k-1}\neq\emptyset$ holds up until $n=n'$. Then there exists $\tilde{x}^{[i]}_{n'}\in\mathcal{B}(x^{[i]}(t^{[i]}_{k,n'}), h_{p_{k-1}}) \cap\mathcal{X}_{\text{safe},k-1}^{[i]}$. By \textsf{AL} and controller execution, we have $u^{[i]}(t^{[i]}_{k,n'})\in\mathcal{U}^{[i]}_{p_{k-1}}(\tilde{x}^{[i]}_{n'})$.
By definition of $\tilde{x}^{[i]}_{n'}$, $\rho(x^{[i]}(t^{[i]}_{k,n'}),\tilde{x}^{[i]}_{n'})\leqslant h_{p_{k-1}}$. By Lemma \ref{lmm: phi in G},
we have
\begin{align*}
  &x^{[i]}\Big(t^{[i]}_{k,n'}, t^{[i]}_{k,n'}+\epsilon^{[i]},u^{[i]}(t^{[i]}_{k,n'})\Big)\\
  &\quad \in G_{\epsilon^{[i]}}(x^{[i]}(t^{[i]}_{k,n'}),u^{[i]}(t^{[i]}_{k,n'})).  
\end{align*}
Then Lemma \ref{lmm: nearest in reach} renders
\begin{align*}
    &\textsf{Nearest}(x^{[i]}\Big(t^{[i]}_{k,n'}, t^{[i]}_{k,n'}+\epsilon^{[i]},u^{[i]}(t^{[i]}_{k,n'})\Big),\mathcal{X}_{p_{k-1}})\\
    &\in\tilde{\textsf{FR}}_{k-1}^{[i]}(\tilde{x}_{n'}^{[i]},u^{[i]}(t^{[i]}_{k,n'}),p_{k-1},\epsilon^{[i]}).
\end{align*}
Note that \eqref{discrete}  implies
\begin{align*}
   &\textsf{Nearest}(x^{[i]}\Big(t^{[i]}_{k,n'}, t^{[i]}_{k,n'}+\epsilon^{[i]},u^{[i]}(t^{[i]}_{k,n'})\Big),\mathcal{X}_{p_{k-1}})\\
   &\in\mathcal{B}(x^{[i]}\Big(t^{[i]}_{k,n'}, t^{[i]}_{k,n'}+\epsilon^{[i]},u^{[i]}(t^{[i]}_{k,n'})\Big),h_{p_{k-1}}). 
\end{align*}

Since $x^{[i]}(t^{[i]}_{k,n'+1})=x^{[i]}(t^{[i]}_{k,n'}, t^{[i]}_{k,n'}+\epsilon^{[i]},u^{[i]}(t^{[i]}_{k,n'}))$, then the above two statements give
\begin{align}\label{eq: nearest in reach x safe}
 &\mathcal{B}(x^{[i]}(t^{[i]}_{k,n'+1}),h_{p_{k-1}})\nonumber\\
 &\cap \tilde{\textsf{FR}}_{k-1}^{[i]}(\tilde{x}_{n'}^{[i]},u^{[i]}(t^{[i]}_{k,n'}),p_{k-1},\epsilon^{[i]})
 \neq\emptyset.   
\end{align}
Lemma \ref{lemma: reach*} implies
$$
\tilde{\textsf{FR}}_{k-1}^{[i]}(\tilde{x}^{[i]}_{n'},u^{[i]}(t^{[i]}_{k,n'}),p_{k-1},\epsilon^{[i]})\subset \textsf{FR}_{k-1}^{[i]}(\tilde{x}^{[i]}_{n'},u^{[i]}(t^{[i]}_{k,n'})).
$$ 
Since $\tilde{x}^{[i]}_{n'}\in\mathcal{X}_{\text{safe},k-1}^{[i]}$,
Lemma \ref{lmm: reach in safe} implies $\textsf{FR}_{k-1}^{[i]}(\tilde{x}^{[i]}_{n'},u^{[i]}(t^{[i]}_{k,n'}))\subset\mathcal{X}^{[i]}_{\text{safe},k-1}$.
Combining these two statements with \eqref{eq: nearest in reach x safe} gives
$ \mathcal{B}(x^{[i]}(t^{[i]}_{k,n'+1}) ,h_{p_{k-1}})\cap\mathcal{X}^{[i]}_{\text{safe},k-1}\neq\emptyset$. The induction is completed.
 $\hfill\blacksquare$
\end{lemma}

The following lemma characterizes a sufficient condition for the trajectory of system \eqref{eq: observation model} under constant control input in $\mathcal{U}^{[i]}_{p_{k-1}}$ to be safe for a duration $\epsilon^{[i]}$.
\begin{lemma}\label{lmm: x(t) in X free}
Suppose Assumption \ref{assmp: model} holds and \eqref{eq: g in interval} is true $\forall x^{[i]}\in\mathcal{X}_{p_{k-1}}$, $i\in\mathcal{V}$. If $\tilde{x}^{[i]}\in  \mathcal{B}(x^{[i]}(t^{[i]}_{k,n}), h_{p_{k-1}}) \cap\mathcal{X}_{\text{safe},k-1}^{[i]}$ for all $i\in\mathcal{V}$ and some  $t^{[i]}_{k,n}\in\{t^{[i]}_{k,n}\}_{n=0}^{\bar{n}^{[i]}-1}$,  it holds that $$x^{[i]}(t^{[i]}_{k,n}, t^{[i]}_{k,n}+\epsilon,u^{[i]})\in\mathcal{X}^{[i]}_F(x^{[\neg i]}(t)), ~\forall \epsilon\in[0,\epsilon^{[i]}],$$ for all $u^{[i]}\in\mathcal{U}^{[i]}_{p_{k-1}}(\tilde{x}^{[i]})$, $t\in[t^{[i]}_{k,n},t^{[i]}_{k,n}+\epsilon^{[i]})$ and $i\in\mathcal{V}$.  

{\bf Proof: }
Recall that
Lemma \ref{lemma: reach*} implies
$\tilde{\textsf{FR}}_{k-1}^{[i]}\big(x^{[i]},u^{[i]}, p_{k-1}, \epsilon^{[i]}\big)\subset \textsf{FR}_{k-1}^{[i]}\big(x^{[i]},u^{[i]}\big)$ for all $x^{[i]}\in\mathcal{X}$ and $u^{[i]}\in\mathcal{U}$. Since $\tilde{x}^{[i]}\in  \mathcal{X}_{\text{safe},k-1}^{[i]}$, Lemma \ref{lmm: reach in safe} renders $\textsf{FR}_{k-1}^{[i]}(\tilde{x}^{[i]}, u^{[i]})\subset \mathcal{X}^{[i]}_{\text{safe},k-1}$.  Thus, $\tilde{\textsf{FR}}_{k-1}^{[i]}\big(x^{[i]},u^{[i]}, p_{k-1}, \epsilon^{[i]}\big)\subset\mathcal{X}^{[i]}_{\text{safe},k-1}$.
Combining these with Lemma \ref{lemma: from obstacle} gives
\begin{align}\label{eq: rho( Reach, x_unsafe)} 
    \rho\big(&x^{[i]},\bar{\mathcal{X}}^{[i]}_O[k-1,k+1)\big)>m^{[i]}\epsilon^{[i]}+h_{p_{k-1}},
\end{align}
for all $x^{[i]}\in \tilde{\textsf{FR}}_{k-1}^{[i]}(\tilde{x}^{[i]}, u^{[i]},{p_{k-1}},\epsilon^{[i]})$.

 Lemma \ref{lmm: phi in G} gives $x^{[i]}(t^{[i]}_{k,n}, t^{[i]}_{k,n}+\epsilon^{[i]},u^{[i]})\in G^{[i]}_{\epsilon^{[i]}}(x^{[i]}(t^{[i]}_{k,n}),u^{[i]})$. Since $\tilde{x}^{[i]}\in  \mathcal{B}(x^{[i]}(t^{[i]}_{k,n}), h_{p_{k-1}})$, we have $x^{[i]}(t^{[i]}_{k,n})\in \mathcal{B}(\tilde{x}^{[i]}, h_{p_{k-1}})$. Combining these two statements and \eqref{eq: rho( Reach, x_unsafe)}  with Lemma \ref{lmm: rho(G,x_unsafe)} renders
\begin{align}\label{eq: rho(phi,x_unsafe)>}
    &\rho(x^{[i]}(t^{[i]}_{k,n}, t^{[i]}_{k,n}+\epsilon^{[i]},u^{[i]}),\bar{\mathcal{X}}^{[i]}_O[k-1,k+1))\nonumber\\
    &>m^{[i]}\epsilon^{[i]}+h_{p_{k-1}}.
\end{align}
Then  by the definition of $m^{[i]}$, we have $ \rho(x^{[i]}(t^{[i]}_{k,n}, t^{[i]}_{k,n}+\epsilon,u^{[i]}), x^{[i]}(t^{[i]}_{k,n}))\leqslant m^{[i]}\epsilon^{[i]}$ for all $\epsilon\in[0,\epsilon^{[i]}]$ Combining this with  \eqref{eq: rho(phi,x_unsafe)>}, it renders that, for all $\epsilon\in[0,\epsilon^{[i]}]$, \begin{align*}
    &\rho(x^{[i]}(t^{[i]}_{k,n}, t^{[i]}_{k,n}+\epsilon,u^{[i]}),\bar{\mathcal{X}}^{[i]}_O[k-1,k+1)>h_{p_{k-1}}.
\end{align*}
Combining this with the definition of 
$
  \bar{\mathcal{X}}^{[i]}_O[\cdot,\cdot)
$
renders that
\begin{align}\label{eq: rho(phi,x_unsafe)> (j<i)}
    &x^{[i]}(t^{[i]}_{k,n}, t^{[i]}_{k,n}+\epsilon,u^{[i]})\nonumber\\
    &\in \mathcal{X}\setminus[\mathcal{X}_O\bigcup\cup_{j< i,t\in[(k-1)\xi,(k+1)\xi)}\mathcal{B}(x^{[j]}(t),2\zeta)].
\end{align}
Note that \eqref{eq: rho(phi,x_unsafe)> (j<i)} holds for all $i\in\mathcal{V}$, it further gives
\begin{align*}
    &x^{[i]}(t^{[i]}_{k,n}, t^{[i]}_{k,n}+\epsilon,u^{[i]})\\
    &\in \mathcal{X}\setminus[\mathcal{X}_O\bigcup\cup_{j\neq i,t\in[(k-1)\xi,(k+1)\xi)}\mathcal{B}(x^{[j]}(t),2\zeta)].
\end{align*}

By definitions, for any $t\in[(k-1)\xi,(k+1)\xi)$,
\begin{align*}
   &\mathcal{X}^{[i]}_F(x^{[\neg i]}(t))=\mathcal{X}\setminus [ \mathcal{X}_O\bigcup \cup_{j\neq i}\mathcal{B}(x^{[j]}(t),2\zeta)].
\end{align*}
Combining the above two statements completes the proof.
$\hfill\blacksquare$
\end{lemma}

The following lemma gives the sufficient conditions such that the robots steered by dSLAP are safe within one iteration.
\begin{lemma}\label{lemma: safe in iteration always}
Suppose Assumption \ref{assmp: model} holds and \eqref{eq: g in interval} is true $\forall x^{[i]}\in\mathcal{X}_{p_{k-1}}, u^{[i]}\in\mathcal{U}_{p_{k-1}}$, $i\in\mathcal{V}$.
If  
$\mathcal{B}(x^{[i]}[k], h_{p_{k-1}}) \cap\mathcal{X}_{\text{safe},k-1}^{[i]}\neq\emptyset$
for all $i\in\mathcal{V}$ for some $k>1$, then $x^{[i]}(t)\in\mathcal{X}^{[i]}_F(x^{[\neg i]}(t))$ for all $t\in[k\xi,k\xi+\xi)$ and  all $i\in\mathcal{V}$.

{\bf Proof: }
Lemma \ref{lemma: near safe states} shows that there exists $\tilde{x}_{n}^{[i]}\in\mathcal{B}(x^{[i]}(t^{[i]}_{k,n}), h_{p_{k-1}}) \cap\mathcal{X}_{\text{safe},k-1}^{[i]}$ for each $n=0,1,\bar{n}^{[i]}-1$, $i\in\mathcal{V}$.
By Lemma \ref{lmm: x(t) in X free}, we have  
$$
x^{[i]}(t^{[i]}_{k,n}, t^{[i]}_{k,n}+\epsilon,u^{[i]}(t^{[i]}_{k,n}))\in\mathcal{X}^{[i]}_F(x^{[\neg i]}(t)),
$$ 
for all $\epsilon\in[0,\epsilon^{[i]})$, $t\in[k\xi,(k+1)\xi)$, and  $u^{[i]}(t^{[i]}_{k,n})\in\mathcal{U}^{[i]}_{p_{k-1}}(\tilde{x}^{[i]}_{n})$ for each $n=0,1,\cdots,\bar{n}^{[i]}-1$. Recall that the definition that $t^{[i]}_{k,0}=k\xi$, $t^{[i]}_{k,n}=t^{[i]}_{k,n-1}+\epsilon^{[i]}$ and $t^{[i]}_{k,\bar{n}^{[i]}}=(k+1)\xi$. Hence, the lemma is proved.
 $\hfill\blacksquare$
\end{lemma}
Now we are ready to prove Theorem \ref{thm: system safe}.

{\bf Proof of Theorem \ref{thm: system safe}:}
Given Assumptions \ref{assmp: model} and \ref{assmp: ||g||_kappa} hold, for each robot $i$, \eqref{eq: g in interval} holds with probability at least $1-\delta_{\gamma,p_{k-1}}$   $\forall x^{[i]}\in\mathcal{X}_{p_{k-1}}, u^{[i]}\in\mathcal{U}_{p_{k-1}}$.  Applying the union bound, this gives \eqref{eq: g in interval} holds with probability at least $1-|\mathcal{V}|\delta_{\gamma,p_{k-1}}$, $\forall x^{[i]}\in\mathcal{X}_{p_{k-1}}, u^{[i]}\in\mathcal{U}_{p_{k-1}}$, $i\in\mathcal{V}$. Notice that  Lemma \ref{lemma: safe in iteration always} is given in the event of \eqref{eq: g in interval} is true $\forall x^{[i]}\in\mathcal{X}_{p_{k-1}}, u^{[i]}\in\mathcal{U}_{p_{k-1}}$, $i\in\mathcal{V}$. Therefore, we have  Lemma \ref{lemma: safe in iteration always} hold with probability at least $1-|\mathcal{V}|\delta_{\gamma,p_{k-1}}$.
$\hfill\blacksquare$

\subsection{Proof of Theorem \ref{thm2: system safe}}\label{appdix: proof of obstacle change}
Recall that  $\bar{\mathcal{X}}^{[i]}_{\text{unsafe},k,j}$ is the set of unsafe states induced by robot $1\leqslant j<i$ and $\bar{\mathcal{X}}^{[i]}_{\text{unsafe},k,0}$ is the set of unsafe states induced by static obstacles.
Given Theorem \ref{thm: system safe}, we can prove Theorem \ref{thm2: system safe} by showing that  if 
$
    \mathcal{B}(x^{[i]}[k], h_{p_{k-1}}) \cap\mathcal{X}^{[i]}_{\text{safe}, k-1}\neq \emptyset
$ holds for iteration $k$, it implies that $
    \mathcal{B}(x^{[i]}[k+1], h_{p_{k}}) \cap\mathcal{X}^{[i]}_{\text{safe}, k}\neq \emptyset
$ also holds, and hence Theorem \ref{thm2: system safe} can be proved by induction.
Therefore,  under the condition of $
    \mathcal{B}(x^{[i]}[k], h_{p_{k-1}}) \cap\mathcal{X}^{[i]}_{\text{safe}, k-1}\neq \emptyset
$, we study: 1) the distance between $x^{[i]}[k+1]$ and $\bar{\mathcal{X}}^{[i]}_{\text{unsafe},k-1,j}$ (Lemma \ref{lmm: rho(phi, X unsafe)>=2hp}); 2) the inclusion of  $\bar{\mathcal{X}}^{[i]}_{\text{unsafe},k,j}$ in terms of $\bar{\mathcal{X}}^{[i]}_{\text{unsafe},k-1,j}$ (Lemma \ref{lemma: invariant obstacle size}). The two results imply the distance between $x^{[i]}[k+1]$ and $\bar{\mathcal{X}}^{[i]}_{\text{unsafe},k,j}$ and further characterize the conditions for $\mathcal{B}(x^{[i]}[k+1], h_{p_{k}}) \cap\mathcal{X}^{[i]}_{\text{safe}, k}\neq\emptyset$ (Lemma \ref{lemma: safe in k+1 multi robot}). Then the proof is concluded by combining these results.

\subsubsection*{D.1) The distance between $x^{[i]}[k+1]$ and $\bar{\mathcal{X}}^{[i]}_{\text{unsafe},k-1,j}$}
\begin{lemma}\label{lmm: rho(phi, X unsafe)>=2hp}
Given Assumption \ref{assmp: model}  and event \eqref{eq: g in interval} hold $\forall x^{[i]}\in\mathcal{X}_{p_{k-1}}, u^{[i]}\in\mathcal{U}_{p_{k-1}}$, $i\in\mathcal{V}$ and $\mathcal{B}(x^{[i]}[k],h_{p_{k-1}})\cap\mathcal{X}_{\text{safe},k-1}^{[i]}\neq\emptyset$, it holds that
$
      \rho(x^{[i]}[k+1],\bar{\mathcal{X}}^{[i]}_{\text{unsafe},k-1,j})\geqslant 2h_{p_{k-1}},
$ for all $j<i$.

\textbf{Proof: }
Lemma \ref{lemma: near safe states} shows that there exists $\tilde{x}_{n}^{[i]}\in\mathcal{B}(x^{[i]}(t^{[i]}_{k,n}), h_{p_{k-1}}) \cap\mathcal{X}_{\text{safe},k-1}^{[i]}$ for each $n=0,1,\cdots,\bar{n}^{[i]}-1$, $i\in\mathcal{V}$.
Therefore, by Lemma \ref{lmm: reach in safe}, for all control inputs $u^{[i]}(t^{[i]}_{k,n})\in\mathcal{U}_{p_{k-1}}(\tilde{x}^{[i]}_n)$, we have 
$
\textsf{FR}_{k-1}^{[i]}\big(\tilde{x}^{[i]}_n,u^{[i]}(t^{[i]}_{k,n})\big)\subset\mathcal{X}_{\text{safe},k-1}^{[i]}.
$ 
Based on \textsf{OCA} and \textsf{ICA}, we have
\begin{align}\label{eq: X safe= Xp/X unsafe}
    \mathcal{X}_{\text{safe},k}^{[i]}=\mathcal{X}_{p_k}\setminus [ \cup_{j=0,\cdots,i-1}\bar{\mathcal{X}}^{[i]}_{\text{unsafe},k,j}].
\end{align}
It implies that $\mathcal{X}_{\text{safe},k-1}^{[i]}\cap \bar{\mathcal{X}}^{[i]}_{\text{unsafe},k-1,j}=\emptyset$ for all $j<i$.
 By \textsf{Discrete}, this renders 
$$
\rho(\textsf{FR}_{k-1}^{[i]}\big(\tilde{x}^{[i]}_n,u^{[i]}(t^{[i]}_{k,n})\big),\bar{\mathcal{X}}^{[i]}_{\text{unsafe},k-1,j})\geqslant h_{p_{k-1}},
$$
for each $n=0,1,\cdots, \bar{n}^{[i]}-1$.
Combining this with Lemma \ref{lemma: reach*}, we have, for each $n=0,1,\cdots,\bar{n}^{[i]}-1$,
\begin{align}\label{eq: rho reach, X unsafe >=2hp}
  &\rho(\tilde{\textsf{FR}}_{k-1}^{[i]}\big(\tilde{x}^{[i]}_n,u^{[i]}(t^{[i]}_{k,n}),p_{k-1},\epsilon^{[i]}\big),\bar{\mathcal{X}}^{[i]}_{\text{unsafe},k-1,j})\geqslant 2h_{p_{k-1}}.
\end{align}
By Lemma \ref{lmm: phi in G}, we have, for each $n=0,1,\cdots,\bar{n}^{[i]}-1$,  
\begin{align}\label{eq: from lemma ii.4}
   &x^{[i]}(t^{[i]}_{k,n}, t^{[i]}_{k,n}+\epsilon^{[i]},u^{[i]}(t^{[i]}_{k,n}))\in G^{[i]}_{\epsilon^{[i]}}(x^{[i]}(t^{[i]}_{k,n}),u^{[i]}(t^{[i]}_{k,n})). 
\end{align}
By definition of $\tilde{x}^{[i]}_n$, we have, for each $n=0,1,\cdots,\bar{n}^{[i]}-1$,
\begin{align}\label{eq: from lemma ii.7}
   \rho(x^{[i]}(t^{[i]}_{k,n}), \tilde{x}^{[i]}_n)\leqslant h_{p_{k-1}}. 
\end{align}
Combining \eqref{eq: rho reach, X unsafe >=2hp}, \eqref{eq: from lemma ii.4} and \eqref{eq: from lemma ii.7} with Lemma \ref{lmm: rho(G,x_unsafe)}, we have
\begin{align*}
    \rho(x^{[i]}(t^{[i]}_{k,n}, t^{[i]}_{k,n}+\epsilon^{[i]},u^{[i]}(t^{[i]}_{k,n})),\bar{\mathcal{X}}^{[i]}_{\text{unsafe},k-1,j})\geqslant2h_{p_{k-1}}, 
\end{align*}
for each $n=0,1,\cdots,\bar{n}^{[i]}-1$. Recall that the definition that $t^{[i]}_{k,0}=k\xi$, $t^{[i]}_{k,n}=t^{[i]}_{k,n-1}+\epsilon^{[i]}$ and $t^{[i]}_{k,\bar{n}^{[i]}}=(k+1)\xi$. Hence, the lemma is proved.
$\hfill\blacksquare$

\end{lemma}

\subsubsection*{D.2) The inclusion of $\mathcal{X}^{[i]}_{\text{unsafe},k,j}$}

Denote  $ \beta^{[i]}_{k,1}\triangleq 2\xi m^{[i]}+2\zeta+m^{[i]}\epsilon^{[i]}+2h_{p_k}$ and  
$\beta^{[i]}_{k,2}\triangleq \epsilon^{[i]}\gamma\bar{\sigma}_k^{[i]}$. Then in \textsf{ICA}, we have 
$\mathcal{X}^{[i]}_{k}=
x^{[i]}[k]+\beta^{[i]}_{k,1}\mathcal{B}$ and 
 $\mathcal{X}^{[i]}_{\text{unsafe},k,j}=[\mathcal{X}^{[j]}_{k}+\beta^{[i]}_{k,2}\mathcal{B}]\cap\mathcal{X}_{p_k}$.
Lemma \ref{lemma: k+1<k} below renders a monotonic property.
\begin{lemma}\label{lemma: k+1<k}
It holds that  $\beta^{[i]}_{k+1,2}\leqslant\beta^{[i]}_{k,2}$.

{\bf Proof:}
 Let $z\in\mathcal{X}\times\mathcal{U}$. Equation (11) in \cite{huber2014recursive} shows that $(\sigma^{[i]}_k(z))^2$ can be expressed recursively as 
\begin{align*}
    (\sigma^{[i]}_{k+1}(z))^2=(\sigma^{[i]}_{k}(z))^2-(\sigma^{[i]}_{k}(z))^2a^{[i]}_{k+1}(\sigma^{[i]}_{k}(z))^2
\end{align*}
where $a^{[i]}_{k+1}\geqslant 0$. Hence, $\sigma^{[i]}_{k+1}(z)\leqslant\sigma^{[i]}_{k}(z)$ and  $\bar{\sigma}^{[i]}_{k+1}\leqslant\bar{\sigma}^{[i]}_{k}$. Then the definition of $\beta^{[i]}_{k,2}$ yields $\beta^{[i]}_{k+1,2}\leqslant\beta^{[i]}_{k,2}$.
$\hfill\blacksquare$

\end{lemma}

	The following lemma characterizes the inclusion of $\mathcal{X}^{[i]}_{\text{unsafe},k,j}$ in terms of $\mathcal{X}^{[i]}_{\text{unsafe},k-1,j}$.
	\begin{lemma}\label{lemma: invariant obstacle size}
		Suppose $\xi\leqslant\frac{h_{p_{\tilde{k}}}}{\max_{j\in\mathcal{V}}m^{[j]}}$ and
		\eqref{eq: g in interval} holds for all $p_k$, $k\geqslant 1$. Suppose Assumption \ref{assmp: model} holds.
				Then, for all $j<i$, dSLAP renders 
		$\mathcal{X}^{[i]}_{\text{unsafe},k,j}\subset[\mathcal{X}^{[i]}_{\text{unsafe},k-1,j}+h_{p_{k}}\mathcal{B}]\cap\mathcal{X}_{p_{k}}$. 
		
		{\bf Proof:}
Consider $j=0$. Let $x^{[i]}\in\mathcal{X}^{[i]}_{\text{unsafe},k,0}$. By definition in $\textsf{OCA}$, it holds that    $x^{[i]}\in\mathcal{X}_{p_{k}}$ and $\rho(x^{[i]},\mathcal{X}_O)\leqslant m\epsilon^{[i]}+h_{p_k}$. 
Since \textsf{Discrete} renders $h_{p_{k-1}}=2h_{p_k}$, there exists $y^{[i]}\in\mathcal{X}_{p_{k-1}}$ satisfying $\rho(y^{[i]},x^{[i]})\leqslant h_{p_{k}}$. Then by triangular inequality, it holds that $\rho(y^{[i]},\mathcal{X}_O)\leqslant m\epsilon^{[i]}+2h_{p_k}=m\epsilon^{[i]}+h_{p_{k-1}}$. This implies $y^{[i]}\in\mathcal{X}^{[i]}_{\text{unsafe},k-1,0}$ and hence $x^{[i]}\in\mathcal{X}^{[i]}_{\text{unsafe},k-1,0}+h_{p_k}\mathcal{B}$. Since $x^{[i]}\in\mathcal{X}_{p_{k}}$, we further have 	$\mathcal{X}^{[i]}_{\text{unsafe},k,0}\subset[\mathcal{X}^{[i]}_{\text{unsafe},k-1,0}+h_{p_{k}}\mathcal{B}]\cap\mathcal{X}_{p_{k}}$.

		Consider $j=1,2,\cdots, i-1$. Let $x^{[i]}\in\mathcal{X}^{[i]}_{\text{unsafe},k,j}$. By definition in $\textsf{ICA}$, it holds that $x^{[i]}\in[x^{[j]}[k]+(\beta^{[j]}_{k,1}+\beta^{[i]}_{k,2})\mathcal{B}]\cap\mathcal{X}_{p_k}$. Since \textsf{Discrete} renders $h_{p_{k-1}}=2h_{p_k}$, we have $\beta^{[i]}_{k,1}<\beta^{[i]}_{k-1,1}-2h_{p_k}$. 
		Recall that Lemma \ref{lemma: k+1<k} renders $\beta_{k,2}^{[i]}\leqslant \beta^{[i]}_{k-1,2}$. Since $\xi\leqslant\frac{h_{p_{\tilde{k}}}}{\max_{j\in\mathcal{V}}m^{[j]}}$, it holds that $\rho(x^{[j]}[k],x^{[j]}[k-1])\leqslant m^{[j]}\xi\leqslant h_{p_{\bar{k}}}\leqslant h_{p_k}$. Combining the above three statements renders 
		\begin{align*}
		x^{[i]}&\in x^{[j]}[k]+(\beta^{[j]}_{k,1}+\beta^{[i]}_{k,2})\mathcal{B}\\
		&\subset x^{[j]}[k-1]+(\beta^{[j]}_{k,1}+\beta^{[i]}_{k,2}+h_{p_k})\mathcal{B}\\
		&\subset x^{[j]}[k-1]+(\beta^{[j]}_{k-1,1}+\beta^{[i]}_{k-1,2}-h_{p_k})\mathcal{B}.
		\end{align*}
		This implies $\rho(x^{[i]},x^{[j]}[k-1])\leqslant\beta^{[j]}_{k-1,1}+\beta^{[i]}_{k-1,2}-h_{p_k}$
		Similar to the logic above, \textsf{Discrete} renders that there exists $y^{[i]}\in\mathcal{X}_{p_{k-1}}$ satisfying $\rho(y^{[i]},x^{[i]})\leqslant h_{p_{k}}$. Triangular inequality further renders  $\rho(y^{[i]},x^{[j]}[k-1])\leqslant\beta^{[j]}_{k-1,1}+\beta^{[i]}_{k-1,2}$, or $$y^{[i]}\in [x^{[j]}[k-1]+(\beta^{[j]}_{k-1,1}+\beta^{[i]}_{k-1,2})\mathcal{B}]\cap\mathcal{X}_{p_{k-1}}=\mathcal{X}^{[i]}_{\text{unsafe},k-1,j}$$ and hence $x^{[i]}\in\mathcal{X}^{[i]}_{\text{unsafe},k-1,j}+h_{p_k}\mathcal{B}$. Since $x^{[i]}\in\mathcal{X}_{p_k}$, we have $x^{[i]}\in[\mathcal{X}^{[i]}_{\text{unsafe},k-1,j}+h_{p_{k}}\mathcal{B}]\cap\mathcal{X}_{p_{k}}$. $\hfill\blacksquare$
	\end{lemma}

\subsubsection*{D.3) Conditions for $ \mathcal{B}(x^{[i]}[k+1], h_{p_{k}})\cap\mathcal{X}^{[i]}_{safe, k}\neq \emptyset$}\label{proof: safe in k+1}
Define a sequence of sets such that  $\mathcal{P}^{[i]}_{k,j,1}\triangleq\mathcal{X}^{[i]}_{\text{unsafe},k,j}$, $j\geqslant 1$, and  $\mathcal{P}^{[i]}_{k,j,l}\triangleq\{x^{[i]}\in\mathcal{X}_{p_k}\setminus[\cup_{l'\leqslant l-1}\mathcal{P}^{[i]}_{k,j,l'}]\mid \forall u^{[i]}\in\mathcal{U}_{p_k}, ~x^{[i]}\in\textsf{BR}^{[i]}_k(\tilde{x}^{[i]},u^{[i]}) \textrm{ for some } \tilde{x}^{[i]}\in[\cup_{l'\leqslant l-1}\mathcal{P}^{[i]}_{k,j,l'}]\bigcup[\cup_{j'\leqslant j-1}\bar{\mathcal{X}}^{[i]}_{\text{unsafe},k,j}]
\}$.  
Lemma \ref{lemma: X=cup P} below characterizes $\bar{\mathcal{X}}^{[i]}_{\text{unsafe},k,j}$ using $\mathcal{P}^{(l)}_{k,i,j}$.

\begin{lemma}\label{lemma: X=cup P}
	Suppose Assumption \ref{assmp: model} holds.
	For all iterations $k$,  it holds that
	$\bar{\mathcal{X}}^{[i]}_{\text{unsafe},k,j}=\cup_{l=1}^{n^{[i]}_{k,j}}\mathcal{P}^{[i]}_{k,j,l}$ for some $n^{[i]}_{k,j}<\infty$. Furthermore, it holds that 
	$\mathcal{P}^{[i]}_{k,j,l}\neq\emptyset$ for all $l=1,\cdots, n^{[i]}_{k,j}$ and  $\mathcal{P}^{[i]}_{k,j,l}=\emptyset$ for all $l>n^{[i]}_{k,j}$.
	
	{\bf Proof:}
	By definition of $\mathcal{P}^{[i]}_{k,j,l}$, if $\mathcal{P}^{[i]}_{k,j,l}=\emptyset$, then
	\begin{align*}
		&\mathcal{P}^{[i]}_{k,j,l+1}=\{x^{[i]}\in\mathcal{X}_{p_k}\setminus[\cup_{l'\leqslant l}\mathcal{P}^{[i]}_{k,j,l'}]\mid \forall u^{[i]}\in\mathcal{U}_{p_k}, \\
		&\quad x^{[i]}\in\textsf{BR}^{[i]}_k(\tilde{x}^{[i]},u^{[i]}) \textrm{ for some }\\
		&\quad \tilde{x}^{[i]}\in\cup_{l'\leqslant l}\mathcal{P}^{[i]}_{k,j,l'}\bigcup[\cup_{j'\leqslant j-1}\bar{\mathcal{X}}^{[i]}_{\text{unsafe},k,j}]
		\}\\
		&= \{x^{[i]}\in\mathcal{X}_{p_k}\setminus[\cup_{l'\leqslant l-1}\mathcal{P}^{[i]}_{k,j,l'}]\mid \forall u^{[i]}\in\mathcal{U}_{p_k},   \\
		&\quad x^{[i]}\in\textsf{BR}^{[i]}_k(\tilde{x}^{[i]},u^{[i]})  \textrm{ for some }\\
		&\quad \tilde{x}^{[i]}\in\cup_{l'\leqslant l-1}\mathcal{P}^{[i]}_{k,j,l'}\bigcup[\cup_{j'\leqslant j-1}\bar{\mathcal{X}}^{[i]}_{\text{unsafe},k,j}]
		\}\\
		&=\mathcal{P}^{[i]}_{k,j,l}=\emptyset.
	\end{align*}
Therefore, we have
	$\mathcal{P}^{[i]}_{k,j,l}=\emptyset$ for all $l>l'$. The definition indicates that $\mathcal{P}^{[i]}_{k,j,l''}$ and $\cup_{l=1}^{l'}\mathcal{P}^{[i]}_{k,j,l}$ are mutually disjoint for any $l''> l'$.  Hence $n^{[i]}_{k,j}$ is finite since $\mathcal{X}_{p_k}$ is finite due to the compactness of $\mathcal{X}$ in Assumption \ref{assmp: model}, and $\mathcal{P}^{[i]}_{k,j,l}\neq\emptyset$, $\forall l=1,\cdots, n^{[i]}_{k,j}$ and  $\mathcal{P}^{[i]}_{k,j,l}=\emptyset$ $\forall l>n^{[i]}_{k,j}$.
	
	Now we show $\cup_{l=1}^{n^{[i]}_{k,j}}\mathcal{P}^{[i]}_{k,j,l}\subset\bar{\mathcal{X}}^{[i]}_{\text{unsafe},k,j}$.  For $j=0,1,\cdots,i-1$,  according to the \textsf{UnsafeUpdate} procedure, we have $\mathcal{P}^{[i]}_{k,j,1}=\mathcal{X}^{[i]}_{\text{unsafe},k,j}\subset\bar{\mathcal{X}}^{[i]}_{\text{unsafe},k,j}$. For any $x^{[i]}$ in non-empty $\mathcal{P}^{[i]}_{k,j,l'}$, $l'>1$, since $x^{[i]}\in\textsf{BR}^{[i]}_k(\tilde{x}^{[i]},u) \textrm{ for some } \tilde{x}^{[i]}\in\cup_{l=1}^{l'-1}\mathcal{P}^{(l)}_{k,i,j}
	$ 
	for all control inputs $u^{[i]}\in\mathcal{U}_{p_k}$, it renders that $\mathcal{U}_{p_k}^{[i]}(x^{[i]})=\emptyset$, and hence $x^{[i]}\in\bar{\mathcal{X}}^{[i]}_{\text{unsafe},k,j}$ according to  $\textsf{UnsafeUpdate}$.  Therefore, we have $\mathcal{P}^{[i]}_{k,j,l}\subset\bar{\mathcal{X}}^{[i]}_{\text{unsafe},k,j}$ for all non-empty $\mathcal{P}^{[i]}_{k,j,l}$, $l=1,\cdots, n^{[i]}_{k,j}$. This shows $\cup_{l=1}^{n^{[i]}_{k,j}}\mathcal{P}^{[i]}_{k,j,l}\subset\bar{\mathcal{X}}^{[i]}_{\text{unsafe},k,j}$.
	
	We show  $\bar{\mathcal{X}}^{[i]}_{\text{unsafe},k,j}\subset\cup_{l=1}^{n^{[i]}_{k,j}}\mathcal{P}^{[i]}_{k,j,l}$ using contradiction. Suppose there exists a state $x^{(1)}\in \bar{\mathcal{X}}^{[i]}_{\text{unsafe},k,j}$ and $x^{(1)}\not\in\mathcal{P}^{[i]}_{k,j,l}$ for all  $\mathcal{P}^{[i]}_{k,j,l}$,  $l=1,\cdots, n^{[i]}_{k,j}$.

		Obviously,  we have $x^{(1)}\not\in \mathcal{X}^{[i]}_{\text{unsafe},k,j}$ because otherwise $x^{(1)}\in\mathcal{P}^{(1)}_{k,i,j}$ according to the definition of $\mathcal{P}^{[i]}_{k,j,1}$, $j\geqslant 0$.
		Then  $x^{(1)}$ can only be added to $\bar{\mathcal{X}}^{[i]}_{\text{unsafe},k,j}$ by \textsf{UnsafeUpdate}. Then there  exists $x^{(2)}\in \bar{\mathcal{X}}^{[i]}_{\text{unsafe},k,j}$ such that $x^{(1)}\in\textsf{BR}^{[i]}_k(x^{(2)},u^{[i]})$, which reduces the control set $\mathcal{U}^{[i]}_{p_k}(x^{(1)})$ to an empty set and leads to the addition of $x^{(1)}$ to $\bar{\mathcal{X}}^{[i]}_{\text{unsafe},k,0}$. 
		Furthermore, we also have $x^{(2)}\not\in \mathcal{P}^{[i]}_{k,j,l}$ for any $l=1,\cdots, n^{[i]}_{k,j}-1$ since otherwise we have $x^{(1)}\in\mathcal{P}^{[i]}_{k,j,l+1}$. 
		By induction, we have a set $\{x^{(l)}\}_{l=1}^{n_l}\subset\bar{\mathcal{X}}^{[i]}_{\text{unsafe},k,j}$ 
		but $\{x^{(l)}\}_{l=1}^{n_l}\not\subset\cup_{l=1}^{n^{[i]}_{k,j}}\mathcal{P}^{[i]}_{k,j,l}$ for any $n_l=1,\cdots, |\mathcal{X}_{p_k}|$.  However, this is impossible because $\mathcal{X}_{p_k}$ is finite and $\cup_{l=1}^{n^{[i]}_{k,j}}\mathcal{P}^{[i]}_{k,j,l}\supset \mathcal{P}^{(1)}_{k,i,j}\neq \emptyset$. This gives  $\bar{\mathcal{X}}^{[i]}_{\text{unsafe},k,j}\subset\cup_{l=1}^{n^{[i]}_{k,j}}\mathcal{P}^{[i]}_{k,j,l}$.
	$\hfill\blacksquare$
	
\end{lemma}

The following lemma shows that the robot after one iteration is near the safe states under the priority assignment in the previous iteration.
\begin{lemma}\label{lemma: safe in k+1 multi robot}
	Suppose Assumption \ref{assmp: model} holds.  Suppose $4\gamma\bar{\sigma}\epsilon^{[i]}\leqslant h_{p_{\tilde{k}}}$ and \eqref{eq: g in interval} holds for all $x^{[i]}\in\mathcal{X}_{p_k}$, $u^{[i]}\in\mathcal{U}_{p_k}$, $k\geqslant 1$. Suppose $\xi\leqslant\frac{h_{p_{\tilde{k}}}}{\max_{j\in\mathcal{V}}m^{[j]}}$. 
	For each $i\in\mathcal{V}$, if 
	$\mathcal{B}(x^{[i]}[k], h_{p_{k-1}}) \cap\mathcal{X}_{safe,k-1}^{[i]}\neq\emptyset$, 
	$k\geqslant1$,  then 
	$
	\mathcal{B}(x^{[i]}[k+1], h_{p_{k}})\cap\mathcal{X}^{[i]}_{\text{safe}, k}\neq \emptyset
	$. 
	
	{\bf Proof: } Let $j=0,1,\cdots, i-1.$ Lemma \ref{lmm: rho(phi, X unsafe)>=2hp} shows that $ \rho(x^{[i]}[k+1],\bar{\mathcal{X}}^{[i]}_{\text{unsafe},k-1,j})\geqslant 2h_{p_{k-1}}$. By \eqref{eq: X safe= Xp/X unsafe},  this implies that $\mathcal{B}(x^{[i]}[k+1], h_{p_{k-1}}) \cap\mathcal{X}_{\text{safe},k-1}^{[i]}\neq\emptyset$. Let $\tilde{x}^{[i]}\in\mathcal{B}(x^{[i]}[k+1], h_{p_{k-1}}) \cap\mathcal{X}_{\text{safe},k-1}^{[i]}$. Then Lemma \ref{lmm: reach in safe} renders that there exists $\tilde{u}^{[i]}\in\mathcal{U}^{[i]}_{p_{k-1}}(\tilde{x}^{[i]})$ such that 
 $\textsf{FR}_{k-1}^{[i]}(\tilde{x}^{[i]},\tilde{u}^{[i]})\subset\mathcal{X}^{[i]}_{\text{safe},k-1}$. By \textsf{Discrete} and \eqref{eq: X safe= Xp/X unsafe},  this implies that, for all $j\in[0,i)$,
 \begin{align}\label{ineq: FR in X k-1 2 hpk}
 	\rho(\textsf{FR}_{k-1}^{[i]}(\tilde{x}^{[i]},\tilde{u}^{[i]}),\bar{\mathcal{X}}^{[i]}_{\text{unsafe},k-1,j})\geqslant h_{p_{k-1}}=2h_{p_k}.
 \end{align}
Based on the \textsf{UnsafeUpdate} procedure, it is obvious that
$\mathcal{X}^{[i]}_{\text{unsafe},k,j}\subset \bar{\mathcal{X}}^{[i]}_{\text{unsafe},k,j}$. Then \eqref{ineq: FR in X k-1 2 hpk} renders
\begin{align}\label{ineq: FR in hat X k-1 2 hpk}
	\rho(\textsf{FR}_{k-1}^{[i]}(\tilde{x}^{[i]},\tilde{u}^{[i]}),\mathcal{X}^{[i]}_{\text{unsafe},k-1,j})\geqslant h_{p_{k-1}}=2h_{p_k}.
\end{align}
Consider the rewriting in \eqref{eq: rewrite reach*} and recall that $\textsf{FR}_{k-1}^{[i]}(\tilde{x}^{[i]},\tilde{u}^{[i]})\subset\mathcal{X}_{p_{k-1}}$ and $\mathcal{X}^{[i]}_{\text{unsafe},k-1,j}\subset\mathcal{X}_{p_{k-1}}$. Then \eqref{ineq: FR in hat X k-1 2 hpk} implies
\begin{align}\label{eq: B tilde x, hat X unsafe}
\rho(\mathcal{B}^{[i]}_{\tilde{x}^{[i]},\tilde{u}^{[i]},k-1},\mathcal{X}^{[i]}_{\text{unsafe},k-1,j})>0.
\end{align}
 Due to the \textsf{Discrete} procedure, there exists $y^{[i]}\in\mathcal{X}_{p_k}$ such that $\rho(x^{[i]}[k+1],y^{[i]})\leqslant h_{p_k}$ and $\rho(\tilde{x}^{[i]},y^{[i]})\leqslant h_{p_k}$. Then combining \eqref{eq: B tilde x, hat X unsafe} with Lemma \ref{lemma: By k+1 in Bx} renders 
\begin{align}\label{eq: Bk hat X unsafe k}
\rho(\mathcal{B}^{[i]}_{y^{[i]},\tilde{u}^{[i]},k},\mathcal{X}^{[i]}_{\text{unsafe},k-1,j})> h_{p_k}.
\end{align}
Recall that Lemma \ref{lemma: invariant obstacle size} renders 
  \begin{align}\label{eq: re prop}
  	\mathcal{X}^{[i]}_{\text{unsafe},k,j}\subset[\mathcal{X}^{[i]}_{\text{unsafe},k-1,j}+h_{p_{k}}\mathcal{B}]\cap\mathcal{X}_{p_{k}}.
  \end{align}
Then combining \eqref{eq: Bk hat X unsafe k} and \eqref{eq: re prop} renders 
\begin{align}\label{ineq: By not in X unsafe}
	\rho(\mathcal{B}^{[i]}_{y^{[i]},\tilde{u}^{[i]},k},\mathcal{X}^{[i]}_{\text{unsafe},k,j})>0.
\end{align}
Note that \eqref{ineq: By not in X unsafe} holds for all $j<i$.  Then it follows that
 \begin{align}\label{eq: FR k y cap hat X unsafe empty}
 \textsf{FR}_{k}^{[i]}(y^{[i]},\tilde{u}^{[i]})\cap[\cup_{j<i}\mathcal{X}^{[i]}_{\text{unsafe},k,j}]=\emptyset.
 \end{align}
  Combining the claim below with \eqref{eq: X safe= Xp/X unsafe} renders that $y^{[i]}\in\mathcal{X}^{[i]}_{\text{safe},k}$. Combining this with  $\rho(x^{[i]}[k+1],y^{[i]})\leqslant h_{p_k}$ concludes the proof. $\hfill\blacksquare$

\begin{claim}\label{claim:FR not intersect bar X}
	It holds that $y^{[i]}\not\in[\cup_{j<i}\bar{\mathcal{X}}^{[i]}_{\text{unsafe},k,j}]$. 
\end{claim}

{\em Proof of Claim \ref{claim:FR not intersect bar X}:}
The proof of the claim is composed of three parts.

\underline{Part (i)}. We show that
\begin{align}\label{eq: y not in X unsafe}
y^{[i]}\not\in[\cup_{j<i}\mathcal{X}^{[i]}_{\text{unsafe},k,j}].
\end{align}
Let $j=0,1,\cdots,i-1$.
Since $\tilde{x}^{[i]}\in\mathcal{X}^{[i]}_{\text{safe},k-1}$, by \eqref{eq: X safe= Xp/X unsafe} and \textsf{Discrete}, it renders that $\rho(\tilde{x}^{[i]}, \bar{\mathcal{X}}^{[i]}_{\text{unsafe},k-1,j})\geqslant h_{p_{k-1}}$.
According to the construction of $\bar{\mathcal{X}}^{[i]}_{\text{unsafe},k-1,j}$ through $\textsf{UnsafeUpdate}$, it holds that $\mathcal{X}^{[i]}_{\text{unsafe},k-1,j}\subset\bar{\mathcal{X}}^{[i]}_{\text{unsafe},k-1,j}$. Combining the above two statements renders 
\begin{align}\label{ineq: tilde x X unsafe k-1}
\rho(\tilde{x}^{[i]}, \mathcal{X}^{[i]}_{\text{unsafe},k-1,j})\geqslant h_{p_{k-1}}=2h_{p_k}.
\end{align}
Next we show that $y^{[i]}\not\in\mathcal{X}^{[i]}_{\text{unsafe},k,j}$ through two cases.

Case 1: $j=0$. By construction of $\mathcal{X}^{[i]}_{\text{unsafe},k,0}$ in \textsf{OCA}, it holds that $x^{[i]}\in\mathcal{X}^{[i]}_{\text{unsafe},k,0}$ if and only if $\rho(x^{[i]},\mathcal{X}_O)\leqslant m\epsilon^{[i]}+h_{p_k}$. Combining this with \eqref{ineq: tilde x X unsafe k-1} renders $\rho(\tilde{x}^{[i]},\mathcal{X}_O)\geqslant m\epsilon^{[i]}+2h_{p_{k-1}}$. Recall that $h_{p_{k-1}}=2h_{p_k}$. Since $\rho(\tilde{x}^{[i]},y^{[i]})\leqslant h_{p_k}$, triangular inequality further gives $\rho(y^{[i]}, \mathcal{X}_O)\geqslant m\epsilon^{[i]}+3h_{p_{k}}$. Hence $y^{[i]}\not\in\mathcal{X}^{[i]}_{\text{unsafe},k,0}$.

Case 2: $j=1,\cdots, i-1$. Recall the definitions of $\beta^{[i]}_{k,1}$ and $\beta^{[i]}_{k,2}$ above Lemma \ref{lemma: k+1<k}. By construction of $\mathcal{X}^{[i]}_{\text{unsafe},k,j}$ in \textsf{ICA}, it holds that $x^{[i]}\in\mathcal{X}^{[i]}_{\text{unsafe},k,j}$ if and only if $\rho(x^{[i]},x^{[j]}[k])\leqslant \beta_{k,1}^{[i]}+\beta^{[i]}_{k,2}$. Combining this with \eqref{ineq: tilde x X unsafe k-1} renders $$
\rho(\tilde{x}^{[i]},x^{[j]}[k-1])\geqslant \beta^{[i]}_{k-1,1}+\beta^{[i]}_{k-1,2}+h_{p_{k-1}}.$$ Since $h_{p_{k-1}}=2h_{p_k}$, we have $\beta^{[i]}_{k,1}<\beta^{[i]}_{k-1,1}-2h_{p_k}$. 
Recall that Lemma \ref{lemma: k+1<k} renders $\beta_{k,2}^{[i]}\leqslant \beta^{[i]}_{k-1,2}$. Then combining the above three statements renders $$
\rho(\tilde{x}^{[i]},x^{[j]}[k-1])\geqslant \beta^{[i]}_{k,1}+\beta^{[i]}_{k,2}+4h_{p_{k}}.$$
Since $\xi\leqslant \frac{h_{p_{\bar{k}}}}{\max_{j\in\mathcal{V}}m^{[j]}}$, it holds that $\rho(x^{[j]}[k-1],x^{[j]}[k])\leqslant\xi m^{[j]}\leqslant h_{p_k}$. 
Combining the above two statements with triangular inequality renders 
$$
\rho(\tilde{x}^{[i]},x^{[j]}[k])\geqslant \beta^{[i]}_{k,1}+\beta^{[i]}_{k,2}+3h_{p_{k}}.
$$
Since $\rho(\tilde{x}^{[i]},y^{[i]})\leqslant h_{p_k}$, triangular inequality further gives $$
\rho(y^{[i]}, x^{[j]}[k])\geqslant \beta^{[i]}_{k,1}+\beta^{[i]}_{k,2}+2h_{p_{k}}.$$
 Hence $y^{[i]}\not\in\mathcal{X}^{[i]}_{\text{unsafe},k,j}$.
The proof of Part (i) is concluded.

\underline{Part (ii)}.
Consider a sequence of states and control inputs pairs $\{(x^{[i]}_{p_k,n},x^{[i]}_{p_{k-1},n},u^{[i]}_{p_k,n})\}$, $n=0,1,2,\cdots$, where 
\begin{align*}
&x^{[i]}_{p_k,0}\triangleq y^{[i]},~ x^{[i]}_{p_{k-1},0}\triangleq \tilde{x}^{[i]},~u^{[i]}_{p_k,0}\triangleq \tilde{u}^{[i]}, ~u^{[i]}_{p_k,n}\in\mathcal{U}_{p_{k-1}},\\
&x^{[i]}_{p_k,n+1}\in\textsf{FR}_{k}^{[i]}(x^{[i]}_{p_k,n},u^{[i]}_{p_k,n}),~ \rho(x_{p_k,n}^{[i]},x_{p_{k-1},n}^{[i]})\leqslant h_{p_k}.
\end{align*}
 We use induction to show that,   for all $n=0,1,2,\cdots$,
\begin{subequations}\label{eq: induction FRk cap hat X unsafe}
\begin{align}
	\textsf{FR}_{k}^{[i]}(x^{[i]}_{p_k,n},u^{[i]}_{p_k,n})\cap[\cup_{j<i}\mathcal{X}^{[i]}_{\text{unsafe},k,j}]=\emptyset\label{eq: induction FRk cap hat X unsafe1}\\
	\textsf{FR}_{k-1}^{[i]}(x^{[i]}_{p_{k-1},n},u^{[i]}_{p_k,n})\cap[\cup_{j<i}\bar{\mathcal{X}}^{[i]}_{\text{unsafe},k-1,j}]=\emptyset.\label{eq: induction FRk cap hat X unsafe2}
\end{align}
\end{subequations}

The base case $n=0$ is obvious by \eqref{eq: FR k y cap hat X unsafe empty} and \eqref{ineq: FR in X k-1 2 hpk} as well as the definitions of $x^{[i]}_{p_k,0}$, $x^{[i]}_{p_{k-1},0}$ and $u^{[i]}_{p_k,0}$.
	
	Now consider \eqref{eq: induction FRk cap hat X unsafe} holds until $n=m$.
By the definition of $\textsf{FR}^{[i]}_k$ and recall the rewriting in \eqref{eq: rewrite reach*}, it holds that
\begin{align*}
\mathcal{B}^{[i]}_{x^{[i]}_{p_k,m},u^{[i]}_{p_k,m},k}\supset &x^{[i]}_{p_k,m}+\epsilon^{[i]} (f^{[i]}(x^{[i]}_{p_k,m},u^{[i]}_{p_k,m})\\
&+\mu^{[i]}_k(x^{[i]}_{p_k,m},u^{[i]}_{p_k,m}))+2h_{p_k}\mathcal{B}.
\end{align*}
 Recall that \textsf{Discrete} renders $h_{p_{k-1}}=2h_{p_k}$.
 Therefore, it holds that $\mathcal{B}^{[i]}_{x^{[i]}_{p_k,m},u^{[i]}_{p_k,m},k}\cap\mathcal{X}_{p_{k-1}}\neq\emptyset$, and for every $x_{p_k,m+1}^{[i]}\in\textsf{FR}_{k}^{[i]}(x^{[i]}_{p_k,m},u^{[i]}_{p_k,m})$, there exists $x_{p_{k-1},m+1}^{[i]}\in\mathcal{X}_{p_{k-1}}$ satisfying $\rho(x_{p_k,m+1}^{[i]},x_{p_{k-1},m+1}^{[i]})\leqslant h_{p_k}$.  
Lemma \ref{lemma: By k+1 in Bx} renders $\mathcal{B}_{x_{p_k,m}^{[i]},u^{[i]}_{p_k,m},k}^{[i]}+h_{p_{k}}\mathcal{B}\subset\mathcal{B}_{x_{p_{k-1},m}^{[i]},u^{[i]}_{p_k,m},k-1}^{[i]}$. Combining the above two statements renders
\begin{align*}
x_{p_{k-1},m+1}^{[i]}\in&[\mathcal{B}^{[i]}_{x_{p_{k-1},m}^{[i]},u^{[i]}_{p_k,m},k-1}\cap\mathcal{X}_{p_{k-1}}]\\
=&\textsf{FR}_{k-1}^{[i]}(x_{p_{k-1},m}^{[i]},u^{[i]}_{p_k,m}).
\end{align*}
 Then it follows from the induction hypothesis that $$\textsf{FR}_{k-1}^{[i]}(x^{[i]}_{p_{k-1},m},u^{[i]}_{p_k,m})\cap[\cup_{j<i}\bar{\mathcal{X}}^{[i]}_{\text{unsafe},k-1,j}]=\emptyset$$ and hence $x_{p_{k-1},m+1}^{[i]}\in\mathcal{X}^{[i]}_{\text{safe},k-1}$. Furthermore, it follows from Lemma \ref{lmm: reach in safe} that there exists $u^{[i]}_{p_{k-1},m+1}\in\mathcal{U}_{p_{k-1}}$ such that $\textsf{FR}_{k-1}^{[i]}(x_{p_{k-1},m+1}^{[i]},u^{[i]}_{p_{k-1},m+1})\subset\mathcal{X}^{[i]}_{\text{safe},k-1}$, which implies $$	\rho(\textsf{FR}_{k-1}^{[i]}(x_{p_{k-1},m+1}^{[i]},u^{[i]}_{p_{k-1},m+1}),\bar{\mathcal{X}}^{[i]}_{\text{unsafe},k-1,j})\geqslant h_{p_{k-1}}.$$ Note that \textsf{Discrete} renders $\mathcal{U}_{p_{k-1}}\subset\mathcal{U}_{p_{k}}$. Hence, we can set $u^{[i]}_{p_k,m+1}=u^{[i]}_{p_{k-1},m+1}\in\mathcal{U}_{p_k}$. 
 Then following the same logic of \eqref{ineq: FR in X k-1 2 hpk} to \eqref{eq: FR k y cap hat X unsafe empty} by replacing $\tilde{x}^{[i]}$ with  $x_{p_{k-1},m+1}^{[i]}$, $\tilde{u}^{[i]}$ with $u^{[i]}_{p_{k-1},m+1}$, and $y^{[i]}$ with $x_{p_{k},m+1}^{[i]}$, we have $$\textsf{FR}_{k}^{[i]}(x_{p_{k},m+1}^{[i]},u^{[i]}_{p_k,m+1})\cap[\cup_{j<i}\mathcal{X}^{[i]}_{\text{unsafe},k,j}]=\emptyset.$$ The induction is completed.
 
 \underline{Part (iii)}.
Recall the definition of $\mathcal{P}^{[i]}_{k,j,l}$.
 Next we use induction to show that
 \begin{align}\label{induction: x not in P}
 x^{[i]}_{p_k,n}\not\in\cup_{l=1}^{n^{[i]}_{k,j}}\mathcal{P}^{[i]}_{k,j,l},\quad \forall j=0,1,\cdots,i-1 \text{ and } n=0,1,\cdots.
 \end{align}

Consider the base case $j=0$ and  all $n=0,1,2,\cdots$. By \eqref{eq: y not in X unsafe} and \eqref{eq: induction FRk cap hat X unsafe1}, we have $x^{[i]}_{p_k,n}\not\in\mathcal{P}^{[i]}_{k,0,1}$.
By \eqref{eq: induction FRk cap hat X unsafe1},  we have $\textsf{FR}_{k}^{[i]}(x^{[i]}_{p_k,n},u^{[i]}_{p_k,n})\cap\mathcal{X}^{[i]}_{\text{unsafe},k,0}]=\emptyset$.
Hence, we have $x^{[i]}_{p_k,n}\not\in\mathcal{P}^{[i]}_{k,0,2}$. Since $x^{[i]}_{p_k,n}\in\textsf{FR}_{k}^{[i]}(x^{[i]}_{p_k,n-1},u^{[i]}_{p_k,n-1})$, we have $x^{[i]}_{p_k,n-1}\not\in\mathcal{P}^{[i]}_{k,0,3}$. Following the same logic, it follows that $x^{[i]}_{p_k,n}\not\in\mathcal{P}^{[i]}_{k,0,l}$ for all $l=1,\cdots,n^{[i]}_{k,0}$.  This  renders   $x^{[i]}_{p_k,n}\not \in  \cup_{l=1}^{n^{[i]}_{k,0}}\mathcal{P}^{[i]}_{k,0,l}$.

Now suppose \eqref{induction: x not in P} holds  until $j=\tilde{j}<i-1$. By \eqref{eq: y not in X unsafe} and \eqref{eq: induction FRk cap hat X unsafe1}, we have $x^{[i]}_{p_k,n}\not\in\mathcal{P}^{[i]}_{k,\tilde{j}+1,1}$. Since the induction hypothesis renders $x^{[i]}_{p_k,n}\not\in\cup_{l=1}^{n^{[i]}_{k,j}}\mathcal{P}^{[i]}_{k,j,l}$ for all $j\leqslant\tilde{j}$ and  \eqref{eq: induction FRk cap hat X unsafe1} implies $\textsf{FR}_{k}^{[i]}(x^{[i]}_{p_k,n},u^{[i]}_{p_k,n})\cap\mathcal{X}^{[i]}_{\text{unsafe},k,\tilde{j}+1}]=\emptyset$,
it renders $x^{[i]}_{p_k,n}\not\in\mathcal{P}^{[i]}_{k,\tilde{j}+1,2}$ for all $n=0,1,2,\cdots$. By similar logic, we have $x^{[i]}_{p_k,n-1}\not\in\mathcal{P}^{[i]}_{k,\tilde{j}+1,3}$ and hence $x^{[i]}_{p_k,n}\not\in\cup_{l=1}^{n^{[i]}_{k,\tilde{j}}}\mathcal{P}^{[i]}_{k,\tilde{j}+1,l}$.  
This concludes the proof of \eqref{induction: x not in P}. 

Since $y^{[i]}=x^{[i]}_{p_k,0}$ and Lemma \ref{lemma: X=cup P} renders  $\bar{\mathcal{X}}^{[i]}_{\text{unsafe},k,j}=\cup_{l=1}^{n^{[i]}_{k,j}}\mathcal{P}^{[i]}_{k,j,l}$, \eqref{induction: x not in P} renders $y^{[i]}\not\in[\cup_{j<i}\bar{\mathcal{X}}^{[i]}_{\text{unsafe},k,j}]$. $\hfill\qedsymbol$
\end{lemma}

\subsubsection*{D.4) Proof of Theorem \ref{thm2: system safe}}
Given \eqref{eq: g in interval} holds  for all $x^{[i]}\in\mathcal{X}_{p_k}$, $u^{[i]}\in\mathcal{U}_{p_k}$,  $k\geqslant 1$, 
Lemma \ref{lemma: safe in k+1 multi robot} implies 
that if
$\mathcal{B}(x^{[i]}[k], h_{p_{k-1}}) \cap\mathcal{X}_{safe,k-1}^{[i]}\neq\emptyset$, 
for some  $k>1$,  then it holds that
$
\mathcal{B}(x^{[i]}[k'], h_{p_{k'-1}})\cap\mathcal{X}^{[i]}_{safe, k'-1}\neq \emptyset
$ for all $k'\geqslant k$.
Then Lemma \ref{lemma: safe in iteration always} implies that $x_q^{[i]}(t)\in\mathcal{X}^{[i]}_F(x_q^{[\neg i]}(t))$,  $\forall t\in[k'\xi, k'\xi+\xi)$, $k'\geqslant k$; i.e., $\forall t\geqslant k\xi$.

Note that \eqref{discrete} renders that $\mathcal{X}_p\subset\mathcal{X}_{p'}$ and $\mathcal{U}_p\subset \mathcal{U}_{p'}$  $\forall p<p'$, and dSLAP renders that $p_k\leqslant p_{\tilde{k}}$ $\forall k\geqslant 1$. Hence, the definition of $\delta_{\gamma,p}$  above \eqref{eq: g in interval} implies $\delta_{\gamma,p_{\tilde{k}}}\geqslant \delta_{\gamma,p_{k}}$, $\forall k\geqslant 1$. Therefore, for each $i\in\mathcal{V}$, for each $k\geqslant1$, \eqref{eq: g in interval} holds for all $x^{[i]}\in\mathcal{X}_{p_k}$, $u^{[i]}\in\mathcal{U}_{p_k}$, with probability at least $1-\delta_{\gamma,p_k}\geqslant 1-\delta_{\gamma,p_{\tilde{k}}}$. Denote $E^{[i]}_k$ as the event of \eqref{eq: g in interval} being violated for some  $x^{[i]}\in\mathcal{X}_{p_{\tilde{k}}}$ and/or $u^{[i]}\in\mathcal{U}_{p_{\tilde{k}}}$ at iteration $k$ by robot $i$. Then we have $Pr\{E^{[i]}_k\}\leqslant \delta_{\gamma,p_{k}}$.  Applying the union bound  (Theorem 2-3, \cite{papoulis2002probability}) renders that 
$Pr\{\cup_{k=1}^{\tilde{k}}E^{[i]}_k\}\leqslant \tilde{k}\delta_{\gamma,p_{\tilde{k}}} $. Note that $Pr\{\cap_{k=1}^{\tilde{k}}E^{[i]}_k\}=1-Pr\{\cup_{k=1}^{\tilde{k}}E^{[i]}_k\}$. 
Hence, we have 
\eqref{eq: g in interval} holds $\forall k\in\{1,\cdots,\tilde{k}\}$ with probability at least $1-\tilde{k}\delta_{\gamma,p_{\tilde{k}}}$.

Further applying the union bound renders that \eqref{eq: g in interval} holds $\forall k\in\{1,\cdots,\tilde{k}\}$ and $i\in\mathcal{V}$ with  probability at least $1-\tilde{k}|\mathcal{V}|\delta_{\gamma,p_{\tilde{k}}}$.  
$\hfill\blacksquare$

\section{Simulation}\label{sec:simulation}
In this section, we conduct a set of Monte Carlo simulations to evaluate the performance of the dSLAP algorithm. The simulations are run in Python, Linux Ubuntu 18.04 on an Intel Xeon(R) Silver 
4112 CPU, 2.60 GHz with 32 GB of RAM. 

{\em Simulation scenarios.}
We evaluate the dSLAP algorithm using Zermelo's navigation problem \cite{Zlobec2001} in a 2D space under the following scenario:  A group of robots are initially placed evenly on the plane and switch their positions at the destinations. The robots are immediately retrieved once they reach the goals. This example is also used in \cite{tanner2012multiagent} \cite{arslan2016coordinated} to demonstrate complicated multi-robot coordination scenarios.

{\em Dynamic models.}
Consider constant-speed boat robots with length $L=1.5$ meters (m) moving at speed $v=0.5$ meters/seconds (m/s).
For each robot $i$, let $x_{q,1}^{[i]}$ and $x_{q,2}^{[i]}$ be the $x$ and $y$ coordinates on a 2D plane, $x_r^{[i]}$ be the angle between the heading and the $x$-axis, and $u^{[i]}$ be the steering angle.
The state space is given by $\mathcal{X}=[0,100]\times[0,100]\times[-\pi,\pi]$. External wind disturbance $\nu$ is applied at  $x^{[i]}_{q,1}$  such that the system dynamics has the following form:
$
    \dot{x}_{q,1}^{[i]}(t)= 0.5\cos x_r^{[i]}(t)$,
    $\dot{x}_{q,2}^{[i]}(t)= 0.5\sin x_r^{[i]}(t)$, $\dot{x}_r^{[i]}(t)= \frac{0.5}{1.5}\tan u^{[i]}(t)$.
The control $u^{[i]}$ takes discrete values and the control space is $\mathcal{U}=\{\pm0.3\pi,\pm0.15\pi,0\}$.

{\em Parameters}. 
The kernel of GPR is configured as $\kappa(z,z')=0.0025\exp(-\frac{\|z-z'\|_2^2}{2})$, which is 0.0025 times the RBF kernel in the sklearn library. The factor 0.0025 is selected such that the supremum of the predictive standard deviation is 0.05, or 10\% of the robots' speed. This can be selected based on  the prior knowledge of the variability of the disturbance. 
Other parameters are selected as $\gamma=1$, $p_{init}=4$,  $\bar{k}=200$, $\bar{\tau}=20$, $\xi=8$, $q=2$, $\psi=1$, $\delta=0.1$, and $r_k^{[i]}=-\sigma_k^{[i]}$, which are determined according to the desired learning confidence level and the computation capability of the robots. To prevent prolonged computation due to unnecessarily fine discretization, we set a maximum such that $p_{k+1}\leftarrow\min\{p_k,5\}$  $\forall k\geqslant1$.

\begin{figure}[thpb]
	\centering
	
	\begin{subfigure}[t]{0.235\textwidth}
		\includegraphics[width=1\textwidth]{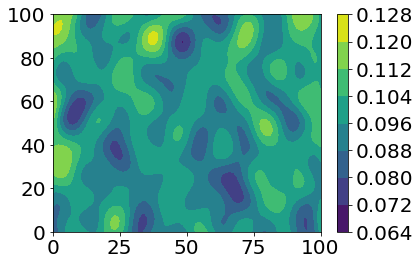}
		\centering
		\caption{A sample of wind field experienced by the robots}
		\label{fig:wind}
	\end{subfigure}
	\hfil
	\begin{subfigure}[t]{0.235\textwidth}
		\includegraphics[width=1\textwidth]{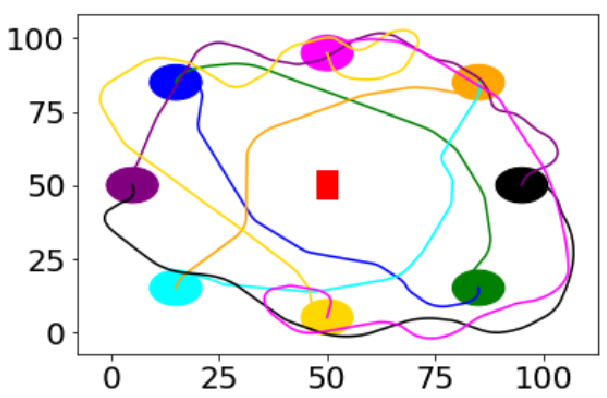}
		\caption{Trajectories of the robots}
		\label{fig:sample trajectories}
	\end{subfigure}
	\caption{A sample of wind fields and robot trajectories}
\end{figure}

{\em Random wind fields with different magnitudes.}
We randomly generate 2D spatial wind fields, with average speed $\nu$ in different ratios of the robots' speed, i.e., $\nu=r_wv$, $r_w>0$, and standard deviation 2\% of the robots' speed,   using  the Von Karman power spectral density function as described in \cite{cole2013impact}. This wind model is used to test multi-robot navigation in  \cite{cole2013impact} \cite{cole2018reactive}.  A sample with $r_w=0.2$ is shown in 
Fig. \ref{fig:wind}. We randomly generate 60 different wind fields for each $r_w\in\{0.1, 0.2, \cdots,1\}$.

\subsection{Safe grid vs. safe region}
\begin{figure*}[t]
	\centering
	\includegraphics[width=0.5\textwidth]{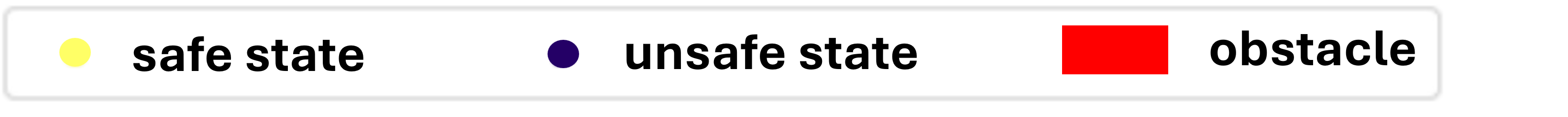}\\
	\begin{subfigure}[t]{0.235\textwidth}
		\centering
		\includegraphics[width=1\textwidth]{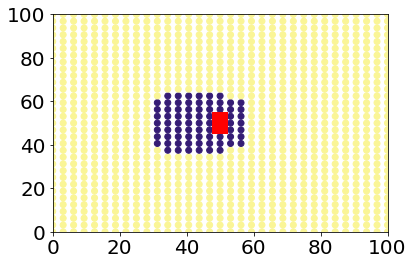}
		\caption{Grid: $\theta^{[i]} = 0$}
		\label{fig: grid theta 0}
	\end{subfigure}
	\hfil
		\begin{subfigure}[t]{0.235\textwidth}
		\centering
		\includegraphics[width=1\textwidth]{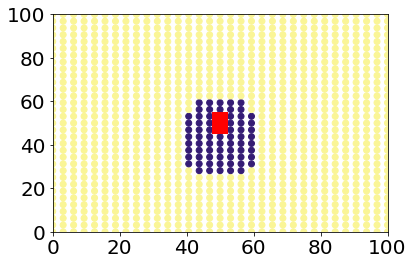}
		\caption{Grid: $\theta^{[i]} = \frac{\pi}{2}$}
	\label{fig: grid theta 90}
	\end{subfigure}
	\hfil
		\begin{subfigure}[t]{0.235\textwidth}
		\centering
		\includegraphics[width=1\textwidth]{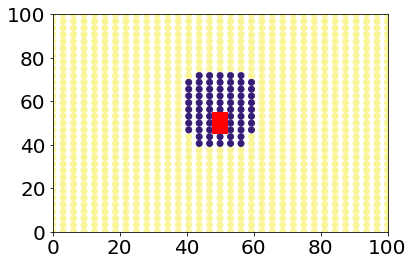}
		\caption{Grid: $\theta^{[i]} = -\frac{\pi}{2}$}
		\label{fig: grid theta -90}
	\end{subfigure}
	\hfil
	\begin{subfigure}[t]{0.235\textwidth}
		\centering
		\includegraphics[width=1\textwidth]{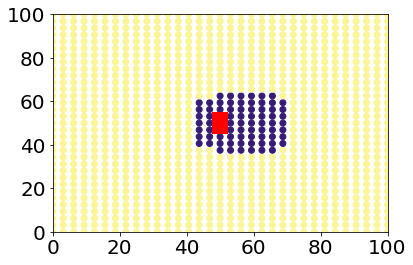}
		\caption{Grid: $\theta^{[i]} = \pi$}
	\label{fig: grid theta 180}
	\end{subfigure}
	\hfil
\begin{subfigure}[t]{0.235\textwidth}
	\centering
	\includegraphics[width=1\textwidth]{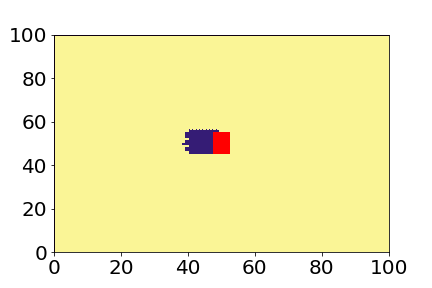}
	\caption{Region: $\theta^{[i]} = 0$}
	\label{fig: region theta 0}
\end{subfigure}
\hfil
\begin{subfigure}[t]{0.235\textwidth}
	\centering
	\includegraphics[width=1\textwidth]{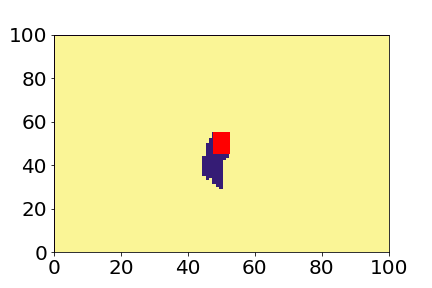}
	\caption{Region: $\theta^{[i]} = \frac{\pi}{2}$}
\label{fig: region theta 90}
\end{subfigure}
\hfil
\begin{subfigure}[t]{0.235\textwidth}
	\centering
	\includegraphics[width=1\textwidth]{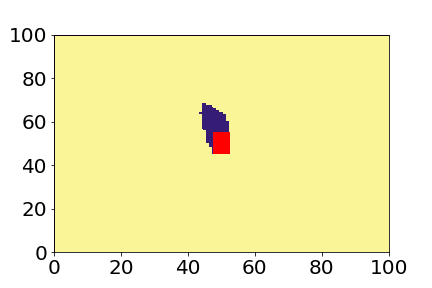}
	\caption{Region: $\theta^{[i]} = -\frac{\pi}{2}$}
\label{fig: region theta -90}
\end{subfigure}
\hfil
\begin{subfigure}[t]{0.235\textwidth}
	\centering
	\includegraphics[width=1\textwidth]{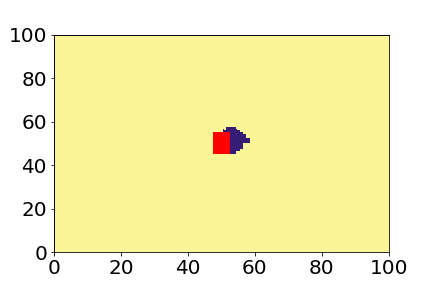}
	\caption{Region: $\theta^{[i]} = \pi$}
	\label{fig: region theta 180}
\end{subfigure}
	\caption{Safe grid computed by dSLAP vs. actual safe region}
	\label{fig: safe grid vs safe region}
\end{figure*}

To visualize how safety is guaranteed by dSLAP, in Fig. \ref{fig: safe grid vs safe region}, we compare the safe grids $\mathcal{X}^{[i]}_{\text{safe},k}$ (see Fig. \ref{fig: grid theta 0} - \ref{fig: grid theta 180}) computed by dSLAP under the wind field in  Fig. \ref{fig:wind} with  the corresponding  safe regions (see Fig. \ref{fig: region theta 0} - \ref{fig: region theta 180}), which are
the sets of initial states that render safe arrival by
applying the control policy returned. The comparison is similar for other wind fields. 
Since the dynamics of the robots has three dimensions,
for simplicity of visualization, we only show the screenshots of the 2D grids/regions with four different heading
angles (i.e., $\theta^{[i]}=0, \frac{\pi}{2} , -\frac{\pi}{2} $ and $\pi$). For the simplicity of illustration, we only compare the safe grids $\mathcal{X}^{[i]}_{\text{safe},k}$ of robot $i$,
 in the presence of only one obstacle, with the
corresponding safe regions. Since the figures are similar for grids with different resolutions, due to space limitation, we only show the comparison for safe grids $\mathcal{X}^{[i]}_{\text{safe},k}$ in one resolution, i.e., $p_k=5$.
In Fig. \ref{fig: region theta 0} - \ref{fig: region theta 180}, the safe regions are approximated by 10,000
evenly distributed initial states.

By comparing Fig. \ref{fig: grid theta 0} - \ref{fig: grid theta 180}  with  Fig. \ref{fig: region theta 0} - \ref{fig: region theta 180},
we can see that the safe grids  are strictly subsets of the safe regions, which verifies  Theorem \ref{thm: system safe} and Theorem \ref{thm2: system safe}.
Furthermore, from the comparison we can see that the identification of safe grids by dSLAP and the safety guarantees by Theorem \ref{thm: system safe} and Theorem \ref{thm2: system safe} are conservative. 
The conservativeness comes from the over-approximation of the one-step forward sets in $\textsf{FR}^{[i]}_k$, the limited knowledge of the dynamics and the errors in discretization. Nevertheless, the conservativeness can be reduced by refining the discretization  in the expense of more computation power.

\subsection{Multi-robot maneuver.}
We evaluate the dSLAP algorithm using 30,000 scenarios generated as follows.

{\em Different initial configurations.} 
We deploy $n$ robots with 10 different initial configurations in the simulation, where $n\in\{1,2,4,6,8\}$.  Fig. \ref{fig:sample trajectories} shows one configuration of 8 robots' initial states and goal regions, and the corresponding trajectories under dSLAP in the  wind field in Fig. \ref{fig:wind}. The circular disks are the goal regions of the robots and the red rectangle is the static obstacle. Other configurations are generated by different permutations and removals of the robots in Fig. \ref{fig:sample trajectories}.

{\em Ablation study.}
To the best of our knowledge, this paper is the first to consider multi-robot motion planning coupled with online learning. Hence, we compare dSLAP with its three variants, Vanilla, Robust and Known, that do not learn the wind disturbances. Vanilla assumes $g^{[i]}=0$, $\forall i\in\mathcal{V}$, whereas Robust assumes an upper bound on $g^{[i]}$: $\sup_{x^{[i]}\in\mathcal{X},u^{[i]}\in\mathcal{U}}|g^{[i]}(x^{[i]},u^{[i]})|\leqslant \hat{r}_wv$ and thus $\dot{x}^{[i]}\in f^{[i]}(x^{[i]},u^{[i]})+\hat{r}_wv\mathcal{B}$, where $\hat{r}_w>0$.  We adopt $\hat{r}_w=0.1$ such that Robust has the same level of conservativeness as dSLAP before collecting any data. 
The benchmark Known is obtained by running the dSLAP framework with the disturbances exactly known, which is the control law obtained by dSLAP when the amount of data of $g^{[i]}$ goes to infinity.

\begin{figure}[thpb]
	\centering
	\begin{subfigure}[t]{0.235\textwidth}
		\centering
		\includegraphics[width=1\textwidth]{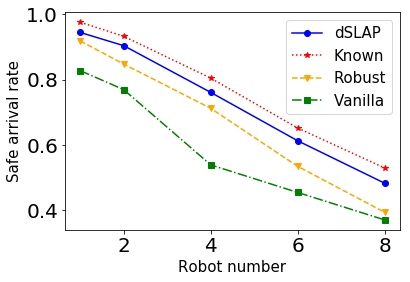}
		\caption{Percentage of safe arrivals}
		\label{fig:cumulative arrival}
	\end{subfigure}
	\hfil
	\begin{subfigure}[t]{0.235\textwidth}
		\centering
		\includegraphics[width=1\textwidth]{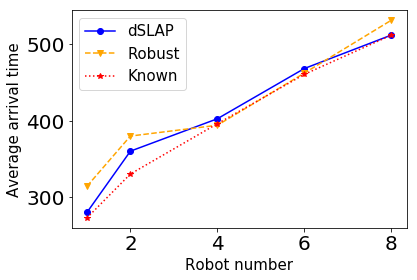}
		\caption{Average time of safe arrivals}
		\label{fig: average time of cumulative arrivals}
	\end{subfigure}
	\caption{Ablation study of dSLAP}
	\label{fig: ablation}
\end{figure}

{\em Results.}
The average safe arrival rates of dSLAP, Robust, Vanilla and Known among the 30,000 cases  are shown in Fig. \ref{fig: ablation}.   From Fig. \ref{fig:cumulative arrival}, we can see that dSLAP's safe arrival rate is superior to those of Robust and Vanilla. This is due to the fact that dSLAP online learns about the unknown disturbances and adjusts the policies accordingly. On the other hand, Robust (or Vanilla) only captures part of (or none of)  the disturbances through the prior estimates, which can be unsafe when the disturbances exceed the estimates. Furthermore, we can observe that the safe arrival rate for dSLAP decreases linearly with respect to the number of robots. This corresponds to the probability  with respect to the number of robots in Theorems \ref{thm: system safe} and \ref{thm2: system safe}. Notice that the gap between Known and dSLAP is small. The cases that are unsafe even in Known are due to the robots being too close to each other or the magnitude of the disturbances being too large to tolerate. Note that the magnitude of the disturbances can be as large as the speed of the robots in the simulation.  This indicates that dSLAP enables safe arrival in most feasible cases.

{\em Arrival time.}
 Fig. \ref{fig: average time of cumulative arrivals} compares the average safe arrival times among dSLAP, Robust and Known. We exclude the comparison with Vanilla since its safe arrival rate is far lower than the other three while safety is this paper's top priority. The arrival times of the three algorithms are comparable. This indicates  dSLAP improves safe arrival rate without sacrificing arrival time, i.e., being more conservative.

\begin{table*}[t]
\begin{center}
\begin{tabular}{ | p{1em} | p{4em}| p{6em} |p{5em} |p{4em} |p{5em} |p{7em} | p{6em} | p{4em} | p{5em} |  } 
\hline
\multirow{2}{*}{ID}& \multirow{2}{*}{Total time}& \multicolumn{2}{|c|}{\textsf{SL}} & \multicolumn{2}{|c|}{\textsf{Discrete}+\textsf{OCA}} &\multicolumn{2}{|c|}{\textsf{ICA} }& \multicolumn{2}{|c|}{\textsf{AL}} \\ 
\cline{3-10}
& & time& Percentage&  time& Percentage& time & Percentage&time & Percentage\\
\hline
1 & 5.71$\pm$0.45  & 0.84$\pm$7.87$e^{-3}$&14.73$\pm$1.08&4.26$\pm$0.12&74.91$\pm$3.95&6.03$e^{-3}\pm$1.85$e^{-3}$ &0.11$\pm$9.44$e^{-3}$&0.61$\pm$0.32&10.26$\pm$5.07 \\ 
\hline
2 & 5.93$\pm$0.47 & 0.81$\pm$0.01&13.70$\pm$1.03&4.21$\pm$0.06&71.32$\pm$5.10&0.06$\pm$0.03&1.08$\pm$0.59&0.85$\pm$0.41&13.90$\pm$6.38 \\ 
\hline
3&5.69$\pm$0.34&0.82$\pm$1.43$e^{-3}$&  14.48$\pm$0.88&4.14$\pm$0.04&73.10$\pm$3.90&0.14$\pm$0.01&2.56$\pm$0.32&0.58$\pm$0.32&9.87$\pm$5.08\\
\hline
4&5.18$\pm$0.14&0.82$\pm$2.03$e^{-3}$&15.85$\pm$0.43&4.01$\pm$0.04&77.37$\pm$1.98&0.16$\pm$0.09&3.07$\pm$1.76&0.20$\pm$0.15&3.70$\pm$2.81\\
\hline
5&5.90$\pm$0.35&0.82$\pm$0.01&13.96$\pm$0.65&4.30$\pm$0.08&73.10$\pm$3.41&0.23$\pm$0.03&3.96$\pm$0.51&0.54$\pm$0.27&8.98$\pm$4.10\\
\hline
6&5.66$\pm$1.29&0.88$\pm$0.16&15.7$\pm$0.65&
4.55$\pm$1.08&80.26$\pm$1.58&0.13$\pm$0.11&  2.57$\pm$2.23&0.10$\pm$0.13&1.46$\pm$1.64\\
\hline
7&5.94$\pm$0.99&0.88$\pm$0.10&14.96$\pm$0.82&4.49$\pm$0.65&75.85$\pm$2.69&0.19$\pm$0.10&3.39$\pm$2.03&0.38$\pm$0.38&5.79$\pm$5.19\\
\hline
8&5.94$\pm$1.14&0.88$\pm$0.13&14.90$\pm$0.55&4.65$\pm$0.83&78.53$\pm$1.30&0.29$\pm$0.07&5.02$\pm$1.62&0.12$\pm$0.22&1.56$\pm$2.45\\

\hline
\end{tabular}
\end{center}
\captionsetup{skip=0pt}
\caption{Computation time (seconds) for each robot in one iteration}
\label{table: computation}
\end{table*}
\subsection{Run-time computation.}
This section shows the wall computation time of dSLAP  when the robots are deployed in the wind field in Fig.  \ref{fig:wind} with the configuration in Fig. \ref{fig:sample trajectories}, as an example. Table \ref{table: computation} presents the average plus/minus one standard deviation of each robot's onboard computation time for one iteration for each component of dSLAP and the corresponding percentages (\%) of the total computation time.
\textsf{Discrete}+\textsf{OCA} consumes most of the computation resources because a discrete set-valued approximation of the continuous dynamics over the entire state-action space is constructed through these two procedures, especially in \textsf{OCA}.    Table \ref{table: computation} shows that the computation costs of the other procedures are mostly sub-second. 
Table \ref{table: wall clock} shows that the average wall time plus/minus one standard deviation per iteration for each robot versus the number of robots deployed. This shows that the computation time within each robot is nearly independent of the number of robots.

\subsection{Hyperparameter tuning}

The parameters in Algorithm \ref{alg:overall},  include mission and system parameters ($\mathcal{X}$, $\mathcal{U}$, $\mathcal{X}_O$,  ${\mathcal{X}^{[i]}_G}$, $\ell^{[i]}$ and $m^{[i]}$)  and tuned parameters (Kernel for GPR: $\kappa$;
Initial discretization parameter: $p_{init}$;
Termination iteration: $\tilde{k}$; 
Number of samples to be obtained in each iteration: $\bar{\tau}$; Discrete time unit: $\xi$; Time horizon for MPC: $\varphi$; Weight in the MPC: $\psi$; Sampling period: $\delta$; Utility function $r_k^{[i]}$). Below we provide an overall guidance on  tuning  these parameters. More detailed guidance can be found in the Appendix.

Parameters $p_{init}$,  $\tilde{k}$, $\bar{\tau}$, $\xi$, and $\varphi$, are referred as the computation parameters, since they are related to the computation power of the machine performing the simulation or the onboard computer of the robots.
In the simulation, these parameters, though can affect performance, determine how much computation power is needed to compute the safe control inputs. Therefore, they can be mainly tuned based on how much computation power is available and how much computation time is desired in one iteration. 
The remaining parameters, referred as the learning parameters, to be tuned are: $\kappa$, $\delta$, $r_k^{[i]}$ and  $\psi$.
Notice that the above parameters are more related to the learning of the unknown dynamics using GPR and active learning. Therefore, standard/common parameters in the related literature are used in the simulation. They can be tuned by following the general guidance of hyperparameter tuning for GPR \cite{williams2006gaussian} and active learning \cite{settles2009active}.

\begin{table}[t]
\begin{center}
\begin{tabular}{ | p{7.1em}| p{3em} |p{3em} |p{3em} |p{3em} |p{3em} |  } 
	\hline
	Number of robots&1&2&4&6&8\\
	\hline
	\multirow{2}{*}{Wall time }&$5.837\pm 0.085$ &$5.843\pm 0.118$&$5.830\pm 0.102$&$5.832\pm 0.129$&$5.839\pm 0.119$\\
	\hline
\end{tabular}
\end{center}
\captionsetup{skip=0pt}
\caption{dSLAP Wall clock time (seconds)  per iteration }
\label{table: wall clock}
\end{table}

\subsection{Comparison}
\begin{figure}[thpb]
	\centering
	\begin{subfigure}[t]{0.24\textwidth}
		\centering
		\includegraphics[width=1\textwidth]{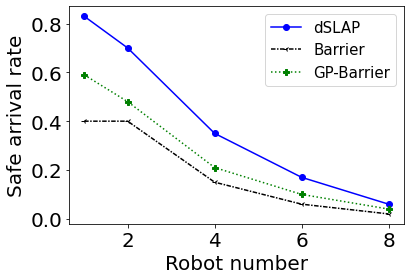}
		\caption{Percentage of safe arrivals}
		\label{fig:cumulative arrival barrier}
	\end{subfigure}
	\hfil
	\begin{subfigure}[t]{0.24\textwidth}
		\centering
		\includegraphics[width=1\textwidth]{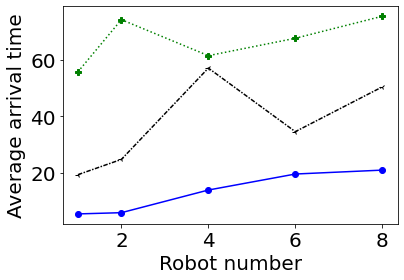}
		\caption{Average time of safe arrivals}
		\label{fig: average time of cumulative arrivals barrier}
	\end{subfigure}
	\caption{Comparison under double integrator dynamics}
	\label{fig: compare double int}
\end{figure}

\begin{figure}[thpb]
	\centering
	\begin{subfigure}[t]{0.24\textwidth}
		\centering
		\includegraphics[width=1\textwidth]{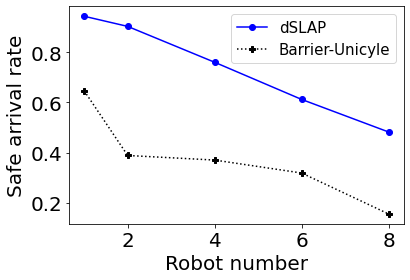}
		\caption{Percentage of safe arrivals}
		\label{fig:cumulative arrival barrier}
	\end{subfigure}
	\hfil
	\begin{subfigure}[t]{0.24\textwidth}
		\centering
		\includegraphics[width=1\textwidth]{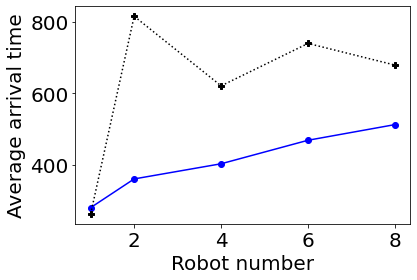}
		\caption{Average time of safe arrivals}
		\label{fig: average time of cumulative arrivals barrier}
	\end{subfigure}
	\caption{Comparison under unicycle dynamics}
	\label{fig: compare panagou}
\end{figure}

To demonstrate the effectiveness of our proposed algorithm, we compare dSLAP with related methods proposed in \cite{wang2017safety},\cite{panagou2015distributed} and \cite{cheng2020safe}.
To the best of our knowledge, this is
the first paper which studies safe online learning and planning of multi-robot systems, and
Table \ref{table: compare} compares the formulations between our work and the aforementioned works. Papers \cite{wang2017safety}\cite{panagou2015distributed}  do not consider uncertainties in dynamic systems. In the comparison, we directly apply their algorithms to the nominal systems (i.e., double integrator and unicycle robot, respectively) but ignore the disturbances.
Paper \cite{cheng2020safe} learns the external disturbance offline using GPR. In  the comparison, we modify the algorithm such that the disturbance is learned online. Specifically, each robot's GPR model is retrained in real time whenever new data is collected on its trajectory. When the new GPR model is being trained, the robot uses the old GPR model for controller synthesis. The experiment results are presented in  Fig. \ref{fig: compare double int} and Fig. \ref{fig: compare panagou} where we denote the algorithm proposed by \cite{wang2017safety} as \textsf{Barrier}, \cite{cheng2020safe} as \textsf{GP-Barrier} and \cite{panagou2015distributed} as \textsf{Barrier-Unicycle}.

In terms of safe arrival rate,  Fig. \ref{fig: compare double int} and Fig. \ref{fig: compare panagou} show that dSLAP outperforms both \textsf{Barrier} and \textsf{Barrier-Unicycle}. This is due to the fact that disturbances are not considered in Barrier and Barrier-Unicycle. In addition,
dSLAP also outperforms \textsf{GP-Barrier}. This can be due to the fact that the modified \textsf{GP-Barrier}'s online retraining of full GPR (for predictions of arbitrary points) can be too time consuming to update the estimates timely.  On the other
hand, the discretization of the state space and control space in dSLAP only requires predictions of the disturbances on a fixed set of points, which allows the adoption of recursive GPR  \cite{huber2014recursive} for online training
with constant computational complexity. 

In terms of average arrival times,	Fig. \ref{fig: compare double int} and Fig. \ref{fig: compare panagou} shows that dSLAP outperforms the three benchmarks.
Papers \cite{wang2017safety}\cite{cheng2020safe}\cite{panagou2015distributed} only consider keeping robots safe but do not consider steering the robots to goal regions. In contrast, our paper wants the robots to safely reach goal regions as soon as possible. This is considered in  objective function \eqref{opt: multi-objective2} in \textsf{AL}.

\begin{figure}[thpb]
	\centering
	\begin{subfigure}[t]{0.24\textwidth}
		\centering
		\includegraphics[width=1\textwidth]{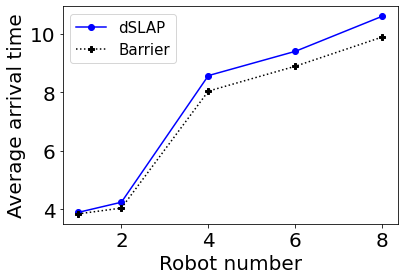}
		\caption{Comparison under double integrator dynamics}
		\label{fig: free disturbance double int}
	\end{subfigure}
	\hfil
	\begin{subfigure}[t]{0.24\textwidth}
		\centering
		\includegraphics[width=1\textwidth]{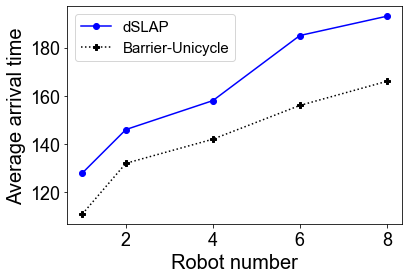}
		\caption{Comparison under unicycle dynamics}
		\label{fig: free disturbance unicycle}
	\end{subfigure}
	\caption{Comparison of average arrival time without disturbance}
	\label{fig: compare time}
\end{figure}

\begin{figure}[thpb]
	\centering
	\begin{subfigure}[t]{0.24\textwidth}
		\centering
		\includegraphics[width=1\textwidth]{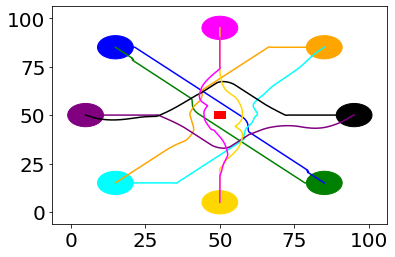}
		\caption{Robot trajectories under \textsf{Barrier}}
		\label{fig: tra free disturbance double int}
	\end{subfigure}
	\hfil
	\begin{subfigure}[t]{0.24\textwidth}
		\centering
		\includegraphics[width=1\textwidth]{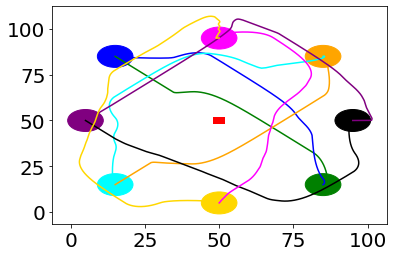}
		\caption{Robot trajectories under \textsf{dSLAP}}
		\label{fig: tra disturbance double int dslap}
	\end{subfigure}
	\caption{Comparison of robot trajectories with double integrator dynamics without disturbance}
	\label{fig: compare doubleInt trajectory}
\end{figure}

\begin{figure}[thpb]
	\centering
	\begin{subfigure}[t]{0.24\textwidth}
		\centering
		\includegraphics[width=1\textwidth]{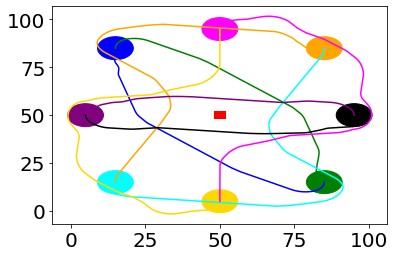}
		\caption{Robot trajectories under \textsf{Barrier-Unicycle}}
		\label{fig: tra free disturbance unicycle}
	\end{subfigure}
	\hfil
	\begin{subfigure}[t]{0.24\textwidth}
		\centering
		
		\includegraphics[width=1\textwidth]{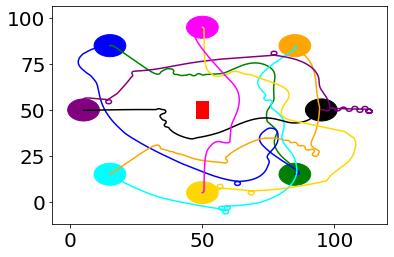}
		\caption{Robot trajectories under \textsf{dSLAP}}
		\label{fig: tra free disturbance unicycle dslap}
	\end{subfigure}
	\caption{Comparison of robot trajectories with unicycle dynamics without disturbance}
	\label{fig: compare unicycle trajectory}
\end{figure}

Fig. \ref{fig: compare time} compares the average arrival time among dSLAP, \textsf{Barrier} and \textsf{Barrier-Unicycle} when there is no disturbance.  Fig. \ref{fig: compare time} shows that dSLAP renders slightly longer arrival time than the two benchmarks. One reason can be that dSLAP is designed for general nonlinear dynamics (especially when there are unknown disturbances) while the other two benchmarks are designed for specific dynamics. dSLAP leverages less information and hence renders conservative solutions. This is a reminiscent of the no-free-lunch property \cite{dohmatob2019generalized}. The safe arrival rates for the three algorithms are all 100\%, which aligns with the results in \cite{wang2017safety}\cite{panagou2015distributed} and a straightforward special case of our Theorem \ref{thm2: system safe} (see the discussion on probabilistic safety in Section \ref{sec: discussion}). Fig. \ref{fig: compare doubleInt trajectory} and Fig. \ref{fig: compare unicycle trajectory} show the corresponding trajectories.



\begin{table*}[t]
	\begin{center}
		\begin{tabular}{ | p{6em} || p{10em}| p{10em} |p{10em} |p{10em} |} 
			\hline
			Algorithms& dSLAP & Barrier \cite{wang2017safety} & GP-Barrier \cite{cheng2020safe} & Barrier-Unicycle \cite{panagou2015distributed}\\
			
			\hline
			Dynamics considered&General nonlinear dynamics&Double integrator &Double integrator& Unicycle\\
			\hline
			Uncertainty & Considered& Not considered&Considered &No considered\\
			\hline
			Learning&Online learning&N/A&Offline learning&N/A\\
			\hline
		\end{tabular}
	\end{center}
	\captionsetup{skip=0pt}
	\caption{Comparison on the capabilities of the dSLAP, Barrier \cite{wang2017safety}, GP-Barrier \cite{cheng2020safe} and Barrier-Unicycle \cite{panagou2015distributed}}
	\label{table: compare}
\end{table*}

\section{Conclusion and future work}
We study the problem where a group of mobile robots subject to unknown external disturbances aim to safely reach goal regions. We propose the dSLAP algorithm that enables the robots to quickly adapt to a sequence of learned models resulting from online Gaussian process regression, and safely reach the goal regions. We provide sufficient conditions to ensure the safety of the system. The developed algorithm is evaluated by Monte Carlo simulations.

Exciting future works include mitigating the computation complexities for synthesizing new controllers in each iteration. Possible approaches include using adaptive gridding \cite{grune2004using} and designing a selective update scheme for updating $\textsf{FR}^{[i]}_k$.

\bibliographystyle{IEEEtran}
\bibliography{Biblio-dataset}
\vspace{-2em}
\vskip 0pt plus -1fil
\begin{IEEEbiography}[{\includegraphics[width=1in,height=1.25in,clip,keepaspectratio]{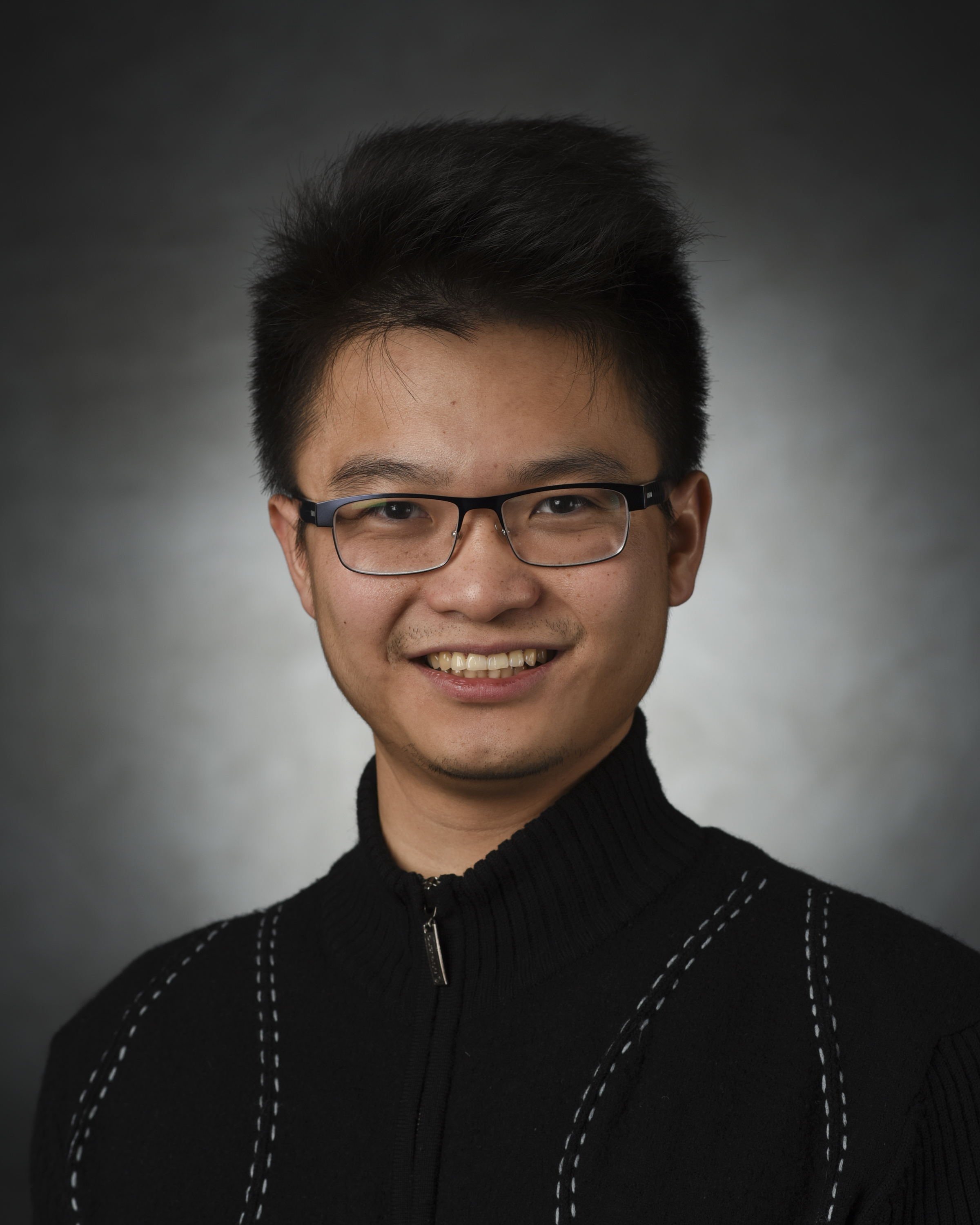}}]{Zhenyuan Yuan}
is  a  Advanced Analytics \& Machine Learning Researcher in the Virginia Tech Transportation Institute at Virginia Polytechnic Institute and State University (Virginia Tech). Prior to that, he was a postdoctoral associate in the Bradley Department of Electrical and Computer Engineering at Virginia Tech. He received Ph.D. in  Electrical  Engineering  and dual B.S. in Electrical Engineering and in Mathematics from the  Pennsylvania  State  University  in 2024 and 2018, respectively. His research interests lie in trustworthy machine learning and motion planning with applications in multi-robot systems. He is a recipient of the Rudolf Kalman Best Paper Award of  the ASME Journal of Dynamic Systems Measurement and Control in 2019 and
the Penn State Alumni Association Scholarship for Penn State Alumni in the Graduate School in 2021.
\end{IEEEbiography}
\vspace{-4em}
\begin{IEEEbiography}[{\includegraphics[width=1in,height=1.25in,clip,keepaspectratio]{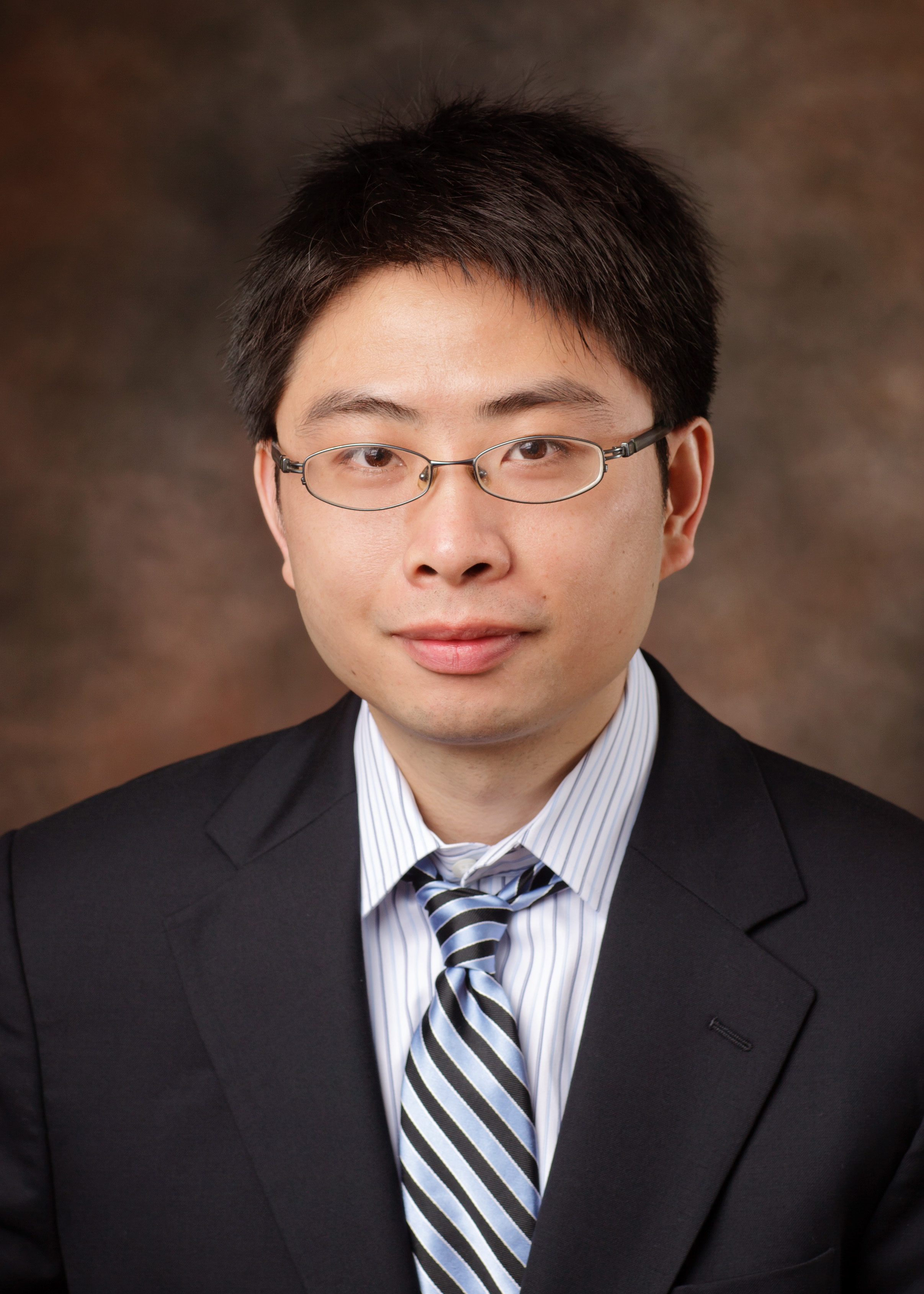}}]{Minghui Zhu}
is a Professor in the School of Electrical Engineering and Computer Science at the Pennsylvania State University. Prior to joining Penn State in 2013, he was a postdoctoral associate in the Laboratory for Information and Decision Systems at the Massachusetts Institute of Technology. He received Ph.D. in Engineering Science (Mechanical Engineering) from the University of California, San Diego in 2011. He was a joint appointee as Senior Engineer in the Optimization and Controls Department at the Pacific Northwest National Laboratory. His research interests lie in distributed control and decision-making of multi-agent networks with applications in robotic networks, security and the smart grid. He is an author of the book “Distributed optimization-based control of multi-agent networks in complex environments” (Springer, 2015). He is a recipient of the Dorothy Quiggle Career Development Professorship in Engineering at Penn State in 2013, the award of Outstanding Reviewer of Automatica in 2013 and 2014, the National Science Foundation CAREER award in 2019 and the Superior Paper Award of the American Society of Agricultural and Biological Engineers in 2024. He is an associate editor of Automatica, the IEEE Transactions on Automatic Control, the IEEE Open Journal of Control Systems and the IET Cyber-systems and Robotics.
\end{IEEEbiography}

\appendix
\section{Hyperparameter tuning}\label{appdx: hyperparameter tuning}
 Regarding the tuning of the parameters, there are actually not too many parameters to be freely tuned for the sake of performance. 
As in Algorithm 1, the parameters, other than those related to the mission or the systems (i.e., State space: $\mathcal{X}$, Control input space: $\mathcal{U}$, Obstacle: $\mathcal{X}_O$, Goal: ${\mathcal{X}^{[i]}_G}, i\in\mathcal{V}$, Lipschitz constant $\ell^{[i]}$, Prior supremum of dynamic model $m^{[i]}$), needed to be tuned are: Kernel for GPR: $\kappa$;
Initial discretization parameter: $p_{init}$;
Termination iteration: $\tilde{k}$; 
Number of samples to be obtained: $\bar{\tau}$; Discrete time unit: $\xi$; Time horizon for MPC: $\varphi$; Weight in the MPC: $\psi$; Sampling period: $\delta$; Utility function $r_k^{[i]}$. Below we provide the guidance on tuning the parameters.

Computation parameters:
\begin{itemize}
	\item Initial discretization parameter $p_{init}$: The larger the discretization parameter, while provides tighter approximation of the one-step forward  sets, the more computation is needed to build up the graph of one-step forward set.
	\item Termination iteration $\tilde{k}$: The larger  termination iteration, the longer the simulation would run if neither arrival nor collision happens before.
	\item Number of samples to be obtained $\bar{\tau}$: While the larger the number of samples to be obtained in each iteration generally provide better learning results of the unknown dynamics, it also requires longer time to recursively train the GP model. 
	\item Discrete time unit $\xi$: The larger the discrete time unit, the long each iteration would last, which can make the safe planning in ICA more conservative. Therefore, each iteration should stop right after all the modules (i.e., SL, Discrete, OCA, ICA and AL)) in Fig. 1 are done computing. 
	\item Time horizon for MPC $\varphi$: The larger the time horizon for MPC, while improves the optimality of the solution, the more computation power is needed.
\end{itemize}

Learning parameters:
\begin{itemize}
	\item Kernel for GPR $\kappa$: In the simulation, we use a square-exponential kernel, which is one of the most common kernels used in GPR-related applications \cite{williams2006gaussian}. 
	\item  Sampling period $\delta$: This parameter is related to how to sample each robot's trajectory in each iteration to obtain data for training the GPR. With larger sampling period, the collected data would be more spatially separated over the trajectories of the robots. In the simulation, $\delta$ is chosen such that the data are temporally uniformly distributed over the trajectories of the robots.
	\item Utility function $r_k^{[i]}$: This determines provide guidance to the robots on where they should go to collect new data. In the simulation, we use predictive variance as the utility function, which is one of the most common utility function in active learning \cite{settles2009active}.
	\item Weight in the MPC $\psi$: The parameter determines the weights between the utility function for active learning and the distance-to-goal for arrival. It is rather not important as eventually as more weight is eventually put on distance-to-goal, according to the definition of $w[k]$. With higher $\psi$, the robots would explore for more iteration, which could lead to a better learned model of the unknown dynamics.
\end{itemize}

\end{document}